\documentclass[prr,twocolumn,superscriptaddress,hidepacs]{revtex4}
\usepackage{amsmath}
\usepackage{amsfonts}
\usepackage{amssymb}
\usepackage{graphicx}
\usepackage{pst-all}
\usepackage{multirow}

\begin{document}

\title{Measuring Topological Order}
\author{Parsa Bonderson}
\affiliation{Microsoft Station Q, Santa Barbara, California 93106-6105, USA}
\date{\today}

\begin{abstract}
The topological order of a $(2+1)$D topological phase of matter is characterized by its chiral central charge and a unitary modular tensor category that describes the universal fusion and braiding properties of its anyonic quasiparticles.
I discuss the topologically invariant quantities associated with these and identify ones that are useful for determining the topological order.
I propose a variety of physical experiments that probe these quantities and detail the relation of the measured data to the topological invariants.
\end{abstract}

\maketitle









%

\section{Introduction}
\label{sec:Intro}

Topological phases of matter may possess emergent quasiparticle excitations that exhibit exotic exchange statistics, such as anyons in $(2+1)$D systems~\cite{Leinaas77,Nayak08}. The topological order of a $(2+1)$D topological phase, i.e. the collection of universal properties associated with the phase, is understood to be fully characterized by the combination of: (1) the chiral central charge, which is associated with the chiral thermal transport of gapless edge modes, and (2) a unitary modular tensor category (UMTC)~\cite{Moore89b,Turaev94,Bakalov01}, which specifies the fusion and braiding properties of the quasiparticle excitations. These quantities are expected to be topological invariants of the phase, which are quantized with high precision and remain unchanged under continuous deformations of the system that do not close the spectral gap. As such, it is important to understand how to experimentally extract information about them from physical systems.

The chiral central charge $c_-$ of a topological phase may be measured via thermal transport experiments, as it is associated with a thermal Hall conductivity of $\kappa_{H} = \frac{\pi}{6}T c_{-}$~\cite{Kane97,Read00,Cappelli02}.
Such thermal Hall transport experiments have recently been performed for fractional quantum Hall states~\cite{Banerjee17,Banerjee18,Srivastav20} and $\alpha$-RuCl$_{3}$~\cite{Kasahara18,Yamashita20}.
While these studies affirm successful extractions of the chiral central charge for the states examined in these challenging experiments, efforts to understand their results~\cite{Mross17,Wang17,Simon18,Feldman_Comment,Simon_Reply,AharonSteinberg19,Ye18,Vinkler18,Ma18,Simon19,Asasi20} suggest the measured quantities may not all be universal and topologically quantized, which muddles their characterization of the bulk topological order.
Even if accurately determined, the chiral central charge is, by itself, not a particularly distinguishing characteristic of a topological phase, because any value of $c_-$ may always be associated with an infinite number of distinct topological orders.

In contrast, the UMTC associated with a topological phase contains significantly more detailed information that can very precisely identify the topological order.
In fact, it even contains $c_{-} \text{ mod }8$ [see Eq.~(\ref{eq:cmod8})], so nearly all of the information about the topological order is contained in the UMTC.
A consequence of this is that a larger variety of experiments is needed to extract this larger set of information embodied by the UMTC.
The ``basic data'' defining a UMTC, which consists of the so-called ``$F$-symbols'' and ``$R$-symbols,'' contain gauge freedom. Mathematically, it is important to determine the quantities associated with a UMTC that are gauge invariant and capable of precisely identifying the UMTC.
From the perspective of designing physical experiments, it is important to identify experiments probing these gauge invariants that both collect a complete (or near-complete) characterization of the UMTC, and which do so with methods that allow the invariants to be carefully separated from non-universal effects.
In particular, it is crucial for the experiments not to rely upon fine-tuning, precise knowledge of the microscopic Hamiltonian, or other unrealistic assumptions.
Prior efforts in these directions have been disjointed, with the mathematical side identifying complete UMTC invariants that are mostly experimentally unapproachable, and the physical side identifying experiments that narrowly access UMTC information about certain topological phases, often with methods that are not robust.
In this paper, I bridge this divide with a systematic treatment that advances both arms of the problem with the other in mind.

I begin, in Sec.~\ref{sec:UMTC}, by reviewing the UMTC formalism and discussing the topological gauge invariant quantities associated with a UMTC, particularly those that may be used to mathematically identify the topological order.
Then, I discuss experimental methods of determining the invariant UMTC data associated with a topological phase.
In Sec.~\ref{sec:Bulk}, I propose and analyze various robust experiments involving deterministic manipulation of localized bulk quasiparticles and topological charge measurements (which have probabilistic outcomes) in detail.
While the potential use of these types of physical operations for performing topological quantum computation has been explored in great detail~\cite{Kitaev03,Preskill98,Freedman02a,Freedman02b,Freedman03b,Mochon03,Mochon04,Bonesteel05,Hormozi07a,Nayak08,Bonderson08a,Bonderson08b,Bonderson12a,Kliuchnikov14,Levaillant15,Bocharov16,Karzig17,Tran19}, their use for experiments characterizing the topological order and identifying the UMTC has been largely neglected.
I find that a substantial portion of the UMTC data can be determined through relatively simple experiments of this type.
However, the characterization of topological order that may be obtained from such experiments is, in general, not complete.
Filling in the full details needed to completely determine the UMTC generally requires more challenging types of experiments.
In particular, experiments are needed that allow for more general superpositions of topological charges or quasiparticle trajectories, as these are needed to access the more elusive phase factors associated with a UMTC's basic data.
In Sec.~\ref{sec:Additional}, I discuss several classes of additional experiments (some of which were previously proposed) that could potentially be used to access this remaining information, and obstacles associated with each of them.
These include experiments that utilize edge modes, topological defects, and mapping class group transformations on higher genus surfaces.

It is worth remarking that most physically realized topological phases are actually ``quasi-topological phases''~\cite{Bonderson12d}, for which some of the expected topological properties lose their topological protection.
Properties that are particularly vulnerable to such effects include overall phases of transformations and ground state degeneracy on higher genus surfaces.
In this regard, it is expected that the properties measured by the bulk quasiparticle experiments of Sec.~\ref{sec:Bulk} will remain topologically-protected.

\section{UMTCs and Topological Invariants}
\label{sec:UMTC}

In this section, I review the mathematical structure of UMTCs and focus on the topological invariants associated with them. The UMTC formalism may be used to represent anyonic states and operators, which encodes the purely topological properties of quasiparticles in a topological phase, independent of any particular physical realization. For additional details, I refer the reader to Refs.~\onlinecite{Bonderson07b,Bonderson07c}.

A UMTC is defined by a set $\mathcal{C}$ of conserved quantum numbers called topological charge, fusion rules specifying what can result from combining or splitting topological charges, associativity of fusion on the state space, and braiding rules specifying what happens when the positions of objects carrying topological charge are exchanged. Each localized quasiparticle carries a definite value of topological charge. There is a unique ``vacuum'' charge, denoted $0$ (or $I$), for which fusion and braiding is trivial, and each charge $a$ has a unique conjugate $\bar{a}$ which can fuse with $a$ to give $0$.

The topological charges obey the UMTC's (commutative and associative) fusion algebra
\begin{equation}
a \times b = \sum_{c} N_{ab}^{c} c
,
\end{equation}
and where $N_{ab}^{c}$ are non-negative integers specifying the number of ways that topological charges $a$ and $b$ can combine to produce charge $c$. The properties of the vacuum charge require that $N_{ab}^{0} = \delta_{\bar{a}b}$ and $N_{a0}^{c}=N_{0a}^{c} = \delta_{ac}$.

These rules prescribe fusion/splitting Hilbert spaces $\mathcal{V}_{ab}^{c}$ and $\mathcal{V}^{ab}_{c}$ with $\dim(\mathcal{V}_{ab}^{c}) = \dim(\mathcal{V}^{ab}_{c}) = N_{ab}^{c}$, which generate the nonlocal state space through repeated fusion/splitting. A charge $a$ is non-Abelian if it does not have a unique fusion channel for every type of charge it is fused with, or, alternatively, if it has $\sum_{c} N^{c}_{aa} > 1$. It is clear that the dimension of the topological state space increases as one includes more non-Abelian anyons.

Diagrammatically, the orthonormal bra/ket vectors in the fusion/splitting spaces are represented by trivalent vertices:
\begin{eqnarray}
\left( d_{c} / d_{a}d_{b} \right) ^{1/4}
\pspicture[shift=-0.6](-0.1,-0.2)(1.5,-1.2)
  \small
  \psset{linewidth=0.9pt,linecolor=black,arrowscale=1.5,arrowinset=0.15}
  \psline{-<}(0.7,0)(0.7,-0.35)
  \psline(0.7,0)(0.7,-0.55)
  \psline(0.7,-0.55) (0.25,-1)
  \psline{-<}(0.7,-0.55)(0.35,-0.9)
  \psline(0.7,-0.55) (1.15,-1)	
  \psline{-<}(0.7,-0.55)(1.05,-0.9)
  \rput[tl]{0}(0.4,0){$c$}
  \rput[br]{0}(1.4,-0.95){$b$}
  \rput[bl]{0}(0,-0.95){$a$}
 \scriptsize
  \rput[bl]{0}(0.85,-0.5){$\mu$}
  \endpspicture
&=&\left\langle a,b;c,\mu \right| \in
\mathcal{V}_{ab}^{c} ,
\label{eq:bra}
\\
\left( d_{c} / d_{a}d_{b}\right) ^{1/4}
\pspicture[shift=-0.65](-0.1,-0.2)(1.5,1.2)
  \small
  \psset{linewidth=0.9pt,linecolor=black,arrowscale=1.5,arrowinset=0.15}
  \psline{->}(0.7,0)(0.7,0.45)
  \psline(0.7,0)(0.7,0.55)
  \psline(0.7,0.55) (0.25,1)
  \psline{->}(0.7,0.55)(0.3,0.95)
  \psline(0.7,0.55) (1.15,1)	
  \psline{->}(0.7,0.55)(1.1,0.95)
  \rput[bl]{0}(0.4,0){$c$}
  \rput[br]{0}(1.4,0.8){$b$}
  \rput[bl]{0}(0,0.8){$a$}
 \scriptsize
  \rput[bl]{0}(0.85,0.35){$\mu$}
  \endpspicture
&=&\left| a,b;c,\mu \right\rangle \in
\mathcal{V}_{c}^{ab},
\label{eq:ket}
\end{eqnarray}
where $\mu = 1, \ldots, N_{ab}^{c}$. The normalization factors involving $d_{a}$, the quantum dimension of the charge $a$, are included so that diagrams are in the isotopy-invariant convention. States and operators involving multiple anyons are constructed by appropriately stacking together diagrams, making sure to conserve charge when connecting endpoints of lines.

In this way, the projection operator of two anyons with topological charges $a_1$ and $a_2$, respectively, onto collective topological charge $c$ is written as
\begin{equation}
\Pi^{(a_1 a_2)}_{c} = \sum_{\mu} \sqrt{\frac{d_{c}} {d_{a_1} d_{a_2} }}
\pspicture[shift=-1.1](-0.1,-0.85)(1.6,1.3)
 \small
  \psset{linewidth=0.9pt,linecolor=black,arrowscale=1.5,arrowinset=0.15}
  \psline{->}(0.7,0)(0.7,0.45)
  \psline(0.7,0)(0.7,0.55)
  \psline(0.7,0.55) (0.2,1.05)
  \psline{->}(0.7,0.55)(0.3,0.95)
  \psline(0.7,0.55) (1.2,1.05)
  \psline{->}(0.7,0.55)(1.1,0.95)
  \rput[bl]{0}(0.88,0.2){$c$}
  \rput[bl]{0}(1.1,1.15){$a_{2}$}
  \rput[bl]{0}(0.1,1.15){$a_{1}$}
  \psline(0.7,0) (0.2,-0.5)
  \psline{-<}(0.7,0)(0.35,-0.35)
  \psline(0.7,0) (1.2,-0.5)
  \psline{-<}(0.7,0)(1.05,-0.35)
  \rput[bl]{0}(1.1,-0.8){$a_{2}$}
  \rput[bl]{0}(0.1,-0.8){$a_{1}$}
 \scriptsize
  \rput[bl]{0}(0.4,0.0){$\mu$}
  \rput[bl]{0}(0.4,0.4){$\mu$}
  \endpspicture
.
\end{equation}
When an operator acts on only a subset of all the anyons, it implicitly means that it acts trivially on the other anyons, e.g. $\Pi^{(12)}_{c}$ really means $\Pi^{(12)}_{c} \otimes \openone^{(3 \ldots n)}$ when there are $n$ anyons.

The projection of three anyons with topological charges $a_1$, $a_2$, and $a_3$, onto collective topological charge $c$ is given by
\begin{equation}
\Pi^{(a_1 a_2 a_3)}_{c} = \sum_{b, \mu, \nu} \sqrt{\frac{d_{c}} {d_{a_1} d_{a_2} d_{a_3} }}
\scalebox{.8}{
\pspicture[shift=-1.7](-1.2,-1.5)(1.2,1.8)
  \small
  \psset{linewidth=0.9pt,linecolor=black,arrowscale=1.5,arrowinset=0.15}
  \psline(0.0,0.5)(0.8,1.5)
  \psline(0.0,0.5)(-0.8,1.5)
  \psline(-0.4,1)(0.0,1.5)
    \psline{->}(-0.4,1)(-0.1,1.375)
    \psline{->}(0.4,1.0)(0.7,1.375)
    \psline{->}(0,0.5)(-0.3,0.875)
    \psline{->}(0,0.5)(-0.7,1.375)
  \psset{linewidth=0.9pt,linecolor=black,arrowscale=1.5,arrowinset=0.15}
  \psline(0.0,0.0)(0.8,-1)
  \psline(0.0,0.0)(-0.8,-1)
  \psline(-0.4,-0.5)(0.0,-1)
    \psline{-<}(-0.4,-0.5)(-0.1,-0.875)
    \psline{-<}(0,0.0)(0.7,-0.875)
    \psline{-<}(0,0.0)(-0.3,-0.375)
    \psline{-<}(0,0.0)(-0.7,-0.875)
  \psline(0,0.0)(0,0.5)
  \psline{->}(0,0.0)(0,0.4)
  \rput[bl]{0}(0.15,0.15){$c$}
  \rput[bl]{0}(-0.45,0.5){$b$}
  \rput[bl]{0}(-0.95,1.6){$a_{1}$}
  \rput[bl]{0}(-0.1,1.6){$a_{2}$}
  \rput[bl]{0}(0.7,1.6){$a_{3}$}
  \rput[bl]{0}(-0.45,-0.2){$b$}
  \rput[bl]{0}(-0.95,-1.4){$a_{1}$}
  \rput[bl]{0}(-0.1,-1.4){$a_{2}$}
  \rput[bl]{0}(0.7,-1.4){$a_{3}$}
 \scriptsize
  \rput[bl]{0}(-0.2,-0.0){$\nu$}
  \rput[bl]{0}(-0.2,0.4){$\nu$}
  \rput[bl]{0}(-0.6,-0.45){$\mu$}
  \rput[bl]{0}(-0.6,0.85){$\mu$}
\endpspicture
}
.
\end{equation}
Similarly, the projection operator for $n$ anyons is given by
\begin{equation}
\Pi_{c}^{(a_1 \ldots a_n)} = \sum_{\substack{ e_{2},\ldots,e_{n-1} \\ \mu_{2},\ldots,\mu_{n} }} \sqrt{ \frac{d_{c}}{ d_{a_{1}} \cdots d_{a_{n}} } }
\psscalebox{.6}{
 \pspicture[shift=-2.05](-0.35,-1.9)(2.5,2.0)
  \small
  \psset{linewidth=0.9pt,linecolor=black,arrowscale=1.5,arrowinset=0.15}
  \psline(0.0,1.75)(1,0.5)
  \psline(2.0,1.75)(1,0.5)
  \psline(0.4,1.25)(0.8,1.75)
   \psline{->}(0.4,1.25)(0.1,1.625)
   \psline{->}(0.4,1.25)(0.7,1.625)
   \psline{->}(1,0.5)(1.9,1.625)
   \psline{->}(1,0.5)(0.5,1.125)
   \rput[bl]{0}(-0.15,1.85){$a_1$}
   \rput[bl]{0}(0.75,1.85){$a_2$}
   \rput[bl]{0}(1.95,1.85){$a_n$}
\rput[bl](1.25,1.85){$\cdots$}
\rput{-45}(0.9,1.05){$\cdots$}
   \rput[bl]{0}(0.25,0.65){$e_2$}
  \psset{linewidth=0.9pt,linecolor=black,arrowscale=1.5,arrowinset=0.15}
  \psline(0.0,-1.45)(1,-0.2)
  \psline(2.0,-1.45)(1,-0.2)
  \psline(0.4,-0.95)(0.8,-1.45)
   \psline{-<}(0.4,-0.95)(0.1,-1.325)
   \psline{-<}(0.4,-0.95)(0.7,-1.325)
   \psline{-<}(1,-0.2)(1.9,-1.325)
   \psline{-<}(1,-0.2)(0.5,-0.825)
   \rput[bl]{0}(-0.15,-1.75){$a_1$}
   \rput[bl]{0}(0.75,-1.75){$a_2$}
   \rput[bl]{0}(1.95,-1.75){$a_n$}
   \rput[bl]{0}(0.25,-0.6){$e_2$}
   \rput[bl](1.25,-1.75){$\cdots$}
   \rput{45}(0.9,-0.75){$\cdots$}
  \psline(1,-0.2)(1,0.5)
  \psline{->}(1,-0.2)(1,0.3)
   \rput[bl]{0}(1.15,0.0){$c$}
 \scriptsize
   \rput[bl]{0}(0.1,1){$\mu_2$}
   \rput[bl]{0}(0.05,-0.95){$\mu_2$}
   \rput[bl]{0}(0.55,0.35){$\mu_n$}
   \rput[bl]{0}(0.6,-0.2){$\mu_n$}
  \endpspicture
}
.
\label{eq:n_line_c_projector}
\end{equation}
The sum of projectors over all fusion channels is, of course, a partition of identity, i.e.
\begin{equation}
\sum_{c} \Pi^{(a_1 \ldots a_n)}_{c} = \openone^{(a_1 \ldots a_n)}
\end{equation}

Associativity of fusion in the state space is encoded by the unitary (change of fusion basis) isomorphisms
$F^{abc}_{d} : \bigoplus_{e} \mathcal{V}^{ab}_{e} \otimes \mathcal{V}^{ec}_{d} \rightarrow \bigoplus_{e} \mathcal{V}^{af}_{d} \otimes \mathcal{V}^{bc}_{f}$. These $F$-symbols are similar to the $6j$-symbols of angular momentum representations. Diagrammatically, these are written as
\begin{equation}
\scalebox{.8}{\pspicture[shift=-1.0](0,-0.45)(1.8,1.8)
  \small
  \psset{linewidth=0.9pt,linecolor=black,arrowscale=1.5,arrowinset=0.15}
  \psline(0.2,1.5)(1,0.5)
  \psline(1,0.5)(1,0)
  \psline(1.8,1.5) (1,0.5)
  \psline(0.6,1) (1,1.5)
   \psline{->}(0.6,1)(0.3,1.375)
   \psline{->}(0.6,1)(0.9,1.375)
   \psline{->}(1,0.5)(1.7,1.375)
   \psline{->}(1,0.5)(0.7,0.875)
   \psline{->}(1,0)(1,0.375)
   \rput[bl]{0}(0.05,1.6){$a$}
   \rput[bl]{0}(0.95,1.6){$b$}
   \rput[bl]{0}(1.75,1.6){${c}$}
   \rput[bl]{0}(0.5,0.5){$e$}
   \rput[bl]{0}(0.9,-0.3){$d$}
 \scriptsize
   \rput[bl]{0}(0.3,0.8){$\alpha$}
   \rput[bl]{0}(0.7,0.25){$\beta$}
  \endpspicture
}
= \sum_{f,\mu,\nu} \left[F_d^{abc}\right]_{(e,\alpha,\beta)(f,\mu,\nu)}
\scalebox{.8}{\pspicture[shift=-1.0](0,-0.45)(1.8,1.8)
  \small
  \psset{linewidth=0.9pt,linecolor=black,arrowscale=1.5,arrowinset=0.15}
  \psline(0.2,1.5)(1,0.5)
  \psline(1,0.5)(1,0)
  \psline(1.8,1.5) (1,0.5)
  \psline(1.4,1) (1,1.5)
   \psline{->}(0.6,1)(0.3,1.375)
   \psline{->}(1.4,1)(1.1,1.375)
   \psline{->}(1,0.5)(1.7,1.375)
   \psline{->}(1,0.5)(1.3,0.875)
   \psline{->}(1,0)(1,0.375)
   \rput[bl]{0}(0.05,1.6){$a$}
   \rput[bl]{0}(0.95,1.6){$b$}
   \rput[bl]{0}(1.75,1.6){${c}$}
   \rput[bl]{0}(1.25,0.45){$f$}
   \rput[bl]{0}(0.9,-0.3){$d$}
 \scriptsize
   \rput[bl]{0}(1.5,0.8){$\mu$}
   \rput[bl]{0}(0.7,0.25){$\nu$}
  \endpspicture
}
.
\end{equation}
If any of $a$, $b$, or $c$ is equal to $0$, then $F_{d}^{abc} = \openone$ (when allowed by the fusion rules), indicating trivial fusion of the vacuum charge.
The $F$-moves can be viewed as changes of bases for the states associated with quasiparticles. To describe topological phases, these are required to be unitary transformations, i.e.
\begin{eqnarray}
\left[ \left( F_{d}^{abc}\right) ^{-1}\right] _{\left( f,\mu
,\nu \right) \left( e,\alpha ,\beta \right) }
&=& \left[ \left( F_{d}^{abc}\right) ^{\dagger }\right]
_{\left( f,\mu ,\nu \right) \left( e,\alpha ,\beta \right) }
\notag \\
&=& \left[ F_{d}^{abc}\right] _{\left( e,\alpha ,\beta \right) \left( f,\mu
,\nu \right) }^{\ast }
.
\end{eqnarray}%

The counterclockwise braiding exchange operator of two anyons is represented diagrammatically by
\begin{equation}
R^{ab}=
\pspicture[shift=-0.55](-0.1,-0.2)(1.3,1.05)
\small
  \psset{linewidth=0.9pt,linecolor=black,arrowscale=1.5,arrowinset=0.15}
  \psline(0.96,0.05)(0.2,1)
  \psline{->}(0.96,0.05)(0.28,0.9)
  \psline(0.24,0.05)(1,1)
  \psline[border=2pt]{->}(0.24,0.05)(0.92,0.9)
  \rput[bl]{0}(-0.02,0.8){$a$}
  \rput[br]{0}(1.2,0.8){$b$}
  \endpspicture
=\sum\limits_{c,\mu ,\nu }\sqrt{\frac{d_{c}}{d_{a}d_{b}}}\left[
R_{c}^{ab}\right] _{\mu \nu }
\pspicture[shift=-0.7](-0.1,-0.6)(1.2,1)
  \small
  \psset{linewidth=0.9pt,linecolor=black,arrowscale=1.5,arrowinset=0.15}
  \psline{->}(0.7,0)(0.7,0.45)
  \psline(0.7,0)(0.7,0.55)
  \psline(0.7,0.55) (0.25,1)
  \psline{->}(0.7,0.55)(0.3,0.95)
  \psline(0.7,0.55) (1.15,1)
  \psline{->}(0.7,0.55)(1.1,0.95)
  \rput[bl]{0}(0.38,0.2){$c$}
  \rput[br]{0}(1.4,0.8){$b$}
  \rput[bl]{0}(0,0.8){$a$}
  \psline(0.7,0) (0.25,-0.45)
  \psline{-<}(0.7,0)(0.35,-0.35)
  \psline(0.7,0) (1.15,-0.45)
  \psline{-<}(0.7,0)(1.05,-0.35)
  \rput[br]{0}(1.4,-0.4){$a$}
  \rput[bl]{0}(0,-0.4){$b$}
\scriptsize
  \rput[bl]{0}(0.85,0.4){$\nu$}
  \rput[bl]{0}(0.85,-0.03){$\mu$}
\endpspicture
,
\end{equation}
where the $R$-symbols are the maps $R^{ab}_{c} : V^{ba}_{c} \rightarrow V^{ab}_{c}$ that result from exchanging two anyons of charges $b$ and $a$, respectively, which are in the charge $c$ fusion channel. If either $a$ or $b$ is equal to $0$, then $R^{ab} = \openone$, indicating trivial braiding with the vacuum charge.

Similarly, the clockwise braiding exchange operator is
\begin{equation}
\left(R^{ab}\right)^{-1}=
\pspicture[shift=-0.55](-0.1,-0.2)(1.3,1.05)
\small
  \psset{linewidth=0.9pt,linecolor=black,arrowscale=1.5,arrowinset=0.15}
  \psline{->}(0.24,0.05)(0.92,0.9)
  \psline(0.24,0.05)(1,1)
  \psline(0.96,0.05)(0.2,1)
  \psline[border=2pt]{->}(0.96,0.05)(0.28,0.9)
  \rput[bl]{0}(-0.01,0.8){$b$}
  \rput[bl]{0}(1.06,0.8){$a$}
  \endpspicture
.
\end{equation}
Unitarity of the braiding operator, $\left(R^{ab}\right)^{-1}=\left(R^{ab}\right)^{\dagger}$, which is a necessary condition to describe topological phases, can be expressed in terms of the $R$-symbols as
\begin{equation}
\left[ \left( R_{c}^{ab}\right) ^{-1}\right] _{\nu
,\mu} = \left[ \left( R_{c}^{ab}\right) ^{\dagger}\right] _{\nu
,\mu} = \left[ R_{c}^{ab}\right]^{\ast}_{\mu
,\nu}
.
\end{equation}

Additional useful ways of representing braiding operations include
\begin{equation}
\pspicture[shift=-0.65](-0.1,-0.2)(1.5,1.2)
  \small
  \psset{linewidth=0.9pt,linecolor=black,arrowscale=1.5,arrowinset=0.15}
  \psline{->}(0.7,0)(0.7,0.43)
  \psline(0.7,0)(0.7,0.5)
 \psarc(0.8,0.6732051){0.2}{120}{240}
 \psarc(0.6,0.6732051){0.2}{-60}{35}
  \psline (0.6134,0.896410)(0.267,1.09641)
  \psline{->}(0.6134,0.896410)(0.35359,1.04641)
  \psline(0.7,0.846410) (1.1330,1.096410)	
  \psline{->}(0.7,0.846410)(1.04641,1.04641)
  \rput[bl]{0}(0.4,0){$c$}
  \rput[br]{0}(1.35,0.85){$b$}
  \rput[bl]{0}(0.05,0.85){$a$}
 \scriptsize
  \rput[bl]{0}(0.82,0.35){$\mu$}
  \endpspicture
=\sum\limits_{\nu }\left[ R_{c}^{ab}\right] _{\mu \nu}
\pspicture[shift=-0.65](-0.1,-0.2)(1.5,1.2)
  \small
  \psset{linewidth=0.9pt,linecolor=black,arrowscale=1.5,arrowinset=0.15}
  \psline{->}(0.7,0)(0.7,0.45)
  \psline(0.7,0)(0.7,0.55)
  \psline(0.7,0.55) (0.25,1)
  \psline{->}(0.7,0.55)(0.3,0.95)
  \psline(0.7,0.55) (1.15,1)	
  \psline{->}(0.7,0.55)(1.1,0.95)
  \rput[bl]{0}(0.4,0){$c$}
  \rput[br]{0}(1.4,0.8){$b$}
  \rput[bl]{0}(0,0.8){$a$}
 \scriptsize
  \rput[bl]{0}(0.82,0.37){$\nu$}
  \endpspicture
\end{equation}
and
\begin{equation}
\scalebox{.8}{
\pspicture[shift=-1.25](0,-0.45)(1.9,2.2)
  \small
  \psset{linewidth=0.9pt,linecolor=black,arrowscale=1.5,arrowinset=0.15}
  \psbezier(1.1,0.5)(1.45,1.25)(1.025,1.5) (1.25,2.0)
  \psbezier[border=2.2pt](1.8,2.0)(1.45,1.25)(1.025,1.5)(0.8,1.0)
  \psline(0.2,2.0)(1.4,0)
   \psline{->}(0.8,1)(0.32858,1.7857)
   \psline{->}(1.183,1.72)(1.18,1.82)
   \psline{->}(1.61,1.68)(1.7,1.7857)
   \psline{->}(1.1,0.5)(0.85,0.916667)
   \psline{->}(1.4,0)(1.2,0.3333)
   \rput[bl]{0}(0.05,2.1){$a$}
   \rput[bl]{0}(1.2,2.1){$b$}
   \rput[bl]{0}(1.75,2.1){${c}$}
   \rput[bl]{0}(0.65,0.5){$g$}
   \rput[bl]{0}(1.35,-0.3){$d$}
 \scriptsize
   \rput[bl]{0}(0.5,0.85){$\gamma$}
   \rput[bl]{0}(0.81,0.2){$\lambda$}
  \endpspicture
}
=
\sum\limits_{ e,\alpha ,\beta }\left[
B_{d}^{abc}\right]
_{\left( g, \gamma, \lambda \right) \left( e,\alpha ,\beta \right) }
\scalebox{.8}{
\pspicture[shift=-1.25](0,-0.45)(1.9,2.2)
  \small
  \psset{linewidth=0.9pt,linecolor=black,arrowscale=1.5,arrowinset=0.15}
  \psline(0.2,2.0)(1.4,0)
  \psline(1.8,2.0) (1.1,0.5)
  \psline(0.8,1.0) (1.25,2.0)
   \psline{->}(0.8,1)(0.32858,1.7857)
   \psline{->}(0.8,1)(1.153565,1.7857)
   \psline{->}(1.1,0.5)(1.7,1.7857)
   \psline{->}(1.1,0.5)(0.85,0.916667)
   \psline{->}(1.4,0)(1.2,0.3333)
   \rput[bl]{0}(0.05,2.1){$a$}
   \rput[bl]{0}(1.2,2.1){$b$}
   \rput[bl]{0}(1.75,2.1){${c}$}
   \rput[bl]{0}(0.6,0.45){$e$}
   \rput[bl]{0}(1.35,-0.3){$d$}
 \scriptsize
   \rput[bl]{0}(0.5,0.85){$\alpha$}
   \rput[bl]{0}(0.85,0.25){$\beta$}
  \endpspicture
}
,
\end{equation}
where
\begin{align}
\left[B_d^{abc}\right]_{(g,\gamma, \lambda)(e,\alpha,\beta)} &=
\sum_{f,\mu,\mu',\nu} \left[F_d^{acb}\right]_{(g,\gamma, \lambda)(f,\mu,\nu)} \notag \\
& \times [R^{bc}_{f}]_{\mu \mu'} \left[(F_d^{abc})^{-1}\right]_{(f,\mu',\nu)(e,\alpha,\beta)}
.
\end{align}

The $F$-symbols and $R$-symbols are subject to consistency conditions~\cite{MacLane98}, known as the pentagon and hexagon equations. The pentagon equation ensures that applying different sequences of $F$-moves that start and end in the same topological configuration yield the same results. The hexagon equations ensure braiding is compatible with fusion, i.e. fusing two anyons and then braiding a third with their fusion is equivalent to first braiding the third anyon around the two anyons and then fusing the first two. Physically, these consistency conditions can be interpreted as ensuring the theory respects locality.

So far, this defines a unitary braided tensor category (UBTC); the ``modularity'' condition that makes it a UMTC is an additional non-degeneracy condition on the braiding, which can be stated as: $R^{ab}R^{ba} = \openone$ for all $b$ only if $a=0$.
Modularity will not play an essential role in identifying or measuring topological invariants, but it will allow some of them to be conveniently expressed as link invariants.
In fact, one should be careful about depending on modularity when interpreting experimental results, because one might be inadvertently restricting to a non-modular sub-theory if the experiment unknowingly accesses only a subset of the topological charge types.

The quantities $\mathcal{C}$, $N_{ab}^{c}$, $[F^{abc}_{d}]_{(e,\alpha, \beta)(f, \mu , \nu)}$, and $[R^{ab}_{c}]_{\mu \nu}$, collectively referred to as the ``basic data,'' fully define a UMTC (in mathematical parlance, they provide the ``skeletonization'' of a UMTC). However, there is gauge freedom in the $F$-symbols and $R$-symbols associated with the choice of vertex basis states. Distinct sets of $F$-symbols and $R$-symbols are considered gauge equivalent, and hence describe the same UMTC, if they can be related by unitary transformations (changes of basis) acting on the fusion/splitting state spaces $V_{c}^{ab}$ and $V^{c}_{ab}$ as
\begin{equation}
\label{eq:gauge}
\widetilde{ \left| a,b;c,\mu \right\rangle } = \sum_{\mu'} \left[\Gamma^{ab}_{c}\right]_{\mu \mu'} \left| a,b;c,\mu' \right\rangle
\end{equation}
where $\Gamma^{ab}_{c}$ is the unitary transformation. Such gauge transformations modify the $F$-symbols and $R$-symbols as
\begin{align}
&\left[\widetilde{F}_d^{abc}\right]_{(e,\alpha,\beta)(f,\mu,\nu)} = \sum_{\alpha', \beta', \mu' ,\nu'} \left[\Gamma^{ab}_{e}\right]_{\alpha \alpha'} \left[\Gamma^{ec}_{d}\right]_{\beta \beta'} \notag\\
& \times \left[F_d^{abc}\right]_{(e,\alpha',\beta')(f,\mu',\nu')} \left[\left(\Gamma^{bc}_{f} \right)^{-1} \right]_{\mu' \mu}  \left[ (\Gamma^{af}_{d} )^{-1} \right]_{\nu' \nu}
\label{eq:gauge_F}
\end{align}
and
\begin{equation}
\label{eq:gauge_R}
\left[ \widetilde{R}_{c}^{ab} \right]_{\mu \nu} = \sum_{\mu',\nu'} \left[\Gamma^{ba}_{c} \right]_{\mu \mu'}  \left[ R_{c}^{ab} \right]_{\mu' \nu'} \left[\left( \Gamma^{ab}_{c} \right)^{-1} \right]_{\nu' \nu}
.
\end{equation}
One must be careful not to use the gauge freedom associated with $\Gamma^{a 0}_{a}$ and $\Gamma^{0 b}_{b}$ to ensure that fusion and braiding with the vacuum $0$ remain trivial. More specifically, one should fix $\Gamma^{a 0}_{a}=\Gamma^{0 b}_{b}=\Gamma^{0 0}_{0}$. (One can think of this as respecting the canonical isomorphisms that allow one to freely add and remove vacuum lines. Alternatively, one could allow the use of these gauge factors and compensate by similarly modifying the canonical isomorphisms.)

A property known as ``Ocneanu rigidity'' states that, apart from gauge equivalences, the solutions of the pentagon and hexagon consistency equations cannot be continuously deformed into each other, and this implies there are a finite number of distinct (gauge equivalence classes of) UBTCS  for a specific fusion algebra~\cite{Etingof05}.

In order to distinguish between gauge equivalence classes of UMTCS, it is useful to consider quantities that are invariant under such gauge transformation. These topological invariants constitute the observables associated with the UMTC. The most relevant gauge invariant quantities are the fusion coefficients $N_{ab}^{c}$; the quantum dimensions
\begin{equation}
\label{eq:qdim}
d_{a}= \left| [F^{a \bar{a} a}_a]_{00} \right|^{-1} =
\pspicture[shift=-0.35](-0.08,0.25)(1.55,1.25)
  \small
  \psarc[linewidth=0.9pt,linecolor=black,arrows=<-,arrowscale=1.5,
    arrowinset=0.15] (0.8,0.7){0.5}{165}{363}
  \psarc[linewidth=0.9pt,linecolor=black,border=0pt]
(0.8,0.7){0.5}{0}{170}
  \rput[bl]{0}(-0.03,0.55){$a$}
 \endpspicture
,
\end{equation}
which are actually equal to the largest eigenvalue of the fusion matrices $N_{a}$, defined as $[N_{a}]_{bc} = N_{ab}^{c}$; the total quantum dimension
\begin{equation}
\mathcal{D} = \sqrt{ \sum_{a} d_{a}^{2}}
;
\end{equation}
and topological twist factors
\begin{equation}
\theta _{a}= \sum\limits_{c,\mu } \frac{d_{c}}{d_{a}}\left[ R_{c}^{aa}\right] _{\mu \mu }
= \frac{1}{d_{a}}
\pspicture[shift=-0.5](-1.3,-0.6)(1.3,0.6)
\small
  \psset{linewidth=0.9pt,linecolor=black,arrowscale=1.5,arrowinset=0.15}
  \psarc[linewidth=0.9pt,linecolor=black] (0.7071,0.0){0.5}{-135}{135}
  \psarc[linewidth=0.9pt,linecolor=black] (-0.7071,0.0){0.5}{45}{315}
  \psline(-0.3536,0.3536)(0.3536,-0.3536)
  \psline[border=2.3pt](-0.3536,-0.3536)(0.3536,0.3536)
  \psline[border=2.3pt]{->}(-0.3536,-0.3536)(0.0,0.0)
  \rput[bl]{0}(-0.2,-0.5){$a$}
  \endpspicture
,
\end{equation}
which are roots of unity.
I note that pure braiding operations can be expressed in terms the topological twist factors through the ``ribbon property''
\begin{equation}
R^{ab}_{c}R^{ba}_{c} = \frac{\theta_{c}}{\theta_{a} \theta_{b}} \openone
.
\end{equation}

For a UMTC, these invariants are equivalent to the ``modular data,'' i.e. the $S$-matrix
\begin{equation}
S_{ab} = \frac{1}{\mathcal{D}} \sum_{c} N_{\bar{a} b }^{c} d_{c} \frac{\theta_{c}}{\theta_{a}\theta_{b}}  = \frac{1}{\mathcal{D}}
\pspicture[shift=-0.4](0.0,0.2)(2.6,1.3)
\small
  \psarc[linewidth=0.9pt,linecolor=black,arrows=<-,arrowscale=1.5,arrowinset=0.15] (1.6,0.7){0.5}{167}{373}
  \psarc[linewidth=0.9pt,linecolor=black,border=3pt,arrows=<-,arrowscale=1.5,arrowinset=0.15] (0.9,0.7){0.5}{167}{373}
  \psarc[linewidth=0.9pt,linecolor=black] (0.9,0.7){0.5}{0}{180}
  \psarc[linewidth=0.9pt,linecolor=black,border=3pt] (1.6,0.7){0.5}{45}{150}
  \psarc[linewidth=0.9pt,linecolor=black] (1.6,0.7){0.5}{0}{50}
  \psarc[linewidth=0.9pt,linecolor=black] (1.6,0.7){0.5}{145}{180}
  \rput[bl]{0}(0.1,0.6){$a$}
  \rput[bl]{0}(0.85,0.6){$b$}
  \endpspicture
.
\label{eqn:mtcs}
\end{equation}
and $T$-matrix
\begin{equation}
T_{ab} = \theta_{a} \delta_{ab}
,
\end{equation}
which provide (projective) representations of the modular transformations of the theory on a torus.
The chiral central charge is related to the data of a UMTC through the relation
\begin{equation}
\label{eq:cmod8}
\frac{1}{\mathcal{D}} \sum_{a} \theta_{a} d_{a}^{2} = e^{i \frac{2 \pi}{8} c_{-}}
,
\end{equation}
so the UMTC contains $c_{-}(\text{mod }8)$.

While the modular data does not generally specify a UMTC completely~\cite{Mignard17}, they are often enough to uniquely specify them for cases of physical interest. Moreover, known examples of UMTCs that require more than the modular data to distinguish between them only require a small amount of additional data to do so~\cite{Bonderson19,Delaney18}. Indeed, the fusion rules alone typically narrow the possible UMTCs to a fairly small set of closely related theories.

The modularity condition, previously expressed as non-degeneracy of braiding, is equivalent to the condition that the $S$-matrix is unitary. When this holds, the relation to the fusion coefficients may be inverted using the Verlinde formula
\begin{equation}
N_{ab}^{c} = \sum_{x} \frac{ S_{ax} S_{bx} S_{cx}^{\ast} }{S_{0x}}
.
\end{equation}
Moreover, the modularity condition allows one to write the topological charge projection operators in terms of link diagrams by using the $S$-matrix to define ``$\omega_{c}$-loops'' as follows
\begin{equation}
\Pi_{c}^{(a_1 \ldots a_n)} =
\psscalebox{.65}{
 \pspicture[shift=-1.5](-0.7,-1.2)(2.8,2.1)
  \small
  \psset{linewidth=0.9pt,linecolor=black,arrowscale=1.5,arrowinset=0.15}
  \psline(0.0,1.75)(0.0,-1.0)
  \psline(2.0,1.75)(2.0,-1.0)
  \psline(0.8,1.75)(0.8,-1.0)
   \psline{->}(0.0,1.25)(0.0,1.625)
   \psline{->}(0.8,1.25)(0.8,1.625)
   \psline{->}(2.0,1.25)(2.0,1.625)
   \rput[bl]{0}(-0.2,1.85){$a_1$}
   \rput[bl]{0}(0.6,1.85){$a_2$}
   \rput[bl]{0}(1.25,1.85){$\ldots$}
   \rput[bl]{0}(1.8,1.85){$a_n$}
  \psellipse[linewidth=0.9pt,linecolor=black,border=0.1](1.0,0.5)(1.5,0.3)
  \psline[linewidth=0.9pt,linecolor=black,border=0.1](0.0,0.5)(0.0,1.0)
  \psline[linewidth=0.9pt,linecolor=black,border=0.1](0.8,0.5)(0.8,1.0)
  \psline[linewidth=0.9pt,linecolor=black,border=0.1](2.0,0.5)(2.0,1.0)
  \psset{linewidth=0.9pt,linecolor=black,arrowscale=1.4,arrowinset=0.15}
  \psline{->}(-0.3,0.37)(-0.4,0.4)
  \rput[bl]{0}(-0.6,0.1){$\omega_c$}
\endpspicture
}
= \sum_{x} S_{0x} S_{cx}^{\ast}
\psscalebox{.65}{
 \pspicture[shift=-1.5](-0.7,-1.2)(2.8,2.1)
  \small
  \psset{linewidth=0.9pt,linecolor=black,arrowscale=1.5,arrowinset=0.15}
  \psline(0.0,1.75)(0.0,-1.0)
  \psline(2.0,1.75)(2.0,-1.0)
  \psline(0.8,1.75)(0.8,-1.0)
   \psline{->}(0.0,1.25)(0.0,1.625)
   \psline{->}(0.8,1.25)(0.8,1.625)
   \psline{->}(2.0,1.25)(2.0,1.625)
   \rput[bl]{0}(-0.2,1.85){$a_1$}
   \rput[bl]{0}(0.6,1.85){$a_2$}
   \rput[bl]{0}(1.25,1.85){$\ldots$}
   \rput[bl]{0}(1.8,1.85){$a_n$}
  \psellipse[linewidth=0.9pt,linecolor=black,border=0.1](1.0,0.5)(1.5,0.3)
  \psline[linewidth=0.9pt,linecolor=black,border=0.1](0.0,0.5)(0.0,1.0)
  \psline[linewidth=0.9pt,linecolor=black,border=0.1](0.8,0.5)(0.8,1.0)
  \psline[linewidth=0.9pt,linecolor=black,border=0.1](2.0,0.5)(2.0,1.0)
  \psset{linewidth=0.9pt,linecolor=black,arrowscale=1.4,arrowinset=0.15}
  \psline{->}(-0.3,0.37)(-0.4,0.4)
  \rput[bl]{0}(-0.5,0.1){$x$}
\endpspicture
}
.
\label{eq:omega_a_projector}
\end{equation}
This will be useful for expressing topological invariants in terms of link diagrams.

The $R$-symbols are fully determined, up to gauge freedom, by the modular data (see, e.g. Ref.~\onlinecite{Bonderson19}). In particular, there is always a choice of gauge such that
\begin{equation}
[\widetilde{R}^{ab}_{c}]_{\mu \nu} = [\widetilde{R}^{ba}_{c}]_{\mu \nu} = \sqrt{\frac{\theta_{c}}{\theta_{a} \theta_{b}}} \delta_{\mu \nu}
,
\end{equation}
and
\begin{equation}
[\widetilde{R}^{aa}_{c}]_{\mu \nu} = \frac{\sqrt{\theta_{c}}}{\theta_{a}} [\Lambda^{aa}_{c}]_{\mu \nu}
,
\end{equation}
where $\Lambda^{aa}_{c}$ is a signature matrix, i.e. a diagonal matrix with values $\pm 1$ on the diagonal. The trace of $\Lambda^{aa}_{c}$ is gauge invariant and, for a UMTC, can be expressed in terms of the modular data, i.e.
\begin{equation}
\sum_{\mu } [\Lambda^{aa}_{c}]_{\mu \mu} = \frac{\sqrt{\theta_{c}}}{\mathcal{D}^{2}} \sum_{x,y,z} N_{xy}^{a} N_{xz}^{c} \frac{\theta_{y}^{2}}{\theta_{x} \theta_{z}} d_{y} d_{z}
.
\end{equation}

Thus, the topological invariants outside of the modular data that are necessary to form a completely specifying set of invariants must be contained in the $F$-symbols. It is known from geometric invariant theory that a fusion category (and hence the $F$-symbols of a MTC) is determined by a finite set of invariants~\cite{Hagge15}. However, these invariants are not given by closed-form expressions nor do they have a clear relation to familiar physical quantities.

On the other hand, there are topological invariants contained in the $F$-symbols that are simple to write and can even be expressed in terms of link invariants. This includes the gauge invariant quantities
\begin{equation}
\label{eq:F_magnitude_invariant}
\sum_{\substack{\alpha, \beta \\ \mu,\nu}} \left| \left[F_d^{abc}\right]_{(e,\alpha,\beta)(f,\mu,\nu)} \right|^{2} = \frac{1}{d_{d}} \!
\psscalebox{.65}{
 \pspicture[shift=-2.5](-0.9,-2)(4.2,3.0)
  \small
  \psellipse[linewidth=0.9pt,linecolor=black](2.0,2.25)(2.017,0.75)
  \psellipse[linewidth=0.9pt,linecolor=black](2.0,2.25)(1.216,0.5)
  \psellipse[linewidth=0.9pt,linecolor=black](2.0,2.25)(0.415,0.25)
  \psellipse[linewidth=0.9pt,linecolor=black](2.0,-1.0)(2.017,0.75)
  \psellipse[linewidth=0.9pt,linecolor=black](2.0,-1.0)(1.216,0.5)
  \psellipse[linewidth=0.9pt,linecolor=black](2.0,-1.0)(0.415,0.25)
  \psframe[fillstyle=solid,fillcolor=white,linecolor=white](-0.1,-0.99)(4.1,2.24)
  \psset{linewidth=0.9pt,linecolor=black,arrowscale=1.5,arrowinset=0.15}
  \psline(0.0,-1.0)(0.0,2.25)
  \psline(0.8,-1.0)(0.8,2.25)
  \psline(1.6,-1.0)(1.6,2.25)
 \psline{->}(0.0,-0.55)(0.0,-0.5)
 \psline{->}(0.8,-0.55)(0.8,-0.5)
 \psline{->}(1.6,-0.55)(1.6,-0.5)
  \psline(2.4,-1.0)(2.4,2.25)
  \psline(3.2,-1.0)(3.2,2.25)
  \psline(4.0,-1.0)(4.0,2.25)
  \psellipse[linewidth=0.9pt,linecolor=black,border=0.1](0.8,0.25)(1.2,0.3)
   \psline{->}(-0.25,0.13)(-0.35,0.18)
    \psline[linewidth=0.9pt,linecolor=black,border=0.1](0.0,0.25)(0.0,0.75)
    \psline[linewidth=0.9pt,linecolor=black,border=0.1](0.8,0.25)(0.8,0.75)
    \psline[linewidth=0.9pt,linecolor=black,border=0.1](1.6,0.25)(1.6,0.75)
  \psellipse[linewidth=0.9pt,linecolor=black,border=0.1](0.4,1.0)(0.8,0.2)
   \psline{->}(-0.25,0.91)(-0.35,0.95)
    \psline[linewidth=0.9pt,linecolor=black,border=0.1](0.0,1.0)(0.0,1.5)
    \psline[linewidth=0.9pt,linecolor=black,border=0.1](0.8,1.0)(0.8,1.5)
  \psellipse[linewidth=0.9pt,linecolor=black,border=0.1](1.2,1.75)(0.8,0.2)
   \psline{->}(0.55,1.66)(0.45,1.71)
    \psline[linewidth=0.9pt,linecolor=black,border=0.1](0.8,1.75)(0.8,2.25)
    \psline[linewidth=0.9pt,linecolor=black,border=0.1](1.6,1.75)(1.6,2.25)
  \rput[bl]{0}(-0.3,-0.75){$a$}
  \rput[bl]{0}(0.5,-0.75){$b$}
  \rput[bl]{0}(1.3,-0.75){$c$}
  \rput[bl]{0}(-0.65,-0.1){$\omega_d$}
  \rput[bl]{0}(-0.65,0.68){$\omega_e$}
  \rput[bl]{0}(0.15,1.4){$\omega_f$}
\endpspicture
}
.
\end{equation}
This represents a large portion of the invariant information contained in the $F$-symbols. Moreover, I will show that these are straightforward quantities to physically measure. For UMTCs with no fusion multiplicities, i.e. $N_{ab}^{c} \in \{ 0 , 1 \}$ for all $a,b,c\in \mathcal{C}$, these invariant are simply the magnitudes of all $F$-symbols.

Notice that setting $c=0$, $f=b$, and $e=d$, Eq.~(\ref{eq:F_magnitude_invariant}) reduces to the fusion coefficients
\begin{equation}
\label{eq:fusion_invariant}
N_{ab}^{e} = \frac{1}{d_{e}}
\psscalebox{.75}{
 \pspicture[shift=-1.25](-0.5,-0.6)(2.8,2.0)
  \small
  \psellipse[linewidth=0.9pt,linecolor=black](0.4,1.5)(0.415,0.25)
  \psellipse[linewidth=0.9pt,linecolor=black](2.0,1.5)(0.415,0.25)
  \psellipse[linewidth=0.9pt,linecolor=black](0.4,0.0)(0.415,0.25)
  \psellipse[linewidth=0.9pt,linecolor=black](2.0,0.0)(0.415,0.25)
  \psframe[fillstyle=solid,fillcolor=white,linecolor=white](-0.1,0.01)(2.5,1.49)
  \psset{linewidth=0.9pt,linecolor=black,arrowscale=1.5,arrowinset=0.15}
  \psline(0.0,0.0)(0.0,1.5)
  \psline(0.8,0.0)(0.8,1.0)
       \psline{->}(0.8,0.25)(0.8,0.4)
  \psline(1.6,0.0)(1.6,1.0)
       \psline{->}(1.6,0.25)(1.6,0.4)
  \psline(2.4,0.0)(2.4,1.5)
  \psellipse[linewidth=0.9pt,linecolor=black,border=0.1](1.2,1.0)(0.8,0.2)
   \psline{->}(0.55,0.91)(0.45,0.95)
    \psline[linewidth=0.9pt,linecolor=black,border=0.1](0.8,1.0)(0.8,1.5)
    \psline[linewidth=0.9pt,linecolor=black,border=0.1](1.6,1.0)(1.6,1.5)
  \rput[bl]{0}(0.5,0.15){$a$}
  \rput[bl]{0}(1.3,0.15){$b$}
  \rput[bl]{0}(0.15,0.68){$\omega_e$}
\endpspicture
}
.
\end{equation}

Some of the invariant phase information of $F$-symbols may be found in the Frobenius-Schur indicators
\begin{equation}
\label{eq:FS-ind}
\kappa_{a} = \frac{\left[F_a^{aaa}\right]_{00}} { \left| \left[F_a^{aaa}\right]_{00} \right| }
,
\end{equation}
for $a=\bar{a}$. (The right hand side of this equation is not gauge invariant for $a \neq \bar{a}$.) Additional invariant phase information may be found in the higher order Frobenius-Schur indicators~\cite{Ng07}, when $a$ is higher order ($a$ is order $n$ when $n$ is the smallest number of copies of $a$ that can fuse together into the vacuum channel $0$). For a UMTC, these can all be expressed in terms of the modular data through the relation
\begin{equation}
\label{eq:FSn-ind}
\kappa_{a}^{(n)} = \frac{1}{\mathcal{D}^{2}} \sum_{x,y} N_{ax}^{y} d_x d_y \left(\frac{\theta_{y}}{\theta_{x}} \right)^{n}
,
\end{equation}
when $a$ is order $n$, i.e. $n$ is the minimal number of $a$ anyons that can be fused together into the vacuum $0$. I note that $\kappa_{a} = \kappa_{a}^{(2)} = \Lambda^{aa}_{0}$. These Frobenius-Shur indicators can also be expressed in terms of link diagrams.

Another significant quantity associated with modular transformations is the punctured torus $S$-matrix
\begin{eqnarray}
\label{eq:def_Sz}
&&S^{(z)}_{(a,\mu)(b,\nu)} =
\frac{1}{\mathcal{D} \sqrt{d_z}}
\pspicture[shift=-1.2](-0.1,-0.6)(2.8,1.8)
\small
  \psellipse[linewidth=0.9pt,linecolor=black](2.05,1.4)(0.465,0.4)
  \psellipse[linewidth=0.9pt,linecolor=black](1.7,0.0)(0.815,0.4)
  \psframe*[linecolor=white](0.8,0.005)(02.6,1.395)
  \psarc[linewidth=0.9pt,linecolor=black,arrows=<-,arrowscale=1.5,arrowinset=0.15] (1.6,0.7){0.5}{167}{373}
  \psarc[linewidth=0.9pt,linecolor=black,border=3pt,arrows=<-,arrowscale=1.5,arrowinset=0.15] (0.9,0.7){0.5}{167}{373}
  \psarc[linewidth=0.9pt,linecolor=black] (0.9,0.7){0.5}{0}{180}
  \psarc[linewidth=0.9pt,linecolor=black,border=3pt] (1.6,0.7){0.5}{45}{150}
  \psarc[linewidth=0.9pt,linecolor=black] (1.6,0.7){0.5}{0}{50}
  \psarc[linewidth=0.9pt,linecolor=black] (1.6,0.7){0.5}{145}{180}
  \psline[linewidth=0.9pt,linecolor=black](1.6,1.2)(1.6,1.4)
  \psline[linewidth=0.9pt,linecolor=black](0.9,0.2)(0.9,0.0)
  \psline[linewidth=0.9pt,linecolor=black](2.5,0.0)(2.5,1.4)
  \psline[linewidth=0.9pt,linecolor=black,arrows=->,arrowscale=1.5,arrowinset=0.15](2.5,0.7)(2.5,0.8)
  \rput[bl]{0}(0.1,0.6){$a$}
  \rput[bl]{0}(0.85,0.6){$b$}
  \rput[bl]{0}(2.65,0.6){$z$}
\rput[bl]{0}(1.4,1.25){$\scriptstyle\nu$}
\rput[bl]{0}(0.65,-0.0){$\scriptstyle\mu$}
\endpspicture
\\
&&= \frac{1}{\mathcal{D}} \sum_{c,\alpha,\beta, \gamma} d_a d_b \frac{\theta_c}{\theta_a \theta_b} \left[F_{a}^{b\bar{b}a}\right]_{0,(c,\alpha,\beta)} \notag \\
&& \times \left[(F_a^{b\bar{b}a})^{-1}\right]_{(c,\alpha,\beta),(\bar{z},\nu,\gamma)} \left[F_{\bar{z}}^{\bar{z} a\bar{a}}\right]_{(a,\gamma,\mu),0}
,
\label{eq:def_Sz_FR}
\end{eqnarray}
where $N_{a\bar{a}}^{z} \neq 0$ and $N_{b\bar{b}}^{z} \neq 0$. This provides a projective representation of the corresponding modular transformation for a torus that has a single boundary which carries the topological charge $z$.

The definition of the punctured torus $S$-matrix is equivalent to the diagrammatic relation
\begin{equation}
\pspicture[showgrid=false,shift=*](-1,-0.8)(0.8,2.5)
\small
\psarc[linewidth=0.9pt,linecolor=black](0,1.2){0.5}{-90}{180}
\psarc[linewidth=0.9pt,linecolor=black,border=3pt](0,0.5){0.5}{-90}{180}
\psarc[linewidth=0.9pt,linecolor=black,border=3pt](0,1.2){0.5}{175}{275}
\psarc[linewidth=0.9pt,linecolor=black,arrows=<-,arrowscale=1.5,arrowinset=0.15] (0,1.2){0.5}{167}{280}
\psarc[linewidth=0.9pt,linecolor=black,arrows=<-,arrowscale=1.5,arrowinset=0.15] (0,0.5){0.5}{167}{373}
\psline[linewidth=0.9pt,linecolor=black,arrows=->,arrowscale=1.5,arrowinset=0.15](0,1.7)(0,2.2)
\psline[linewidth=0.9pt,linecolor=black](0,2.0)(0,2.4)
\psline[linewidth=0.9pt,linecolor=black,arrows=-<,arrowscale=1.5,arrowinset=0.15](0,0)(0,-0.5)
\psline[linewidth=0.9pt,linecolor=black](0,-0.4)(0,-0.7)
\rput[bl]{0}(-0.8,0.35){$a$}
\rput[bl]{0}(-0.8,1.15){$b$}
\rput[bl]{0}(0.1,2.1){$\bar{z}$}
\rput[bl]{0}(0.1,-0.4){$\bar{z}$}
\rput[bl]{0}(-0.2,1.75){$\scriptstyle\nu$}
\rput[bl]{0}(-0.2,-0.2){$\scriptstyle\mu$}
\endpspicture
= \frac{\mathcal{D}}{\sqrt{d_z}} S^{(z)}_{(a,\mu)(b,\nu)}
\pspicture[showgrid=false,shift=*](-0.3,-0.8)(0.4,2.5)
\small
\psline[linewidth=0.9pt,linecolor=black,arrows=->,arrowscale=1.5,arrowinset=0.15](0,1.2)(0,1.25)
\psline[linewidth=0.9pt,linecolor=black](0,-0.7)(0,2.4)
\rput[bl]{0}(0.1,1.2){$\bar{z}$}
\endpspicture
\end{equation}

The punctured torus $S$-matrix also yields the link invariants
\begin{equation}
\label{eq:Sz_magnitude_invariant}
\sum_{\mu,\nu } \left| S^{(z)}_{(a,\mu) (b,\nu)} \right|^{2} = \frac{d_{a} d_{b}}{\mathcal{D}^{2} d_{z}}
\psscalebox{.65}{
 \pspicture[shift=-2.0](-0.9,-0.5)(2.8,4.0)
  \small
  \psellipse[linewidth=0.9pt,linecolor=black](0.4,3.5)(0.415,0.25)
  \psellipse[linewidth=0.9pt,linecolor=black](2.0,3.5)(0.415,0.25)
  \psellipse[linewidth=0.9pt,linecolor=black](0.4,0.0)(0.415,0.25)
  \psellipse[linewidth=0.9pt,linecolor=black](2.0,0.0)(0.415,0.25)
  \psframe[fillstyle=solid,fillcolor=white,linecolor=white](-0.1,0.01)(2.5,3.49)
  \psset{linewidth=0.9pt,linecolor=black,arrowscale=1.5,arrowinset=0.15}
  \psline(0.0,0.0)(0.0,3.5)
     \psline{->}(0.0,0.25)(0.0,0.4)
  \psline(0.8,0.0)(0.8,0.5)
  \psline(1.6,0.0)(1.6,0.5)
       \psline{->}(1.6,0.25)(1.6,0.4)
  \psline(2.4,0.0)(2.4,3.5)
  \psbezier(1.6,0.5)(1.6,0.8)(0.8,0.7)(0.8,1.0)
  \psbezier[linewidth=0.9pt,linecolor=black,border=0.1](0.8,0.5)(0.8,0.8)(1.6,0.7)(1.6,1.0)
  \psbezier(1.6,1.0)(1.6,1.3)(0.8,1.2)(0.8,1.5)
  \psbezier[linewidth=0.9pt,linecolor=black,border=0.1](0.8,1.0)(0.8,1.3)(1.6,1.2)(1.6,1.5)
  \psline(0.8,1.5)(0.8,2.5)
  \psline(1.6,1.5)(1.6,2.5)
  \psbezier(0.8,2.5)(0.8,2.8)(1.6,2.7)(1.6,3.0)
  \psbezier[linewidth=0.9pt,linecolor=black,border=0.1](1.6,2.5)(1.6,2.8)(0.8,2.7)(0.8,3.0)
  \psbezier(0.8,3.0)(0.8,3.3)(1.6,3.2)(1.6,3.5)
  \psbezier[linewidth=0.9pt,linecolor=black,border=0.1](1.6,3.0)(1.6,3.3)(0.8,3.2)(0.8,3.5)
  \psellipse[linewidth=0.9pt,linecolor=black,border=0.1](0.4,2.0)(0.8,0.2)
   \psline{->}(-0.25,1.91)(-0.35,1.95)
    \psline[linewidth=0.9pt,linecolor=black,border=0.1](0.0,2.0)(0.0,2.5)
    \psline[linewidth=0.9pt,linecolor=black,border=0.1](0.8,2.0)(0.8,2.5)
  \rput[bl]{0}(-0.3,0.15){$a$}
  \rput[bl]{0}(1.3,0.15){$b$}
  \rput[bl]{0}(-0.65,1.68){$\omega_z$}
\endpspicture
}
\end{equation}
and
\begin{equation}
\label{eq:Sz_diag_invariant}
\sum_{\mu } S^{(z)}_{(a,\mu) (a,\mu)} = \frac{d_a}{\mathcal{D} d_{z}}
\psscalebox{.65}{
 \pspicture[shift=-1.5](-0.9,-0.5)(2.8,2.5)
  \small
  \psellipse[linewidth=0.9pt,linecolor=black](1.2,0.0)(0.415,0.25)
  \psellipse[linewidth=0.9pt,linecolor=black](1.2,0.0)(1.216,0.5)
  \psellipse[linewidth=0.9pt,linecolor=black](1.2,2.0)(0.415,0.25)
  \psellipse[linewidth=0.9pt,linecolor=black](1.2,2.0)(1.216,0.5)
  \psframe[fillstyle=solid,fillcolor=white,linecolor=white](-0.1,0.01)(2.5,1.99)
  \psarc[linewidth=0.9pt,linecolor=black,border=0.1](0.4,1.5){0.4}{270}{360}
  \psarc[linewidth=0.9pt,linecolor=black,border=0.1](0.4,1.0){0.4}{0}{180}
  \psarc[linewidth=0.9pt,linecolor=black,border=0.1](0.4,1.5){0.4}{180}{271}
  \psset{linewidth=0.9pt,linecolor=black,arrowscale=1.5,arrowinset=0.15}
  \psline(0.0,0.0)(0.0,1.01)
  \psline(0.8,0.0)(0.8,1.01)
  \psline(1.6,0.0)(1.6,2.0)
  \psline(2.4,0.0)(2.4,2.0)
  \psline(0.8,1.5)(0.8,2.0)
  \psline(0.0,1.5)(0.0,2.0)
    \psline{->}(0.0,1.75)(0.0,1.9)
  \psellipse[linewidth=0.9pt,linecolor=black,border=0.1](0.4,0.5)(0.8,0.2)
   \psline{->}(-0.25,0.41)(-0.35,0.45)
    \psline[linewidth=0.9pt,linecolor=black,border=0.1](0.0,0.5)(0.0,1.0)
    \psline[linewidth=0.9pt,linecolor=black,border=0.1](0.8,0.5)(0.8,1.0)
  \rput[bl]{0}(-0.3,1.65){$a$}
  \rput[bl]{0}(-0.65,0.18){$\omega_{\bar{z}}$}
\endpspicture
}
.
\end{equation}
The latter can be expressed in terms of a colored Whitehead link and has proven useful for mathematically distinguishing different UMTCs with identical modular data~\cite{Bonderson19}.

A useful generalization of Eq.~(\ref{eq:Sz_magnitude_invariant}) is given by a link invariant that I denote as~\footnote{Letting $n=-n'=2$, $e=0$, $g=z$, $b\rightarrow \bar{a}$, $c \rightarrow b$, and $d \rightarrow b$ in Eq.~(\ref{eq:Braiding_invariant}) yields the invariant in Eq.~(\ref{eq:Sz_magnitude_invariant}), up to the quantum dimension factors.}
\begin{equation}
\label{eq:Braiding_invariant}
L^{(n,n')}_{a,b,c;d}(e,g) = \frac{1}{d_{d}} \!
\psscalebox{.65}{
 \pspicture[shift=-3.25](-0.9,-2)(4.2,4.5)
  \small
  \psellipse[linewidth=0.9pt,linecolor=black](2.0,3.75)(2.017,0.75)
  \psellipse[linewidth=0.9pt,linecolor=black](2.0,3.75)(1.216,0.5)
  \psellipse[linewidth=0.9pt,linecolor=black](2.0,3.75)(0.415,0.25)
  \psellipse[linewidth=0.9pt,linecolor=black](2.0,-1.0)(2.017,0.75)
  \psellipse[linewidth=0.9pt,linecolor=black](2.0,-1.0)(1.216,0.5)
  \psellipse[linewidth=0.9pt,linecolor=black](2.0,-1.0)(0.415,0.25)
  \psframe[fillstyle=solid,fillcolor=white,linecolor=white](-0.1,-0.99)(4.1,3.74)
  \psset{linewidth=0.9pt,linecolor=black,arrowscale=1.5,arrowinset=0.15}
  \psline(0.0,-1.0)(0.0,3.75)
  \psline(0.8,-1.0)(0.8,3.75)
  \psline(1.6,-1.0)(1.6,3.75)
 \psline{->}(0.0,-0.55)(0.0,-0.5)
 \psline{->}(0.8,-0.55)(0.8,-0.5)
 \psline{->}(1.6,-0.55)(1.6,-0.5)
  \psline(2.4,-1.0)(2.4,3.75)
  \psline(3.2,-1.0)(3.2,3.75)
  \psline(4.0,-1.0)(4.0,3.75)
  \psframe[fillstyle=solid,fillcolor=white,linewidth=0.9pt,linecolor=black,border=0](0.6,1.5)(1.8,2.0)
  \rput[bl]{0}(1.0,1.63){$R^{n}$}
  \psframe[fillstyle=solid,fillcolor=white,linewidth=0.9pt,linecolor=black,border=0](0.6,3.0)(1.8,3.5)
  \rput[bl]{0}(1.0,3.13){$R^{n'}$}
  \psellipse[linewidth=0.9pt,linecolor=black,border=0.1](0.8,0.25)(1.2,0.3)
   \psline{->}(-0.25,0.13)(-0.35,0.18)
    \psline[linewidth=0.9pt,linecolor=black,border=0.1](0.0,0.25)(0.0,0.75)
    \psline[linewidth=0.9pt,linecolor=black,border=0.1](0.8,0.25)(0.8,0.75)
    \psline[linewidth=0.9pt,linecolor=black,border=0.1](1.6,0.25)(1.6,0.75)
  \psellipse[linewidth=0.9pt,linecolor=black,border=0.1](0.4,1.0)(0.8,0.2)
   \psline{->}(-0.25,0.91)(-0.35,0.95)
    \psline[linewidth=0.9pt,linecolor=black,border=0.1](0.0,1.0)(0.0,1.5)
    \psline[linewidth=0.9pt,linecolor=black,border=0.1](0.8,1.0)(0.8,1.5)
  \psellipse[linewidth=0.9pt,linecolor=black,border=0.1](0.4,2.5)(0.8,0.2)
   \psline{->}(-0.25,2.41)(-0.35,2.45)
    \psline[linewidth=0.9pt,linecolor=black,border=0.1](0.0,2.5)(0.0,3)
    \psline[linewidth=0.9pt,linecolor=black,border=0.1](0.8,2.5)(0.8,3)
  \rput[bl]{0}(-0.3,-0.75){$a$}
  \rput[bl]{0}(0.5,-0.75){$b$}
  \rput[bl]{0}(1.3,-0.75){$c$}
  \rput[bl]{0}(-0.65,-0.1){$\omega_d$}
  \rput[bl]{0}(-0.65,0.68){$\omega_e$}
  \rput[bl]{0}(-0.65,2.18){$\omega_g$}
\endpspicture
}
,
\end{equation}
where the $R^n$ box represents $n$ braiding exchanges on the lines entering the bottom of the box.
When $b=c$, this invariant can be written as
\begin{align}
L^{(n,n')}_{a,b,b;d}(e,g) &= \sum_{\alpha,\beta,\mu,\nu} \left[(B_{d}^{abb})^{n}\right]_{(e,\alpha,\beta)(g,\mu,\nu)}
\notag \\
& \qquad \qquad \times \left[(B_{d}^{abb})^{n'}\right]_{(g,\mu,\nu)(e,\alpha,\beta)}
.
\end{align}
The case when $b \neq c$ is straightforward to compute, but cumbersome to write out, so I simply note that , this invariant will automatically be zero if $n+n'$ is odd, due to conservation of charge.
When $n=2m$ and $n'=2m'$ are even, this can be written as
\begin{align}
&L^{(2m,2m')}_{a,b,c;d}(e,g) = \sum_{\substack{\alpha,\beta,\mu,\nu \\ f , \lambda, \sigma \\ f' , \lambda', \sigma'}} \left[F_{d}^{abc}\right]_{(e,\alpha,\beta)(f,\lambda,\sigma)} \left(\frac{\theta_{f}}{\theta_{b}\theta_{c}} \right)^{m}
\notag \\
& \qquad \qquad \times \left[F_{d}^{abc}\right]^{\ast}_{(g,\mu,\nu)(f,\lambda, \sigma)} \left[F_{d}^{abc}\right]_{(g,\mu,\nu)(f',\lambda', \sigma')}
\notag \\
& \qquad \qquad \times \left(\frac{\theta_{f'}}{\theta_{b}\theta_{c}} \right)^{m'} \left[F_{d}^{abc}\right]^{\ast}_{(e,\alpha,\beta)(f',\lambda', \sigma')}
.
\label{eq:purebraid_invariant}
\end{align}
When $n=2m+1$ and $n'=2m'+1$ are odd, this can be written as
\begin{align}
& L^{(2m+1,2m'+1)}_{a,b,c;d}(e,g) = \sum_{\substack{\alpha,\beta,\mu,\nu \\ f , \lambda, \sigma, \eta \\ f' , \lambda', \sigma', \eta'}} \left[F_{d}^{abc}\right]_{(e,\alpha,\beta)(f,\lambda,\sigma)}
\notag \\
& \qquad \qquad \times  \left(\frac{\theta_{f}}{\theta_{b}\theta_{c}} \right)^{m} [R^{cb}_{f}]_{\lambda,\eta} \left[F_{d}^{acb}\right]^{\ast}_{(g,\mu,\nu)(f,\eta, \sigma)}
\notag \\
& \qquad \qquad \times \left[F_{d}^{acb}\right]_{(g,\mu,\nu)(f',\lambda', \sigma')}
 \left(\frac{\theta_{f'}}{\theta_{b}\theta_{c}} \right)^{m'}
\notag \\
& \qquad \qquad \times [R^{bc}_{f'}]_{\lambda',\eta'} \left[F_{d}^{abc}\right]^{\ast}_{(e,\alpha,\beta)(f',\eta', \sigma')}
.
\label{eq:exchangebraid_invariant}
\end{align}
The case where $n' = -n$ will be relevant for the experiments in Sec.~\ref{sec:exp_braiding}.

It is an open question whether all the invariant data contained in $F$-symbols can be expressed in terms of link diagrams or whether they can be expressed in terms of the modular data supplemented by the punctured torus $S$-matrix. For example, this is not known for the gauge invariant quantity
\begin{equation}
\label{eq:F_cur_invariant}
\sum_{\alpha, \beta} \left[F_b^{abc}\right]_{(b,\alpha,\beta)(b,\beta,\alpha)} = \frac{1}{\sqrt{d_a d_b^2 d_c}}\sum_{\alpha, \beta}
\scalebox{.8}{
\pspicture[shift=-2](-0.8,-2)(1.1,2)
\small
  \psellipse[linewidth=0.9pt,linecolor=black](0.5,1.5)(0.515,0.4)
  \psellipse[linewidth=0.9pt,linecolor=black](0.5,-1.5)(0.515,0.4)
  \psframe*[linecolor=white](-0.2,-1.495)(1.2,1.495)
  \psline[linewidth=0.9pt,linecolor=black](1,-1.5)(1,1.5)
  \psline[linewidth=0.9pt,linecolor=black](0,-1.5)(0,1.5)
  \psline[linewidth=0.9pt,linecolor=black](0,-1)(0.505,0.01)
  \psline[linewidth=0.9pt,linecolor=black](0,-0.5)(-0.505,0.01)
  \psline[linewidth=0.9pt,linecolor=black](0.51,-0.01)(0,0.5)
  \psline[linewidth=0.9pt,linecolor=black](-0.505,-0.01)(0,1)
  \psline[linewidth=0.9pt,linecolor=black,arrows=->,arrowscale=1.5,arrowinset=0.15](0.0,0)(0.0,0.1)
  \psline[linewidth=0.9pt,linecolor=black,arrows=->,arrowscale=1.5,arrowinset=0.15](0.0,0.85)(0.0,0.86)
  \psline[linewidth=0.9pt,linecolor=black,arrows=->,arrowscale=1.5,arrowinset=0.15](0.0,-0.55)(0.0,-0.54)
  \psline[linewidth=0.9pt,linecolor=black,arrows=->,arrowscale=1.5,arrowinset=0.15](0.0,1.35)(0.0,1.36)
  \psline[linewidth=0.9pt,linecolor=black,arrows=->,arrowscale=1.5,arrowinset=0.15](0.0,-1.15)(0.0,-1.14)
  \psline[linewidth=0.9pt,linecolor=black,arrows=->,arrowscale=1.5,arrowinset=0.15](0,-1)(0.45,-0.1)
  \psline[linewidth=0.9pt,linecolor=black,arrows=->,arrowscale=1.5,arrowinset=0.15](-0.5,0)(-0.3,0.4)
  \rput[bl]{0}(-0.65,0.3){$a$}
  \rput[bl]{0}(0.5,-0.4){$c$}
  \rput[bl]{0}(0.1,0.7){$b$}
  \rput[bl]{0}(-0.2,0.0){$b$}
  \rput[bl]{0}(-0.25,-0.85){$b$}
  \rput[bl]{0}(-0.25,-1.3){$b$}
  \rput[bl]{0}(-0.25,1.2){$b$}
  \rput[bl]{0}(-0.3,0.95){$\scriptstyle\alpha$}
  \rput[bl]{0}(-0.3,-0.55){$\scriptstyle\alpha$}
  \rput[bl]{0}(0.1,0.4){$\scriptstyle\beta$}
  \rput[bl]{0}(0.1,-1.1){$\scriptstyle\beta$}
\endpspicture
}
\end{equation}
which exists when $N_{ab}^{b} \neq 0$ and $N_{cb}^{b} \neq 0$.

I end this section with a few relations between UMTC quantities, some of which are useful in the study of physical experiments.

A straightforward diagrammatic manipulation of Eq.~(\ref{eq:F_magnitude_invariant}) reveals the relation
\begin{equation}
\label{eq:F_mag_symm}
\sum_{\substack{\alpha, \beta \\ \mu,\nu}} \left| \left[F_d^{abc}\right]_{(e,\alpha,\beta)(f,\mu,\nu)} \right|^{2} = \sum_{\substack{\alpha, \beta \\ \mu,\nu}} \left| \left[F_d^{cba}\right]_{(f,\mu,\nu) (e,\alpha,\beta)} \right|^{2}
.
\end{equation}

Applying $R$-moves to the $a$ line in Eq.~(\ref{eq:F_cur_invariant}) allows the $F$-symbol invariant to be equated with a $B$-symbol invariant
\begin{equation}
\sum_{\alpha, \beta} \left[F_b^{abc}\right]_{(b,\alpha,\beta)(b,\beta,\alpha)} =
\sum_{\alpha, \beta} \left[B_b^{bca}\right]_{(b,\alpha,\beta)(b,\beta,\alpha)}
.
\end{equation}
This shows that, when $N_{ab}^{b} \neq 0$ and $N_{cb}^{b} \neq 0$, there is nontrivial information about braiding that is entirely determined by the fusion $F$-symbols.

Application of a Hexagon equation allows one to rewrite the expression of $B$-symbols in terms of $F$-symbols and $R$-symbols as
\begin{eqnarray}
&& \left[B_d^{abc}\right]_{(g,\gamma, \lambda)(e,\alpha,\beta)}
\\
&&\quad =
\sum_{\mu,\nu} [(R^{ac}_{g})^{-1}]_{\gamma \mu} \left[F_d^{cab}\right]_{(g,\mu, \lambda)(e,\alpha,\nu)} [R^{ec}_{d}]_{\nu \beta}
.
\notag
\end{eqnarray}
Taking the trace of the magnitude square of both sides of this expression reveals the relation
\begin{equation}
\label{eq:magB_magF}
\sum_{\substack{ \gamma,\lambda \\ \alpha, \beta }} \left| \left[B_d^{abc}\right]_{(g,\gamma, \lambda)(e,\alpha,\beta)} \right|^{2} = \sum_{\substack{\mu,\lambda \\ \alpha, \nu }} \left| \left[F_d^{cab}\right]_{(g,\mu,\lambda) (e,\alpha,\nu)} \right|^{2}
.
\end{equation}
This relation indicates that, in general, there are significant braiding properties that are entirely determined by the fusion properties. Indeed, I will show that this relation is relevant to experiments that directly probe the non-Abelian nature of braiding, and that it demonstrates the fact that having multiple possible fusion channels implies the existence of non-Abelian braiding.

\section{Experiments for Measuring Topological Invariants Using Localized Bulk Quasiparticles}
\label{sec:Bulk}

The main class of experiments that I consider in this paper involves creating, manipulating, and measuring quasiparticles that are localized in the bulk of the system. While I discuss these at a level of generality and abstraction applicable to any topological phase, they correspond to physically realistic experiments. Though their implementation will depend on the details of the physical system involved, they require neither unrealistically precise knowledge of the microscopic Hamiltonian nor fine-tuned control of the system and operations.

A localized quasiparticle carries a definite value of topological charge, as superpositions of different localized topological charge values will rapidly decohere due to interaction with local noise (assuming there is no additional symmetry protecting such superpositions). As such, this class of experiments crucially relies on superpositions of nonlocal topological charge values. In other words, the localized quasiparticles in these experiments must be non-Abelian anyons in order to yield nontrivial information about the $F$-symbols and $R$-symbols.

The set of basic operations for performing these experiments are:
\begin{enumerate}
  \item Localization of a quasiparticle with specific topological charge value.
  \item Measurement of the collective topological charge of pairs of quasiparticles.
  \item Moving localized quasiparticles through the bulk.
  \item Splitting one localized quasiparticle into two separate quasiparticles with specific topological charge values.
\end{enumerate}
Localization of a quasiparticle in the bulk may be implemented using some (point-like) local pinning potential that energetically favors one particular topological charge value.
Moving quasiparticles though the bulk may be done via adiabatic transport that moves the locations of pinning potentials. Splitting a quasiparticle into two quasiparticles may be done using an adiabatic process where the initial pinning potential is adiabatically transformed into two separate, appropriately chosen pinning potentials. The topological charge values involved in a splitting processes must respect the fusion rules in order for the process not to spawn stray quasiparticles. Splitting operations include pair-creation from vacuum, in which case the initial ``quasiparticle'' is trivial. Appendix~\ref{sec:Splitting_Fusing} describes details of how such operations may be carried out for an idealized toy-model of pinning potentials.

Measurement of the topological charge of pairs of quasiparticles must be able to distinguish between all distinct fusion channels possible. This will generally involve a calibration of the measurements to identify the signatures of the possible topological charge values, which involves a separate, prior set of experiments to establish. Such measurements may be performed using the measurement of some local quantity that is correlated with the topological charge, e.g. of a localized energy density, charge distribution, etc. of the system. This requires first moving the two quasiparticles into sufficiently close proximity of each other and then performing the local measurement. The proximal distance between quasiparticles required for the measurement is set by the reach of the measurement device in the system.

Alternatively, the measurements could potentially be performed via nonlocal methods, such as interferometry or measurement devices capable of coherently coupling across distances that are long compared to the correlation length. Such nonlocal methods may also enable measurement of the collective fusion channels of not just pairs of quasiparticles, but clusters of multiple quasiparticles. Nonlocal measurements may be useful, but are not essential for the class of experiments considered in this section.

Another natural operation to consider is the fusion of two quasiparticles. This will not be considered a basic operation, as it is essentially the combination of operations from the above list: moving two localized quasiparticles into proximity of each other and performing a local measurement of their collective topological charge. The main distinction one may wish to make is that one typically thinks of only a single quasiparticle remaining after the fusion of two. In terms of the basic operations, this may be achieved simply by treating the pair of quasiparticles as a single object following the measurement. From the perspective of local pinning potentials, this may require modifying the potentials so that the measurement outcome is made to be the energetically favored topological charge value. Since such a process can be performed locally, the post-measurement topological charge value will be conserved, even if it involves level-crossings. (See Appendix~\ref{sec:Splitting_Fusing} for more details.)

The operations of moving and splitting quasiparticles are not essential for performing these experiments. Localization of quasiparticles and pairwise measurements of their topological charge are sufficient, and tunable interactions may also be utilized as an alternate operation. This is possible because anyonic teleportation and ``measurement-only'' methods~\cite{Bonderson08a,Bonderson08b,Bonderson12a} may be used instead of adiabatic transport to generate transformations that have the same effect as moving the quasiparticles. Similar methods may be used instead of splitting quasiparticles for the state initializations used in these experiments. However, these substitutions for splitting require either the ability to perform collective topological charge measurement of multiple quasiparticles, requiring interferometrical methods~\cite{Bonderson07c}, or the restriction to initial states created by pairwise measurements in Abelian fusion channels. The translation of how the experiments would be implemented with the full set of operations described above to these restricted sets of operations is straightforward, so I will only describe the former. It is, however, important to recognize that the use of measurement-only methods does not always provide access to as much information about the braiding properties as actual transport does, in particular with respect to braiding distinct topological charge types with each other. These limitations are discussed in detail in Appendix~\ref{sec:MObraiding}.

In some situations, it may only be possible to localize quasiparticles without knowing their topological charge values precisely (e.g. one only knows the electric charge carried by the quasiparticles), or it may only be possible to perform collective topological charge measurements that distinguish between different fusion channels without specifying their corresponding topological charge values. In such cases, the experiments of this section may still be performed, though they may provide less complete information about the topological order.

An important feature of the class of experiments discussed in this section is their robustness to non-universal physics, when operating in the topological limit. As long as the localized quasiparticles are kept far apart (as compared to the correlation length), except during splitting and measurement steps, these experiments do not require precise knowledge of the Hamiltonian, fine-tuned control of operations, nor much care to avoid the introduction of geometric and dynamical phases. This is because the localized topological charges (which are never superposed) are moved and split deterministically, while the nontrivial operations being probed are effected on the nonlocal topological state space, which is protected. In other words, non-universal effects will naturally drop out of the measurement outcome data, which, also being independent of gauge choices, are associated with universal topological invariants of the phase of matter. The only corrections to this that will not be exponentially suppressed are diabatic corrections associated with performing transport in finite time~\cite{Knapp2016}. However, such corrections should not pose a substantial problem, since (1) they can be reduced, though with polynomial suppression, by increasing transport time scales, and (2) they are, at least in principle, detectable and correctable~\cite{Knapp2016}. Moreover, Ocneanu rigidity~\cite{Etingof05} ensures that, for a given set of fusion rules, there are a finite number of possible UMTCs whose corresponding topological invariants will differ by discrete amounts. In other words, it is not unreasonable to expect that the error bars on the measured quantities can be made small enough to allow sufficient confidence in the ability to distinguish between the possible values of the corresponding invariants.

I note that one can and should use similar, but more mundane experiments to those described in this section to verify that the experiments are indeed operating in the topological limit. Verifying this should be considered a prerequisite step to be performed prior to or in conjunction with the experiments described here, as it will otherwise be unclear whether the measured results are actually universal topological invariants. I describe such an experiment for verifying operation in the topological limit in Appendix~\ref{sec:Rabi_experiment}.

In discussing the detailed predictions of these experiments, I will focus on UMTCs with no fusion multiplicities, i.e. $N_{ab}^{c} \in \{ 0,1 \}$ for all $a,b,c\in \mathcal{C}$, since it simplifies the discussion, yet covers essentially all examples that are likely to be of physical relevance. As such, the fusion multiplicity indices ($\alpha, \beta, \mu, \nu$) will be left implicit in the following. It is straightforward to generalize the discussion and analysis for UMTCs that do have fusion multiplicities.

\subsection{Fusion Rules and Quantum Dimensions}
\label{sec:FusionExp}

\begin{figure}[t!]
\begin{center}
  \includegraphics[scale=0.15]{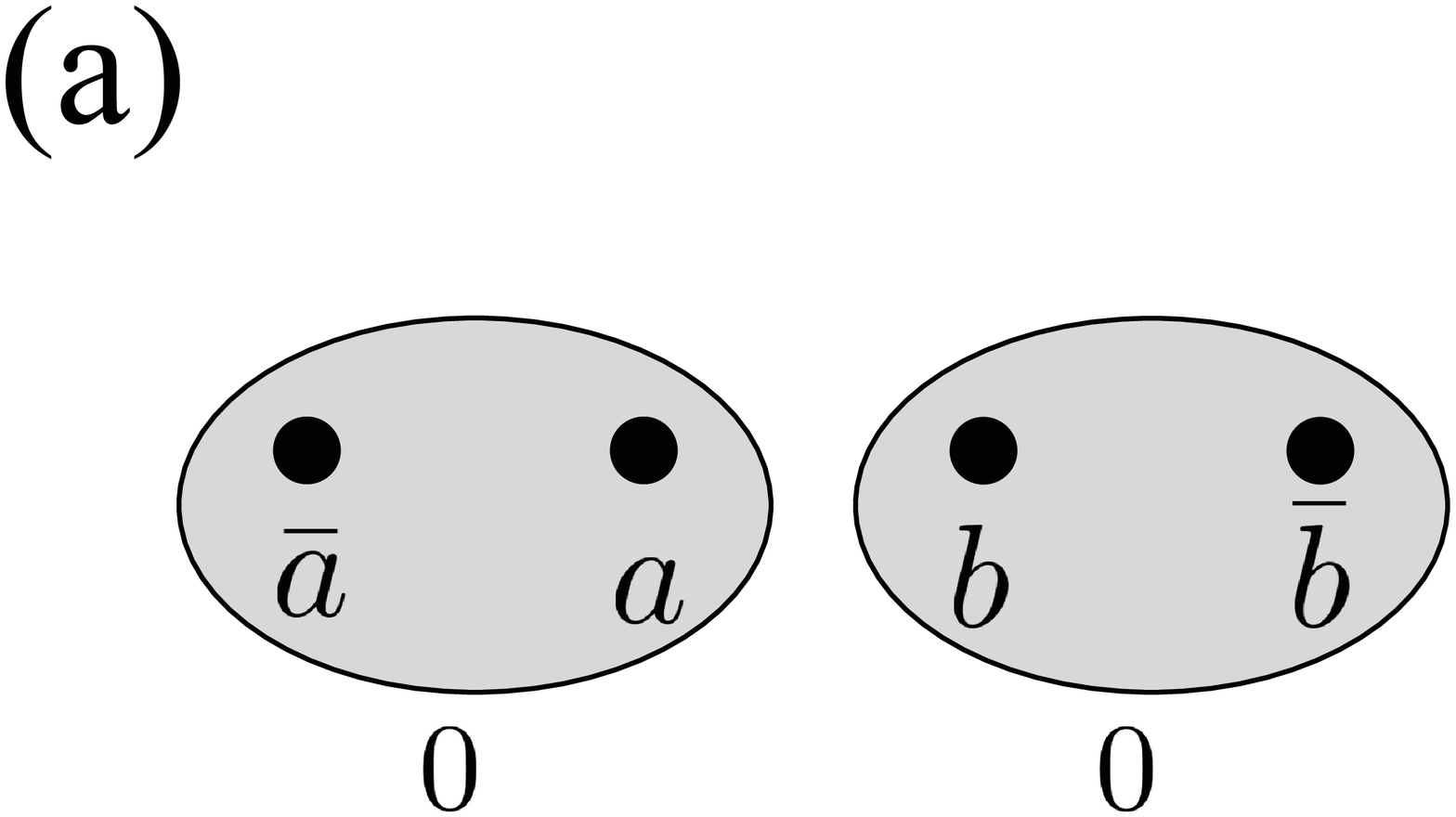}
  \includegraphics[scale=0.15]{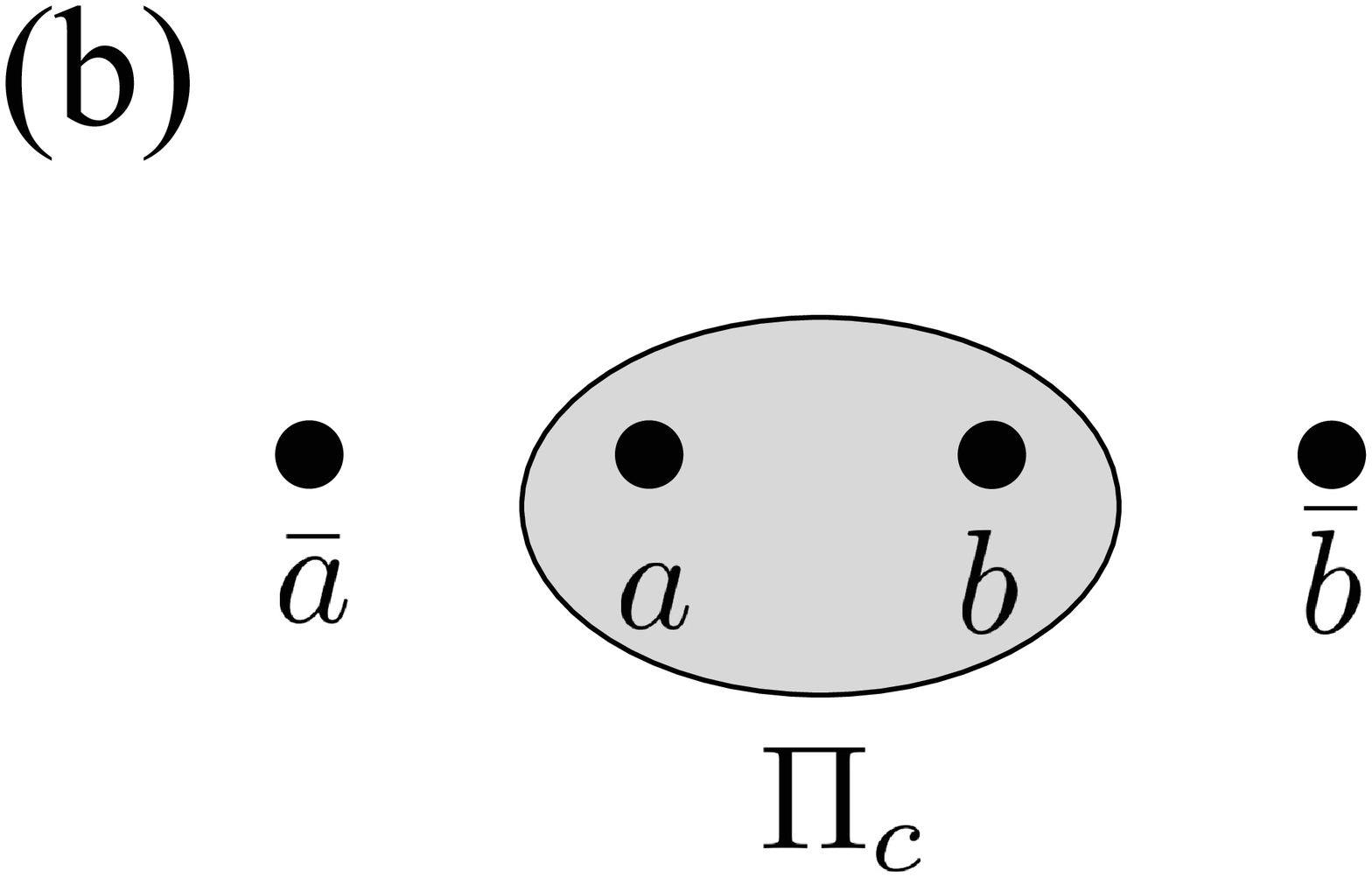}
  \caption{An experiment that can determine the fusion rules and quantum dimensions. (a) The system is initialized to have a $\bar{a}$-$a$ pair of quasiparticles in the vacuum channel and a $b$-$\bar{b}$ pair of quasiparticles in the vacuum channel. (b) After initialization, the joint topological charge (fusion channel) of the $a$-$b$ pair of quasiparticles is measured. The measured fusion channel is found to have topological charge $c$ with probabilities $p_{ab}(c)$ given in Eq.~(\ref{eq:fusion_experiment}).}
  \label{fig:fusion}
\end{center}
\end{figure}

The simplest experiment in the class of bulk quasiparticle experiments is performed by the following steps, shown schematically in Fig.~\ref{fig:fusion}:
\begin{enumerate}
  \item Pair-create quasiparticles carrying topological charges $\bar{a}$ and $a$  from vacuum, and move them apart.
  \item Pair-create quasiparticles carrying topological charges $b$ and $\bar{b}$  from vacuum, and move them apart.
  \item Measure the collective topological charge of the $a$-$b$ pair of quasiparticles.
\end{enumerate}
The initialization steps 1 and 2 can alternatively be implemented by other means, such as by localizing the individual quasiparticles and then using measurements to obtain the desired initial state.
The measurement outcome of this experiment will find the fusion channel of the $a$-$b$ pair to have topological charge $c$ with probability
\begin{eqnarray}
p_{ab}(c) &=& \left| \Pi_{c}^{(ab)} \left| \bar{a}, a ; 0 \right\rangle \left| b, \bar{b} ; 0 \right\rangle \right|^{2}
\notag \\
&=& \left\langle \bar{a}, a ; 0 \right| \left\langle b, \bar{b} ; 0 \right| \Pi_{c}^{(ab)} \left| \bar{a}, a ; 0 \right\rangle \left| b, \bar{b} ; 0 \right\rangle
\notag \\
&=&  \sqrt{\frac{d_{c}}{d_{a}^3 d_{b}^3}}
 \pspicture[shift=-1.2](-0.7,-0.9)(1.95,1.3)
  \small
  \psset{linewidth=0.9pt,linecolor=black,arrowscale=1.5,arrowinset=0.15}
  \psline{->}(0.7,0)(0.7,0.45)
  \psline(0.7,0)(0.7,0.55)
  \psline(0.7,0.55) (0.2,1.05)
  \psline{->}(0.7,0.55)(0.3,0.95)
  \psline(0.7,0.55) (1.2,1.05)
  \psline{->}(0.7,0.55)(1.1,0.95)
  \rput[bl]{0}(0.45,0.25){$c$}
  \rput[bl]{0}(-0.25,0.25){$\bar{a}$}
  \rput[bl]{0}(1.5,0.25){$\bar{b}$}
  \rput[bl]{0}(1.15,0.7){$b$}
  \rput[bl]{0}(0.1,0.7){$a$}
  \rput[bl]{0}(1.15,-0.25){$b$}
  \rput[bl]{0}(0.1,-0.25){$a$}
  \psline(0.7,0) (0.2,-0.5)
  \psline{-<}(0.7,0)(0.35,-0.35)
  \psline(0.7,0) (1.2,-0.5)
  \psline{-<}(0.7,0)(1.05,-0.35)
 \psarc[linewidth=0.9pt,linecolor=black,border=0pt](1.445,0.8){0.35}{0}{135}
 \psarc[linewidth=0.9pt,linecolor=black,border=0pt](1.445,-0.25){0.35}{225}{360}
 \psarc[linewidth=0.9pt,linecolor=black,border=0pt](-0.045,0.8){0.35}{45}{180}
 \psarc[linewidth=0.9pt,linecolor=black,border=0pt](-0.045,-0.25){0.35}{180}{315}
 \psline(1.795,-0.25)(1.795,0.8)
\psline{->}(1.795,0.4)(1.795,0.45)
\psline(-0.395,-0.25)(-0.395,0.8)
\psline{->}(-0.395,0.4)(-0.395,0.45)
  \endpspicture
= N_{ab}^{c} \frac{d_{c}}{d_{a} d_{b}}
.
\label{eq:fusion_experiment}
\end{eqnarray}
For a UMTC, the diagrammatic representation of this process used in this calculation is equivalent to Eq.~(\ref{eq:fusion_invariant}), with appropriate normalizations. Note that the property $d_a d_b = \sum_{c} N_{ab}^{c} d_c$ ensures that the probabilities of all possible outcomes sum to $\sum_{c} p_{ab}(c) = 1$, as they should.

Repeating this experiment many times for all possible values of $a$ and $b$ produces statistics that allow one to infer the fusion coefficients $N_{ab}^{c}$ and the quantum dimensions $d_a$. This can be done by first identifying which values of charge have $c=0$ as one of its possible outcomes, and then recognizing that $p_{ab}(0) = \delta_{\bar{a}b} d_{a}^{-2}$. This identifies the topological charge conjugate $\bar{a}$ of each charge $a$, and yields its quantum dimension $d_{a}$. In turn, this allows one to factor out the quantum dimensions from the results for general $a$, $b$, and $c$, to find the fusion coefficients.

It is worth noting that the existence of multiple fusion channels ($N_{ab}^{c} \neq 0$ for more than one value of $c$) or nontrivial quantum dimensions ($d_a > 1$) implies the existence of non-Abelian braiding~\cite{Barkeshli19,Rowell16} (see also the discussion in Sec.~\ref{sec:exp_nonAbelian_braiding}). Thus, even though it would not be as direct as actually performing non-Abelian braiding experiments, the fusion rules experiments described here constitute perhaps the easiest way to verify the existence of non-Abelian statistics.

The setup for the fusion rules experiments may also be used to verify that the system is operating in the topological limit. For this, one simply needs to repeat the measurement in step 3, with a variable time interval between the repeated measurements. This can be thought of as an anyonic version of a Rabi oscillation experiment, which can be used to detect error rates due to the breaking of topological degeneracies. Indeed, if one also has the ability to vary the distances between quasiparticles in such experiments, then one can also use such experiments to extract the correlation lengths. More details are given in Appendix~\ref{sec:Rabi_experiment}.

\subsection{Associativity}
\label{sec:AssociativityExp}

\begin{figure}[t!]
\begin{center}
  \includegraphics[scale=0.15]{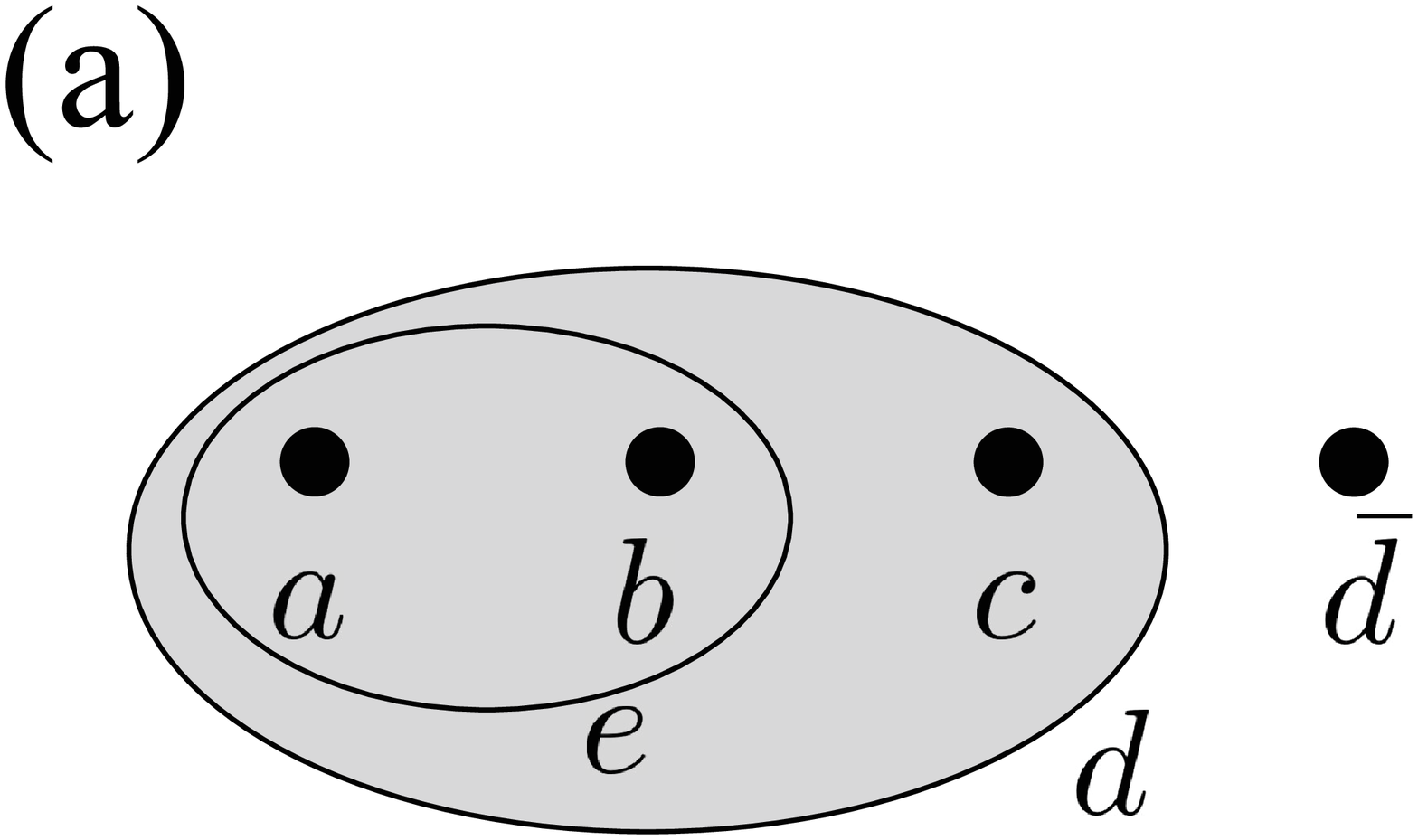}
  \includegraphics[scale=0.15]{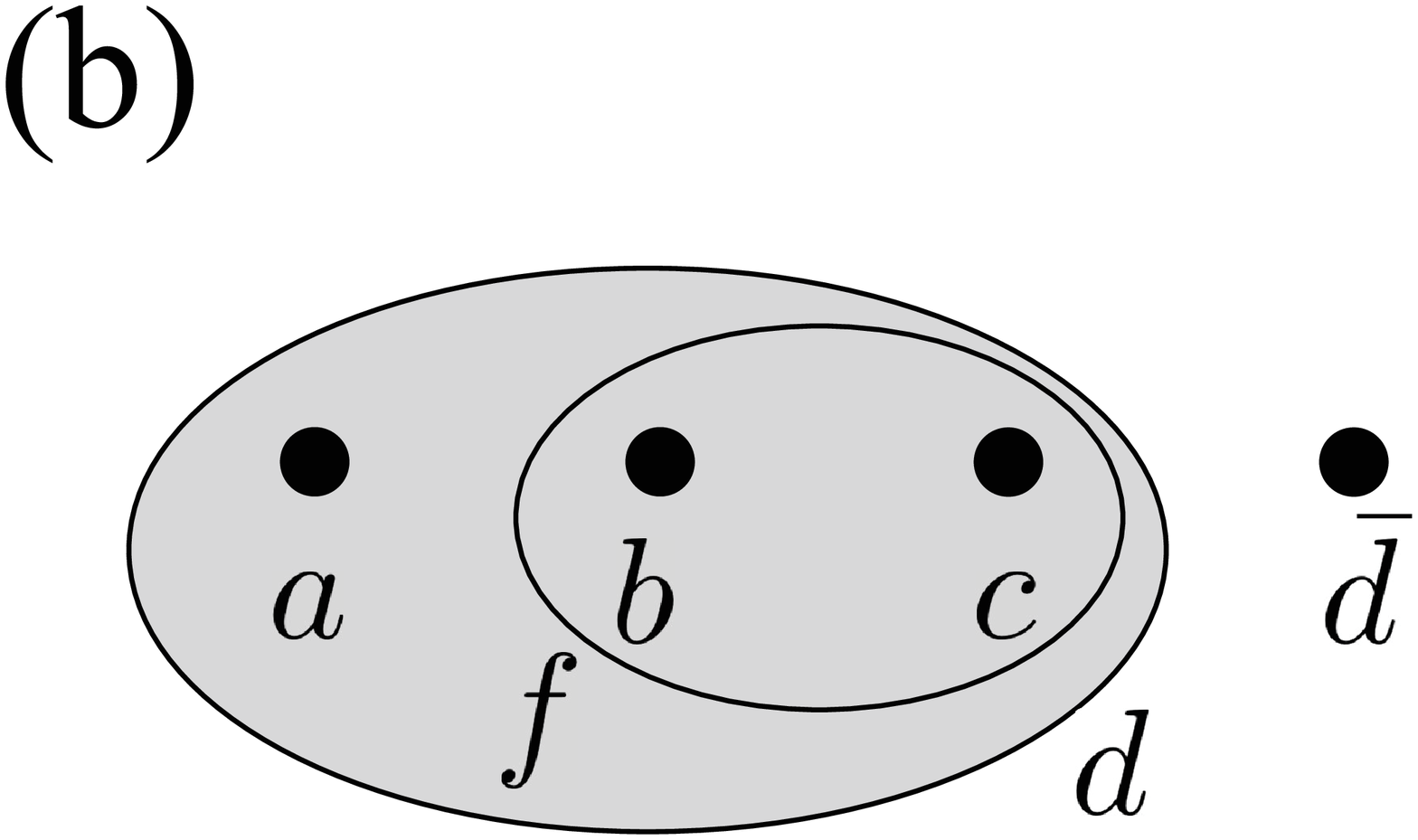}
  \caption{A system of four quasiparticles carrying topological charges $a$, $b$, $c$, and $\bar{d}$, with collective topological charge $0$, can be produced to have different fusion channels. (a) The state with quasiparticles $a$ and $b$ having joint charge $e$. (b) The state with quasiparticles $b$ and $c$ having joint charge $f$. These states can be generated through a sequence of quasiparticle splitting operations or by using measurements.}
  \label{fig:initialization}
\end{center}
\end{figure}

\begin{figure}[t!]
\begin{center}
  \includegraphics[scale=0.15]{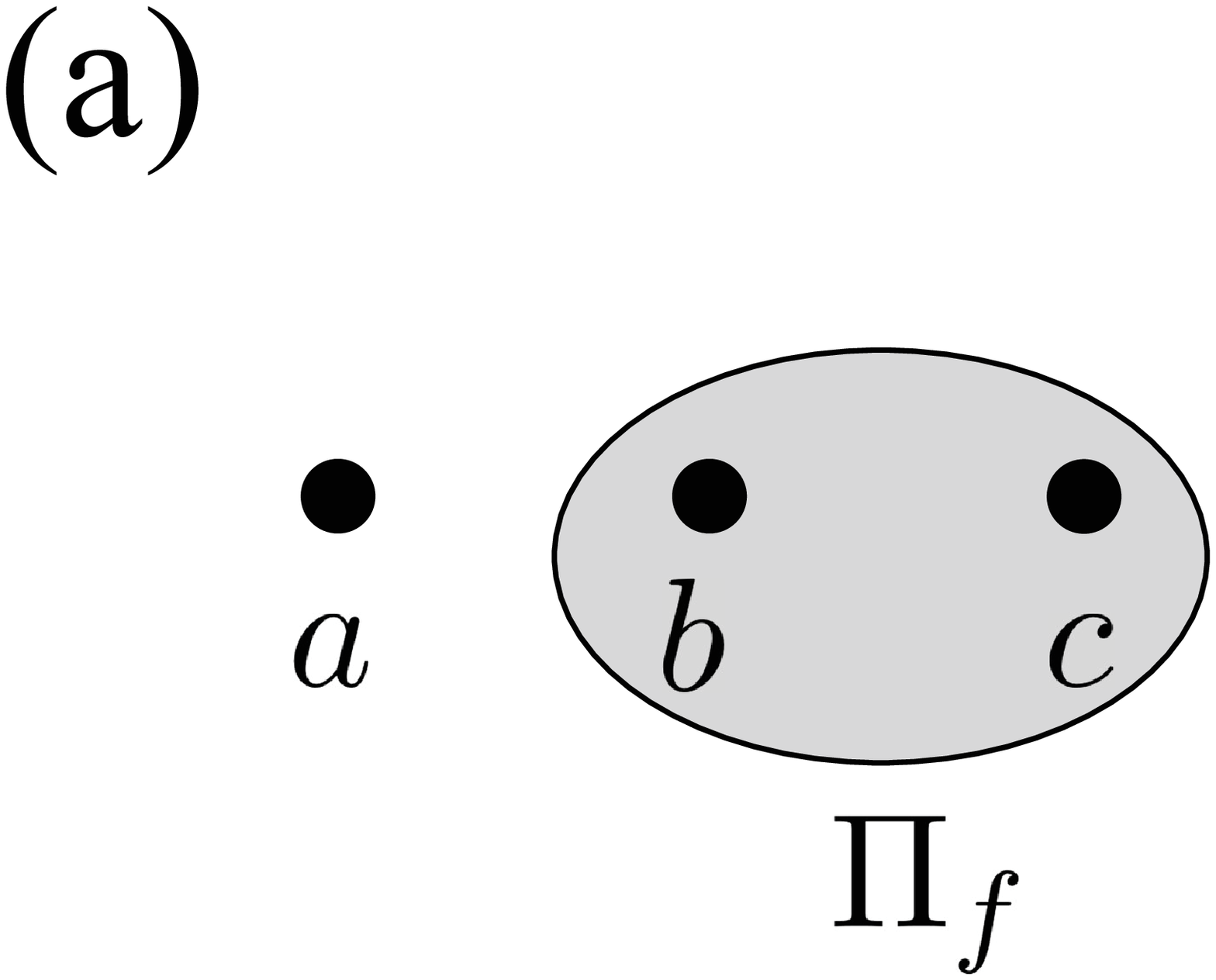}
  \includegraphics[scale=0.15]{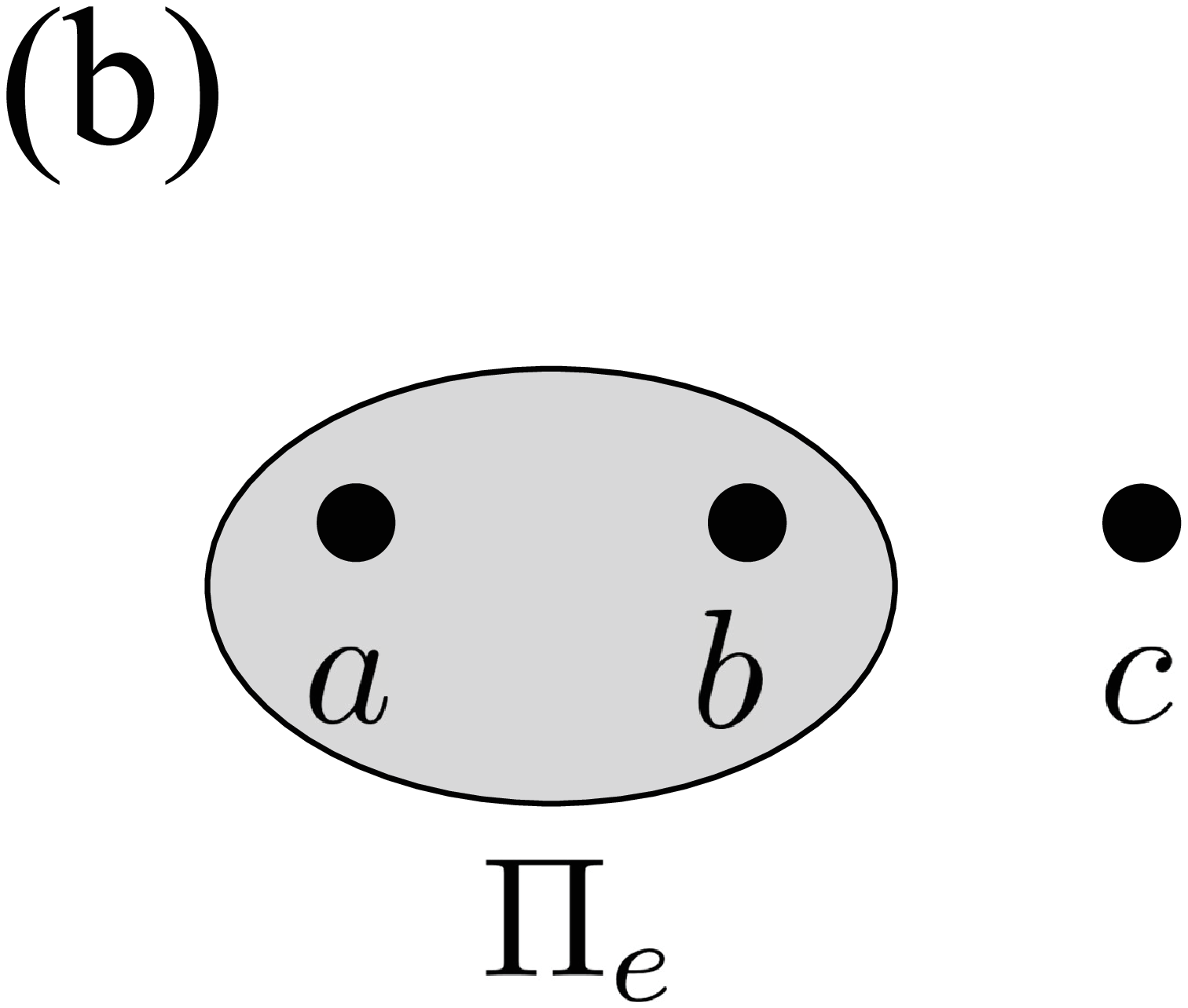}
  \caption{An experiment that can determine the magnitudes of the associativity $F$-symbols. (a) After initializing the system in the state of Fig.~\ref{fig:initialization}(a), the joint topological charge of the $b$-$c$ pair of quasiparticles is measured. This measured fusion channel is found to have topological charge $f$ with probabilities $p_{a(bc);d}(f|e)$ given in Eq.~(\ref{eq:associativity_experiment}). (b) After initializing the system in the state of Fig.~\ref{fig:initialization}(b), the joint topological charge of the $a$-$b$ pair of quasiparticles is measured. This measured fusion channel is found to have topological charge $e$ with probabilities $p_{(ab)c;d}(e|f)$ given in Eq.~(\ref{eq:associativity_experiment_2}).}
  \label{fig:associativity}
\end{center}
\end{figure}

The next experiment involves an initialization setup that localizes quasiparticles of topological charges $a$, $b$, and $c$, in the collective fusion channel $d$, as shown in Fig.~\ref{fig:initialization}(a). This can be realized by the steps:
\begin{enumerate}
  \item Pair-create quasiparticles carrying topological charges $d$ and $\bar{d}$  from vacuum, and move them apart.
  \item Split quasiparticle $d$ into quasiparticles carrying topological charges $e$ and $c$, and move them apart.
  \item Split quasiparticle $e$ into quasiparticles carrying topological charges $a$ and $b$, and move them part.
\end{enumerate}
This creates the three quasiparticles with charges $a$, $b$, and $c$, as well as a fourth quasiparticle of charge $\bar{d}$ that is used to force the fusion channel of the other three to be $d$, but which otherwise will not participate in the experiment~\footnote{There are alternative way of achieving the same desired three quasiparticle setup; for example, one could pair-create quasiparticles carrying topological charges $a$ and $\bar{a}$, $b$ and $\bar{b}$, and $c$ and $\bar{c}$, and then bring together quasiparticles $\bar{a}$, $\bar{b}$, and $\bar{c}$ and measure their collective topological charge.}. With this initial setup, the experiment is performed by the following steps, shown schematically in Fig.~\ref{fig:associativity}:
\begin{enumerate}
\setcounter{enumi}{3}
  \item Measure the collective topological charge of the $b$-$c$ pair of quasiparticles.
  \item Measure the collective topological charge of the $a$-$b$ pair of quasiparticles.
  \item Go to step 4.
\end{enumerate}
The post-measurement state of step 4, which is shown in Fig.~\ref{fig:initialization}(b), becomes the initial (pre-measurement) state for step 5.

The probability that the measurement of the $b$-$c$ pair (at step 4) will have outcome $f$, given that the $a$-$b$ pair previously had collective charge $e$ is given by
\begin{eqnarray}
p_{a(bc);d}(f|e) &=& \left|  \Pi_{f}^{(bc)} \left| a,b ; e \right\rangle \left| e, c ; d \right\rangle \left| d, \bar{d} ; 0 \right\rangle \right|^{2}
\notag \\
&=& \left| \left[F_d^{abc}\right]_{ef} \right|^{2}
.
\label{eq:associativity_experiment}
\end{eqnarray}
This probability may also be computed using Eq.~(\ref{eq:F_magnitude_invariant}), whose diagram is related to the process in question.

Similarly, the probability that the measurement of the $a$-$b$ pair (at step 5) will have outcome $e$ (possibly different from the previous value of $e$), given that the $b$-$c$ pair previously had collective charge $f$ is given by
\begin{eqnarray}
p_{(ab)c;d}(e|f) &=& \left|  \Pi_{e}^{(ab)} \left| b,c ; f \right\rangle \left| a, f ; d \right\rangle \left| d, \bar{d} ; 0 \right\rangle \right|^{2}
\notag \\
&=& \left| \left[F_d^{abc}\right]_{ef} \right|^{2}
.
\label{eq:associativity_experiment_2}
\end{eqnarray}
I note that $p_{(ab)c;d}(e|f) = p_{a(bc);d}(f|e)$.

Repeating steps 4 and 5 many times will yield different measurement outcomes, i.e. different values of $f$ and $e$, and generate statistics that can determine these conditional probabilities for all possible values of the fusion channels of the $b$-$c$ pair and $a$-$b$ pair, respectively.

This set of experiments is seen to contain the fusion and quantum dimension experiments by setting $a=\bar{b}$ and $d = c$, and noting that $p_{\bar{b}(bc);c}(f|0) = p_{bc}(f)$ from Eq.~(\ref{eq:fusion_experiment}). The only reason to perform the fusion and quantum dimension experiments separately from these would be to allow for an experimental setup that is only capable of measuring the joint fusion channel of one fixed pair of quasiparticles. Otherwise, the associativity experiments provide a more efficient method of collecting the data, since the two pair-creation steps for initialization are not required for each round of measurement.

Repeating this entire experiment many times for all possible values of $a$, $b$, $c$, and $d$ will allow one to infer all the $F$-symbols invariants of Eq.~(\ref{eq:F_magnitude_invariant}). This gives the magnitudes of all the $F$-symbols, when the UMTC has no fusion multiplicities. I note that the relation in Eq.~(\ref{eq:F_mag_symm}) makes it clear that the physical positioning of the quasiparticles is unimportant. The primary determining aspect of this experiment is that the joint fusion measurements of quasiparticles are alternating between the $a$-$b$ pair and $b$-$c$ pair. As such, it is not necessary to repeat the experiments for different topological charge configurations that are related by symmetries that leave this property unchanged.

While this set of experiments is (in principle) able to extract the magnitudes of all the $F$-symbols, it does not access the their phases. Some of this phase information may potentially be determined via consistency constraints, such as the pentagon equations and unitarity. Experimental determination of some of the phase information may also be possible using braiding experiments, as I will show.

\subsection{Braiding}
\label{sec:exp_braiding}

In this subsection, I consider bulk quasiparticle experiments that involve braiding operations. Even though these experiments represent direct probes the braiding properties of the quasiparticles, in contrast with the fusion and associativity experiments, not all of these braiding experiments actually reveal additional information about the topological order of the phase. This is because the algebraic consistency conditions of UMTCs impose constraints that relate braiding properties to the fusion and associativity properties, as emphasized at the end of Sec.~\ref{sec:UMTC}. I will point out such relations for the experiments described in the following.

\subsubsection{$S$-matrix}

\begin{figure}[t!]
\begin{center}
  \includegraphics[scale=0.15]{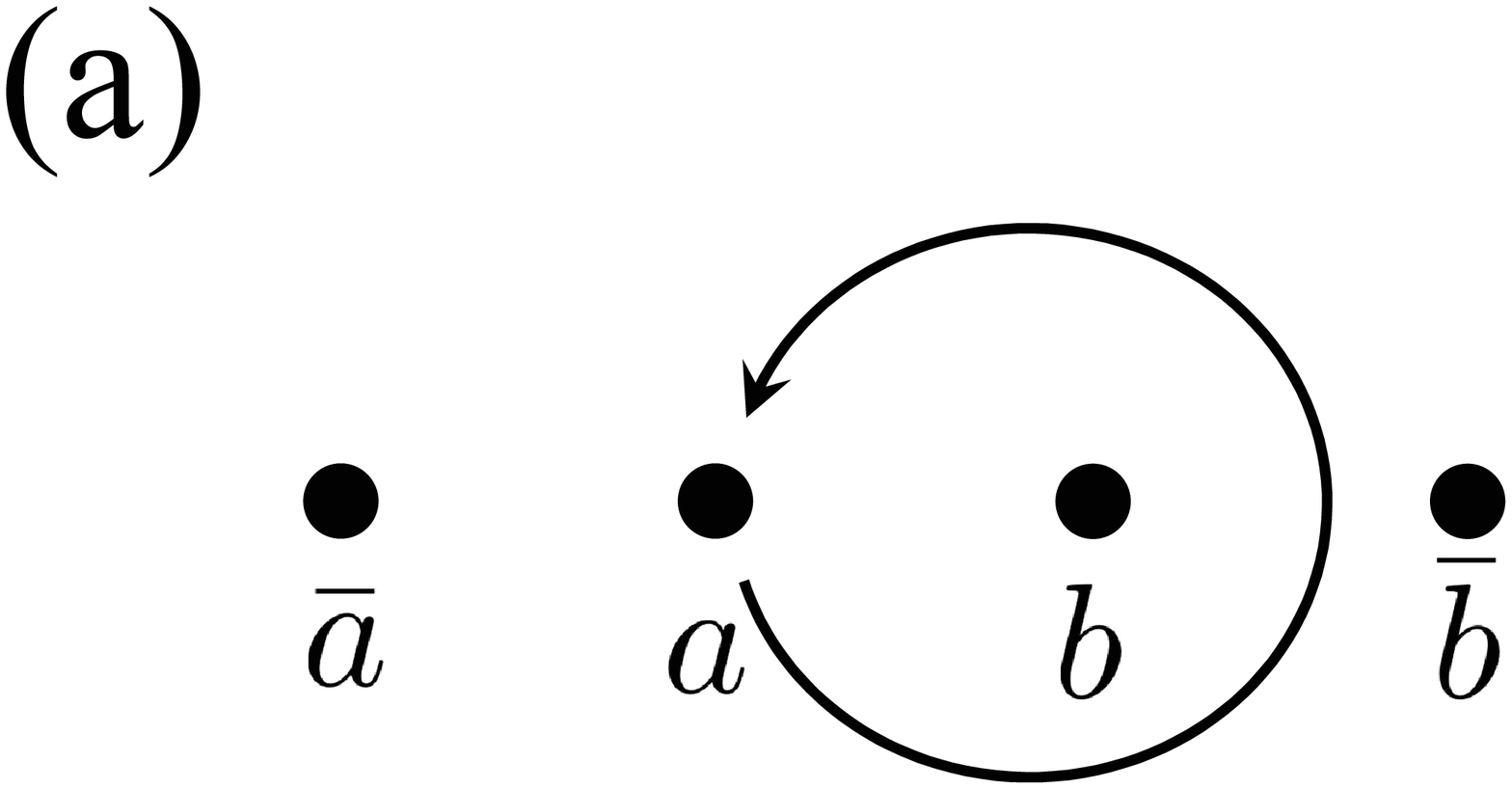}
  \includegraphics[scale=0.15]{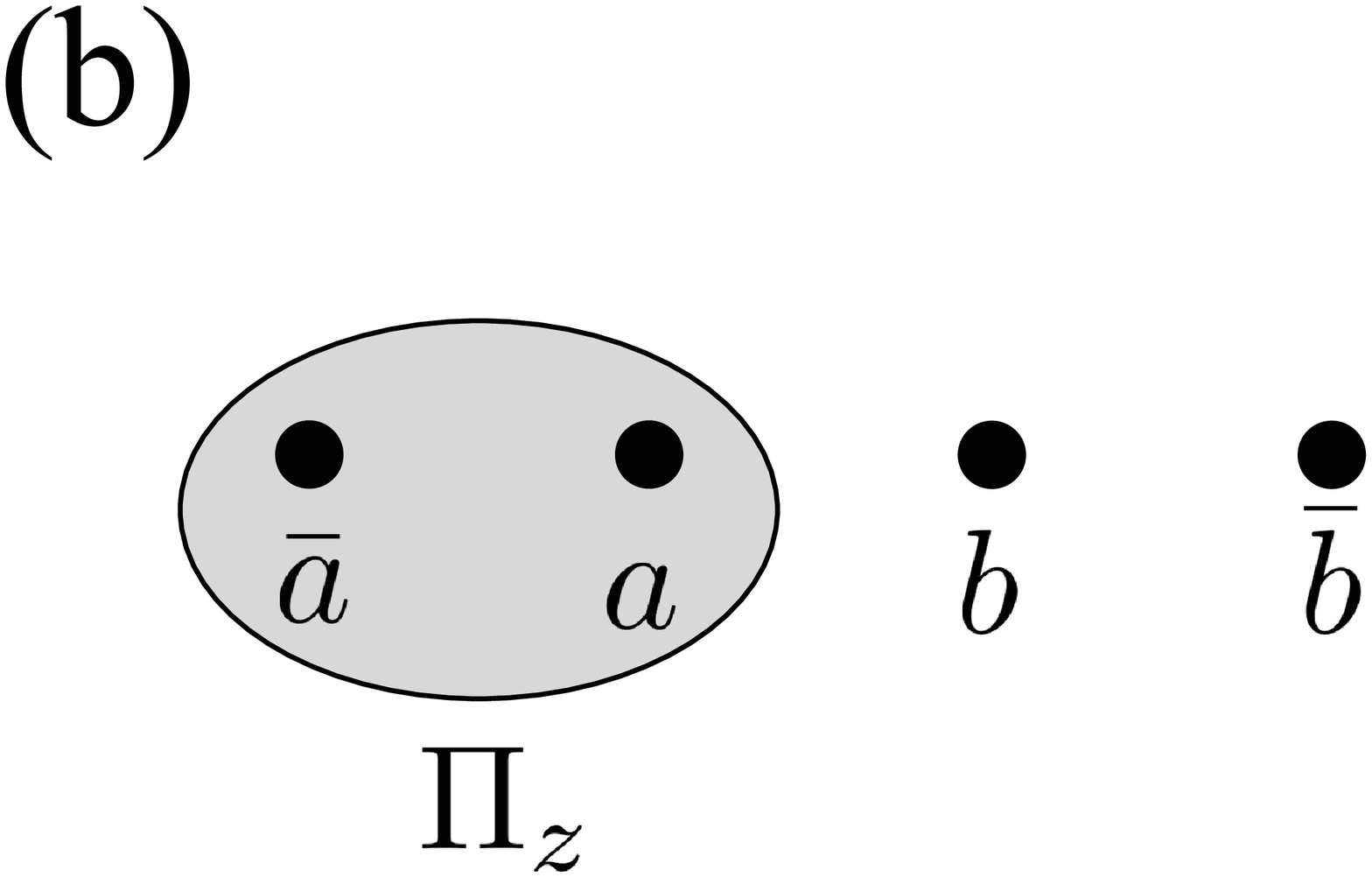}
  \caption{An experiment that can determine the magnitudes of the components of the topological $S$-matrices. (a) After initializing the system in the state of Fig.~\ref{fig:fusion}(a), the quasiparticles $a$ and $b$ are transported one full revolution around each other. (b) After the pure braid of quasiparticles $a$ and $b$, the joint topological charge of the $\bar{a}$-$a$ pair of quasiparticles is measured. This measured fusion channel is found to have topological charge $z$ with probabilities $p_{ab}^{(2)}(z|0)$ given in Eq.~(\ref{eq:S_experiment}).}
  \label{fig:Smatrix}
\end{center}
\end{figure}

The simplest experiment that involves braiding is performed by the following steps, shown schematically in Fig.~\ref{fig:Smatrix}:
\begin{enumerate}
  \item Pair-create quasiparticles carrying topological charges $\bar{a}$ and $a$  from vacuum, and move them apart.
  \item Pair-create quasiparticles carrying topological charges $b$ and $\bar{b}$  from vacuum, and move them apart.
  \item Move quasiparticle $a$ around quasiparticle $b$ once in the counterclockwise direction.
  \item Measure the collective topological charge of the $\bar{a}$-$a$ pair of quasiparticles.
\end{enumerate}
Steps 1 and 2 are the same initialization steps for the fusion experiment in Sec.~\ref{sec:FusionExp}.
The pure braid operation in step 3 is shown in Fig.~\ref{fig:Smatrix}(a). It is equivalent to two successive braiding exchanges of quasiparticles $a$ and $b$, and it can equivalently be implemented by moving quasiparticle $b$ once around $a$ in the counterclockwise direction, or both of them around each other in a way that amounts to a single counterclockwise revolution.
In step 4, the measurement of the $\bar{a}$-$a$ pair, shown in Fig.~\ref{fig:Smatrix}(b), can alternatively be replaced by a measurement of the collective charge of the $b$-$\bar{b}$ pair, the outcome of which will correspond to the conjugate of the outcome of the measurement of the $\bar{a}$-$a$ pair. The measurement of the $\bar{a}$-$a$ pair in this experiment will find topological charge $z$ with probability
\begin{eqnarray}
p_{ab}^{(2)}(z) &=& \left| \Pi_{z}^{(\bar{a} a)}  R^{a b} R^{b a}  \left| \bar{a}, a; 0 \right\rangle \left| b, \bar{b} ; 0 \right\rangle  \right|^{2} \notag \\
&=& \frac{ \mathcal{D}^{2} d_z }{d_{a}^{2} d_{b}^{2}} \left| S^{(z)}_{\bar{a} b} \right|^{2}
\label{eq:S_experiment}
\end{eqnarray}
This probability is computed using Eq.~(\ref{eq:Sz_magnitude_invariant}), whose diagram (when normalized appropriately) represents the process in question.

\subsubsection{Pure Braiding}
\label{sec:exp_purebraiding}

The $S$-matrix braiding experiment can be modified to make it more general and provide additional information. For this, the initial setup can start in the same manner as the associativity experiment by creating four quasiparticles from vacuum that respectively carry topological charge $a$, $b$, $c$, and $\bar{d}$, following the steps 1-3 from Sec.~\ref{sec:AssociativityExp}.

\begin{figure}[t!]
\begin{center}
  \includegraphics[scale=0.15]{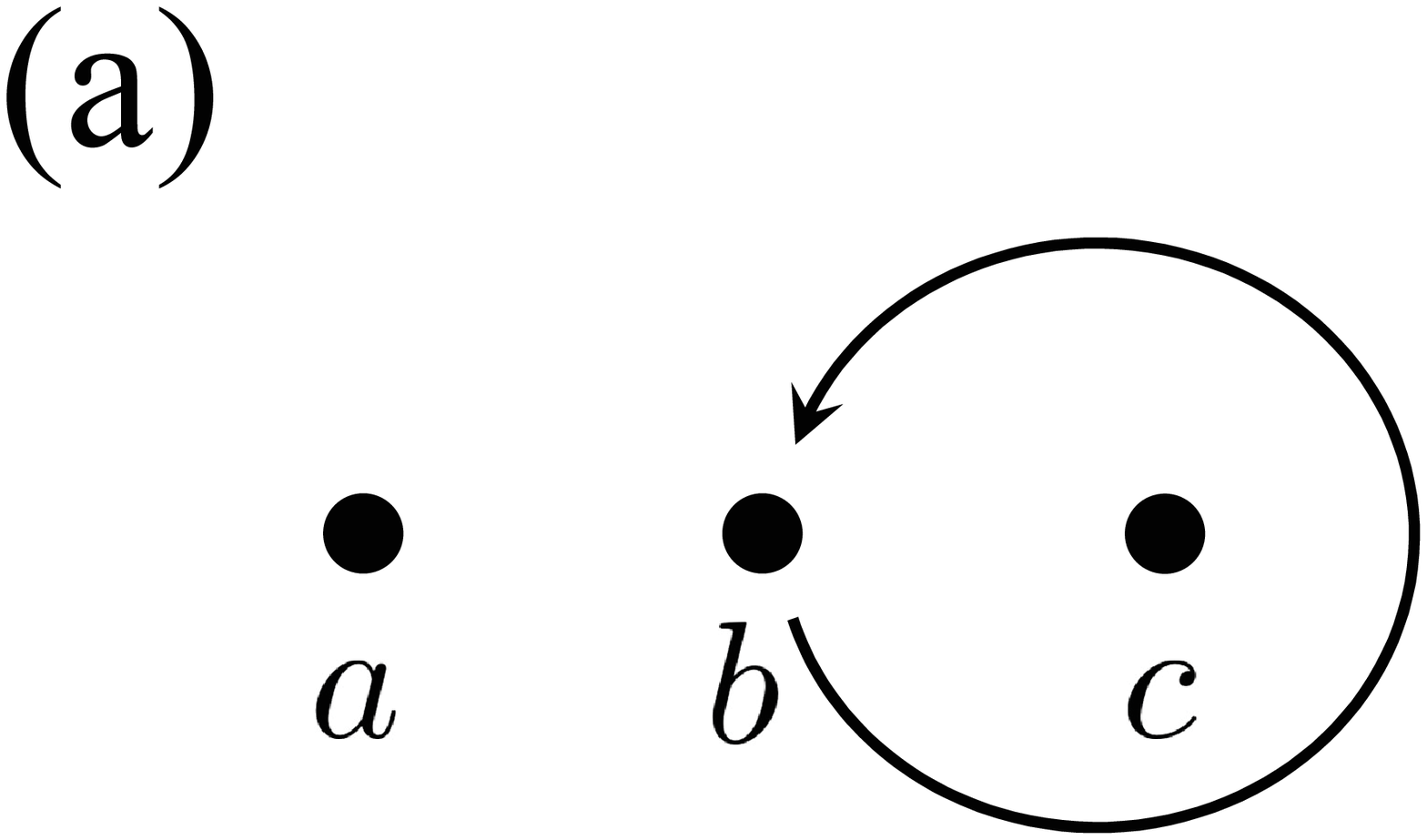}
  \includegraphics[scale=0.15]{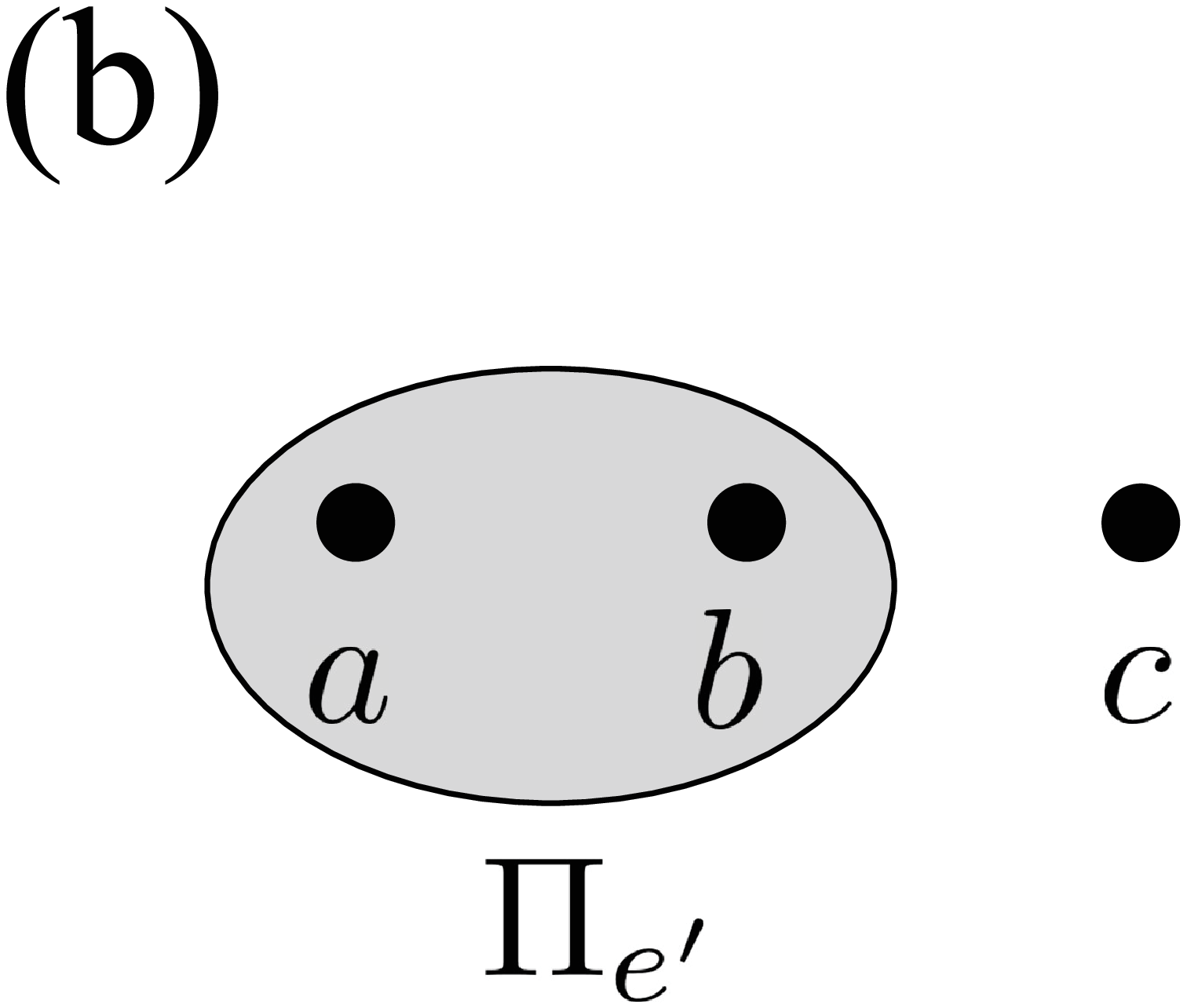}
  \caption{An experiment that can determine braiding properties. (a) After initializing the system in the state of Fig.~\ref{fig:initialization}(a), the quasiparticles $b$ and $c$ are transported around each other for $m$ full revolutions. (b) After the $m$ pure braid operations on quasiparticles $b$ and $c$, the joint topological charge of the $a$-$b$ pair of quasiparticles is measured. This measured fusion channel is found to have topological charge $e'$ with probabilities $p_{abc;d}^{(2m)}(e'|e)$ given in Eq.~(\ref{eq:purebraid_experiment}).}
  \label{fig:braiding}
\end{center}
\end{figure}

With this initial setup, the braiding experiment can be performed by the following repeatable steps, shown schematically in Fig.~\ref{fig:braiding}:
\begin{enumerate}
\setcounter{enumi}{3}
  \item Move quasiparticle $b$ around quasiparticle $c$ in the counterclockwise direction $m$ times.
  \item Measure the collective topological charge of the $a$-$b$ pair of quasiparticles.
  \item Go to step 4.
\end{enumerate}
The $m$ pure braids in step 3 are equivalent to $2m$ braiding exchanges of quasiparticles $b$ and $c$, and can equivalently be implemented by moving quasiparticle $c$ around $b$ in the counterclockwise direction $m$ times, or both of them around each other in a way that amounts to $m$ counterclockwise revolutions.

For a given round of applying steps 4 and 5, where the $a$-$b$ pair initially had collective charge $e$, the probability of the measurement of $a$-$b$ at step 5 having outcome $e'$ will be
\begin{eqnarray}
p_{abc;d}^{(2m)}(e'|e) &=& \left| \Pi_{e'}^{(ab)}  \left(R^{b c} R^{c b} \right)^{m}   \left| a ,b; e \right\rangle \left| e, c ; d \right\rangle \left| d ,\bar{d} ; 0 \right\rangle \right|^{2}
\notag \\
&=& \left| \sum_{f} \left[F_{d}^{a b c}\right]_{ef} \theta_{f}^{m} \left[F_{d}^{a b c}\right]^{\ast}_{e' f}  \right|^{2}
.
\label{eq:purebraid_experiment}
\end{eqnarray}
This probability can be computed using Eq.~(\ref{eq:purebraid_invariant}) with $m' = -m$, as the corresponding diagram in Eq.~(\ref{eq:Braiding_invariant}) is related to the process in question.
This set of experiments is seen to contain the $S$-matrix experiments by setting $a=\bar{b}$ and $d = c$, and noting that $p^{(2)}_{\bar{b}bc;c}(e'|0) = p^{(2)}_{bc}(e')$ from Eq.~(\ref{eq:S_experiment}).

Repeating these experiments many times for all possible values of $a$, $b$, $c$, and $d$, as well as $m$ will allow one to infer information about the twist factors of charges $f$ that are allowed fusion channels of the pairs of charges $b$ and $c$, as well as possibly some information about the $F$-symbols involved. In particular, Eq,~(\ref{eq:purebraid_experiment}) can be written in terms of a sum of the real parts of relative twist and $F$-symbol phase factors, weighted by the magnitudes of $F$-symbols, which can be determined from the associativity experiments. For example, when $e=e'$, the probabilities can be written as
\begin{align}
p_{abc;d}^{(2m)}&(e|e)
= \sum_{f} \left| \left[F_{d}^{a b c}\right]_{ef} \right|^{4} \notag \\
& + 2 \sum_{f < f'} \left| \left[F_{d}^{a b c}\right]_{ef} \left[F_{d}^{a b c}\right]_{ef'} \right|^{2} \text{Re}\left( \frac{\theta_{f}^{m}}{\theta_{f'}^{m}} \right)
,
\end{align}
and $F$-symbol phase factors enter the expressions when $e \neq e'$. (Here, I have assumed some arbitrary ordering on the topological charge set.)

\subsubsection{Exchange Braiding}
\label{sec:exp_exchange}

\begin{figure}[t!]
\begin{center}
  \includegraphics[scale=0.15]{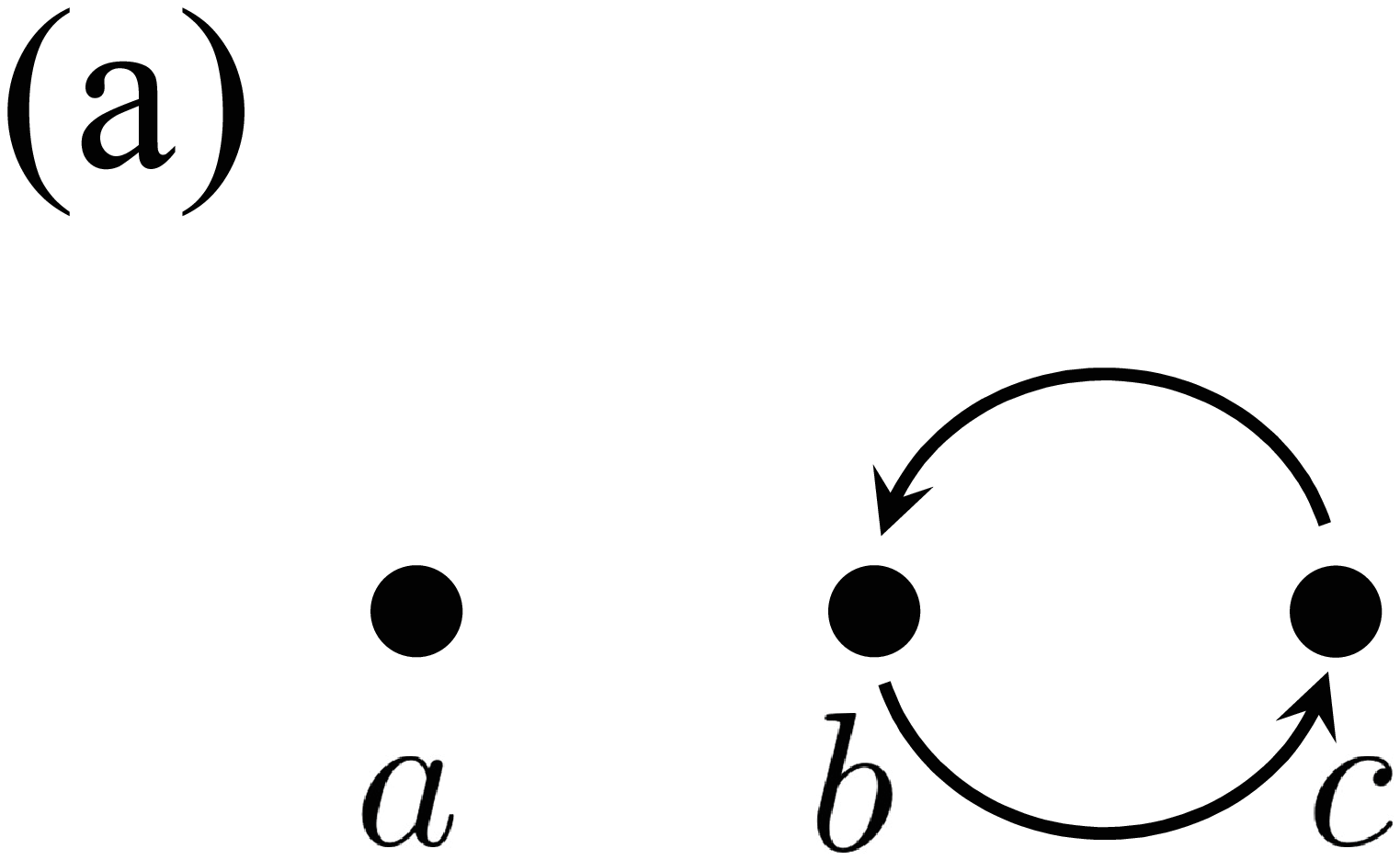}
  \includegraphics[scale=0.15]{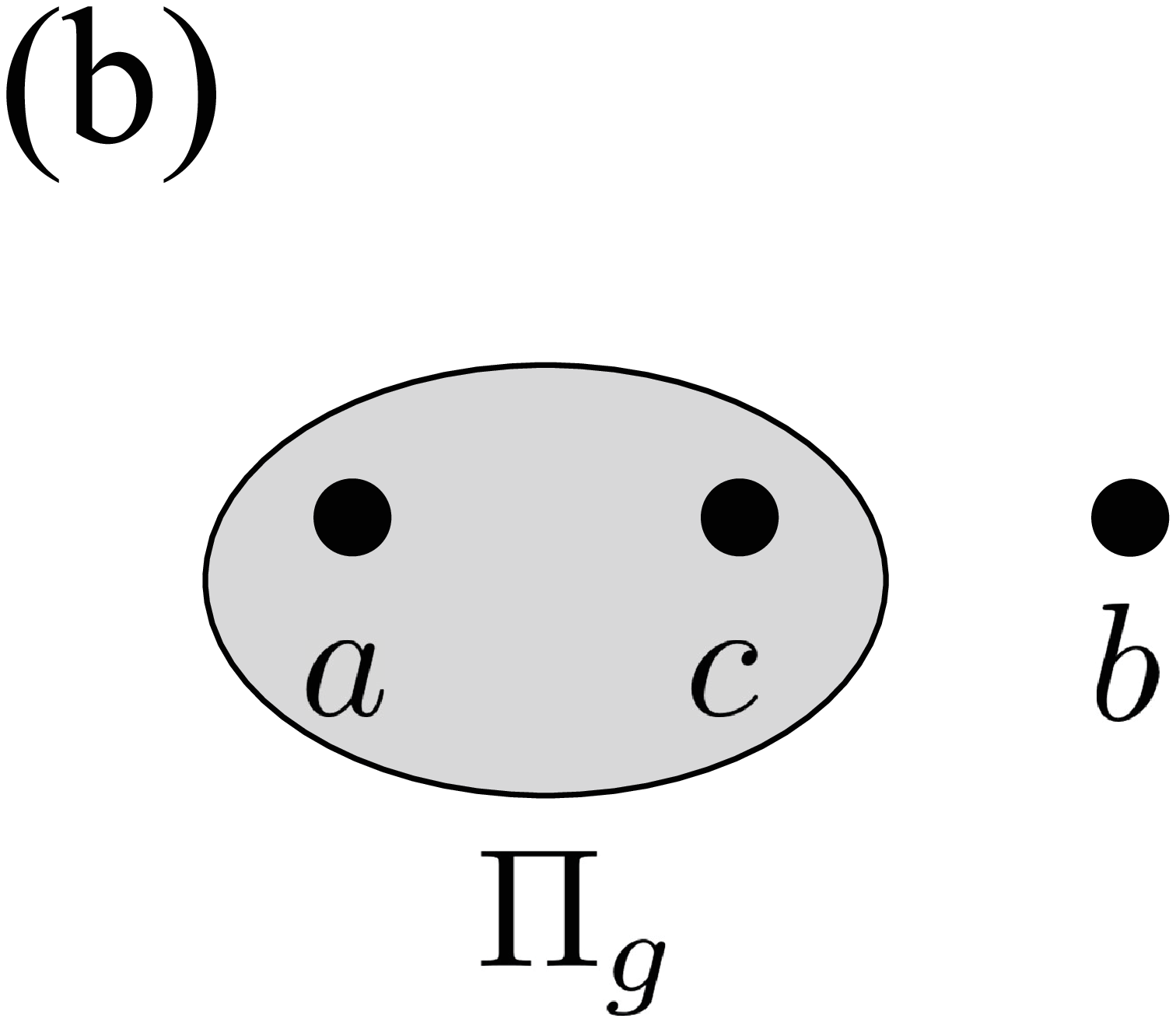}
  \caption{An experiment that can determine braiding properties. (a) After initializing the system in the state of Fig.~\ref{fig:initialization}(a), a braiding exchange of quasiparticles $b$ and $c$ is performed an odd number of times $n$. (b) After the $n$ braiding exchange operations on quasiparticles $b$ and $c$, the joint topological charge of the $a$-$c$ pair of quasiparticles is measured. This measured fusion channel is found to have topological charge $g$ with probabilities $p_{abc;d}^{(n)}(g|e)$ given in Eq.~(\ref{eq:braidexchange_experiment}).}
  \label{fig:exchange}
\end{center}
\end{figure}

A similar braiding experiment can be performed involving an odd number $n$ of braiding exchange operations (allowing braiding exchanges in addition to pure braids) of quasiparticles $b$ and $c$. For this, one starts with the same initial setup in steps 1-3, and then follows the repeatable steps, shown schematically in Fig.~\ref{fig:exchange}:
\begin{enumerate}
\setcounter{enumi}{3}
  \item Exchange the positions of quasiparticle $b$ and quasiparticle $c$ in the counterclockwise direction $n$ times (where $n$ is odd).
  \item Measure the collective topological charge of the $a$-$c$ pair of quasiparticles.
  \item Exchange the positions of quasiparticle $c$ and quasiparticle $b$ in the counterclockwise direction $n$ times (where $n$ is odd).
  \item Measure the collective topological charge of the $a$-$b$ pair of quasiparticles.
  \item Go to step 4.
\end{enumerate}
For a given round of applying steps 4-7, where the $a$-$b$ pair initially had collective charge $e$, the probability of the measurement of $a$-$c$ at step 5 having outcome $g$ will be
\begin{eqnarray}
&& p_{abc;d}^{(n)}(g|e) \notag \\
&&  =\left| \Pi_{g}^{(ac)}  R^{c b} \left(R^{b c} R^{c b} \right)^{\frac{n-1}{2}}   \left| a ,b; e \right\rangle \left| e, c ; d \right\rangle \left| d ,\bar{d} ; 0 \right\rangle \right|^{2}
\notag \\
&& = \left| \sum_{f} \left[F_{d}^{a b c}\right]_{ef} R^{c b}_{f} \theta_{f}^{\frac{n-1}{2}} \left[F_{d}^{a c b}\right]^{\ast}_{g f}  \right|^{2}
,
\label{eq:braidexchange_experiment}
\end{eqnarray}
and given the measurement outcome $g$ at step $5$, the probability of measurement outcome $e$ at step $7$ will be
\begin{equation}
p_{acb;d}^{(n)}(e|g) = p_{abc;d}^{(-n)}(g|e)
.
\end{equation}
These probabilities can be computed using Eq.~(\ref{eq:exchangebraid_invariant}) with $n' = -n$, as the corresponding diagram in Eq.~(\ref{eq:Braiding_invariant}) is related to the process in question.
While $R^{c b}_{f}$ are not a gauge invariant quantities when $b \neq c$, the probabilities expressed in Eq.~(\ref{eq:braidexchange_experiment}) are invariants.

Repeating these experiments many times for all possible values of $a$, $b$, $c$, and $d$, as well as $m$ and $n$ will allow one to infer information about the relative braiding phases and twist factors between the topological charges that show up as fusion channels $f$ of the pairs of charges $b$ and $c$.

I note that, even though these experiments provide direct probes of the braiding properties, not all of the variations of these experiments provide additional information about the topological order of the phase. For example, the information about the topological order that may be gained from the $n=1$ braiding exchange experiments is already provided by the associativity experiments, since
\begin{equation}
\label{eq:n1exchange_relation}
p_{abc;d}^{(1)}(g|e) = \left| \left[B_{d}^{a c b}\right]_{e g}  \right|^{2} = \left|  \left[F_{d}^{b a c}\right]_{e g}  \right|^{2} = p_{b(ac);d}(g|e)
.
\end{equation}
This relation is obtained by first applying the definition of the $B$-symbol, and then using Eq.~(\ref{eq:magB_magF}).

In the case where $c=b$, steps 6 and 7 do not need to be distinguished from steps 4 and 5, and the corresponding measurement outcome probabilities become
\begin{equation}
p_{abb;d}^{(n)}(e'|e)  = \left| \sum_{f} \left[F_{d}^{a b b}\right]_{ef} \left(R^{b b}_{f}\right)^{n} \left[F_{d}^{a b b}\right]^{\ast}_{e' f}  \right|^{2}
.
\label{eq:braidexchange_experiment_3}
\end{equation}
In fact, this expression holds for all integer values of $n$. I note that $R^{b b}_{f} = \frac{\sqrt{\theta_{f}}}{\theta_{b}} \Lambda^{bb}_{f}$ is a gauge invariant root of unity when $N_{bb}^{f}=1$.

%

\subsubsection{Non-Abelian Braiding}
\label{sec:exp_nonAbelian_braiding}

\begin{figure*}[t!]
\begin{center}
  \includegraphics[scale=0.15]{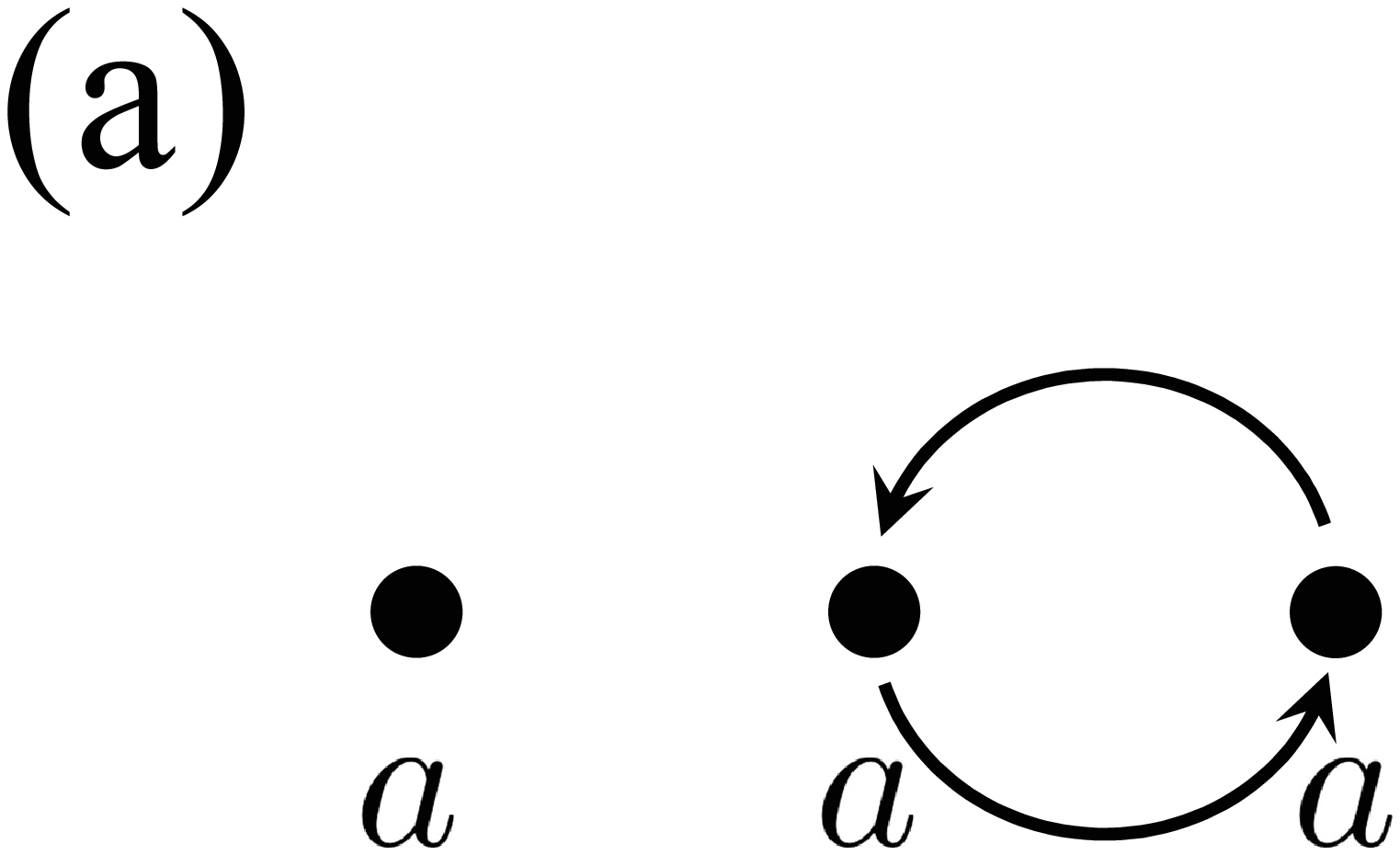}
  \includegraphics[scale=0.15]{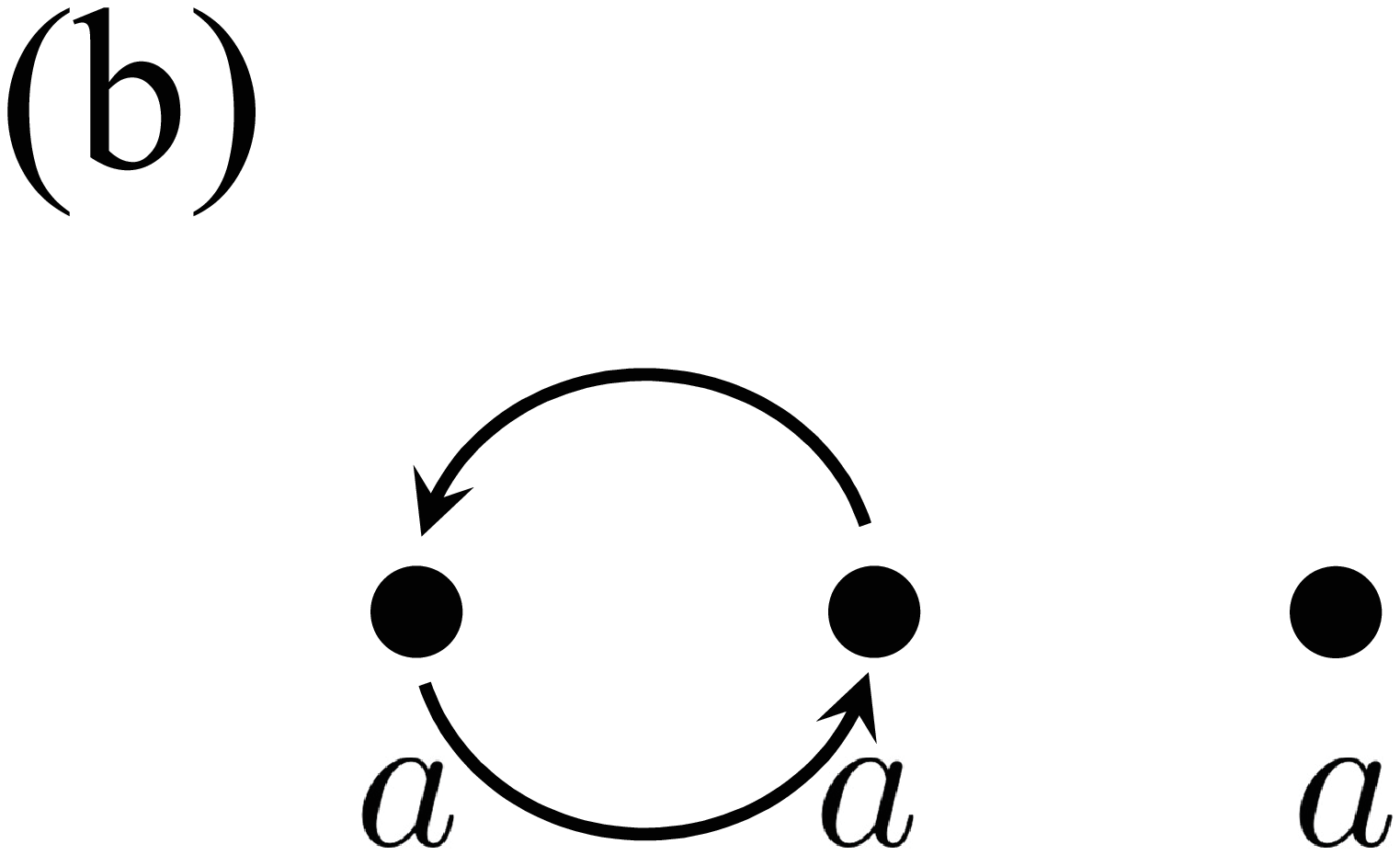}
  \includegraphics[scale=0.15]{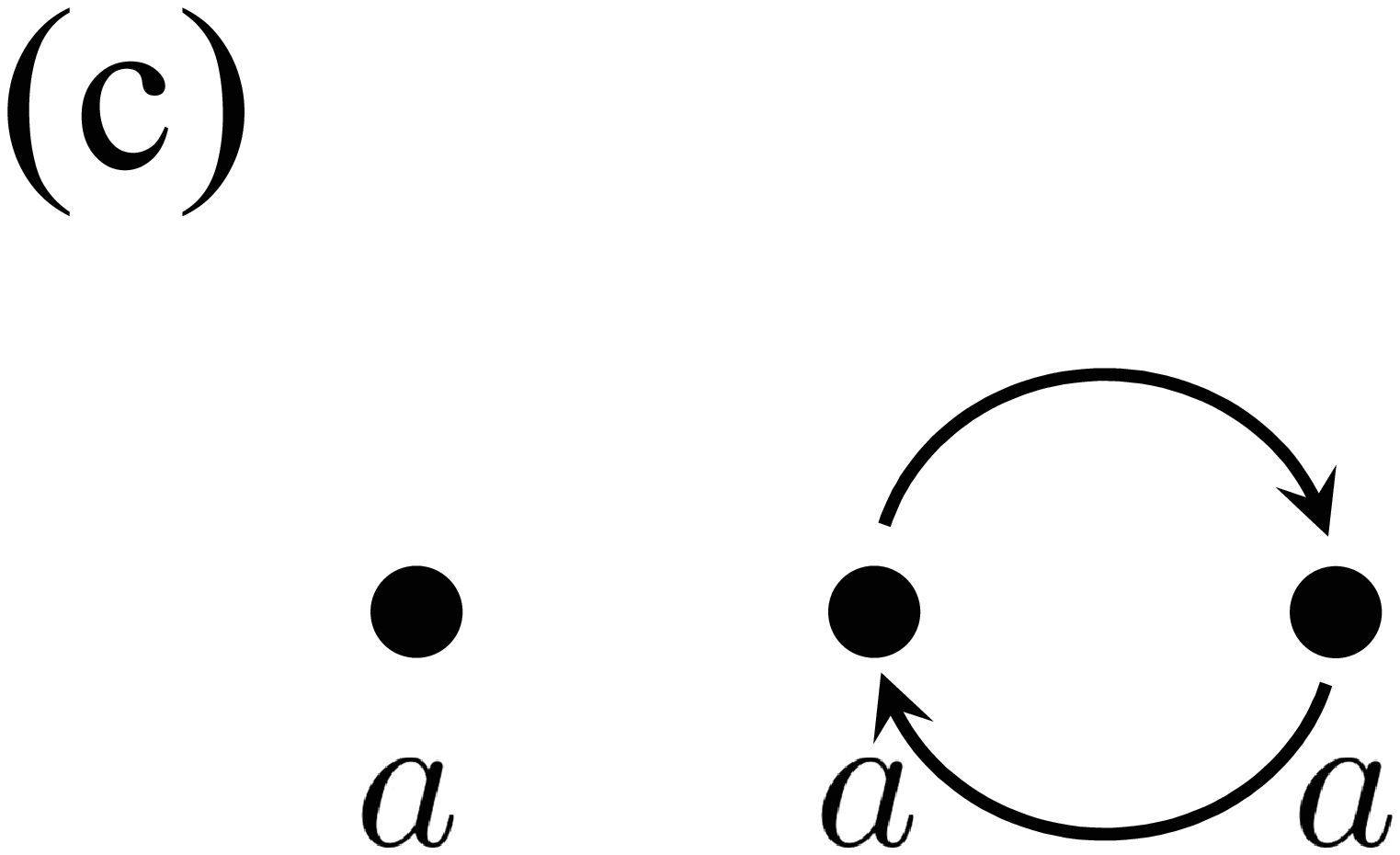}
  \includegraphics[scale=0.15]{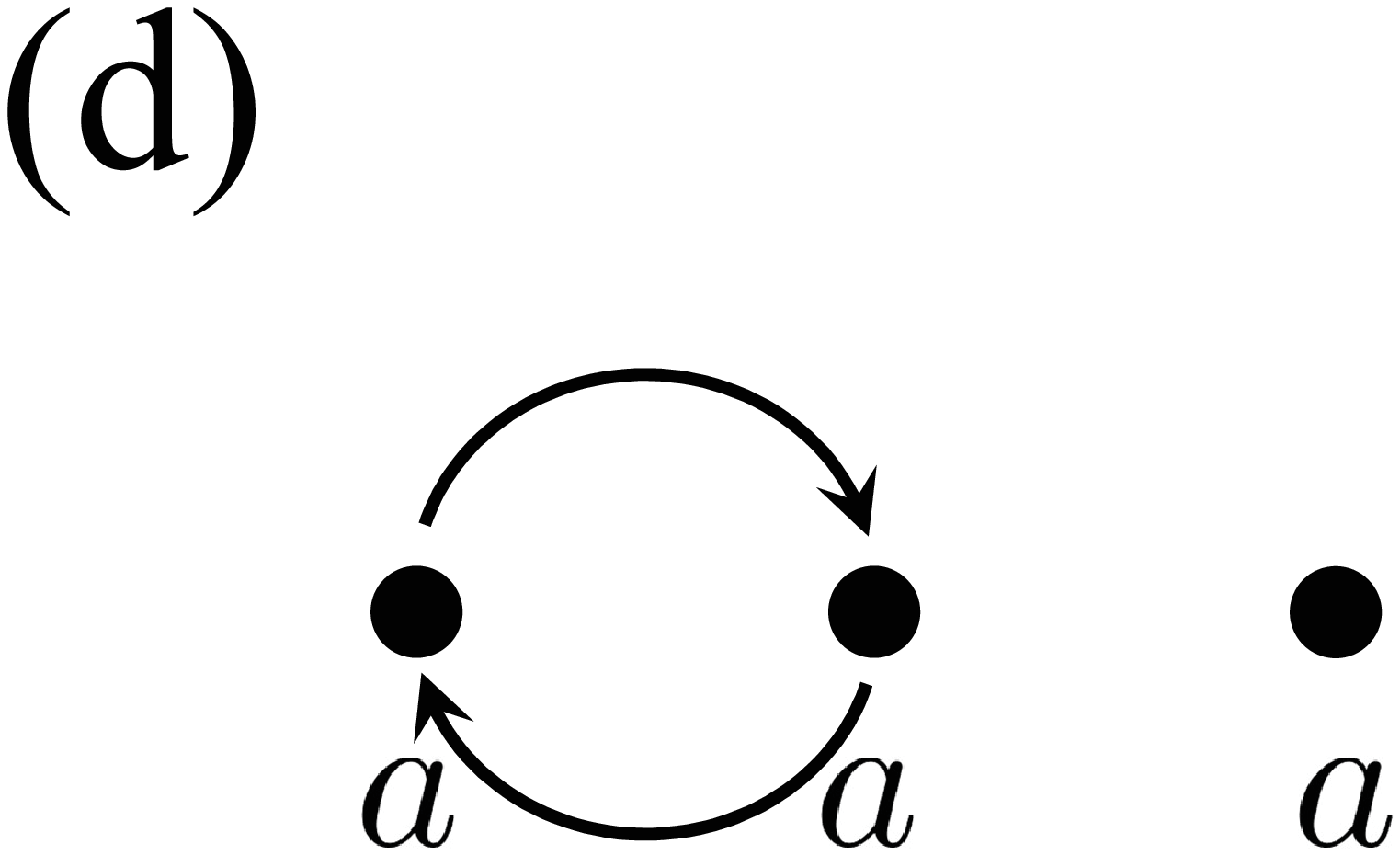}
  \includegraphics[scale=0.15]{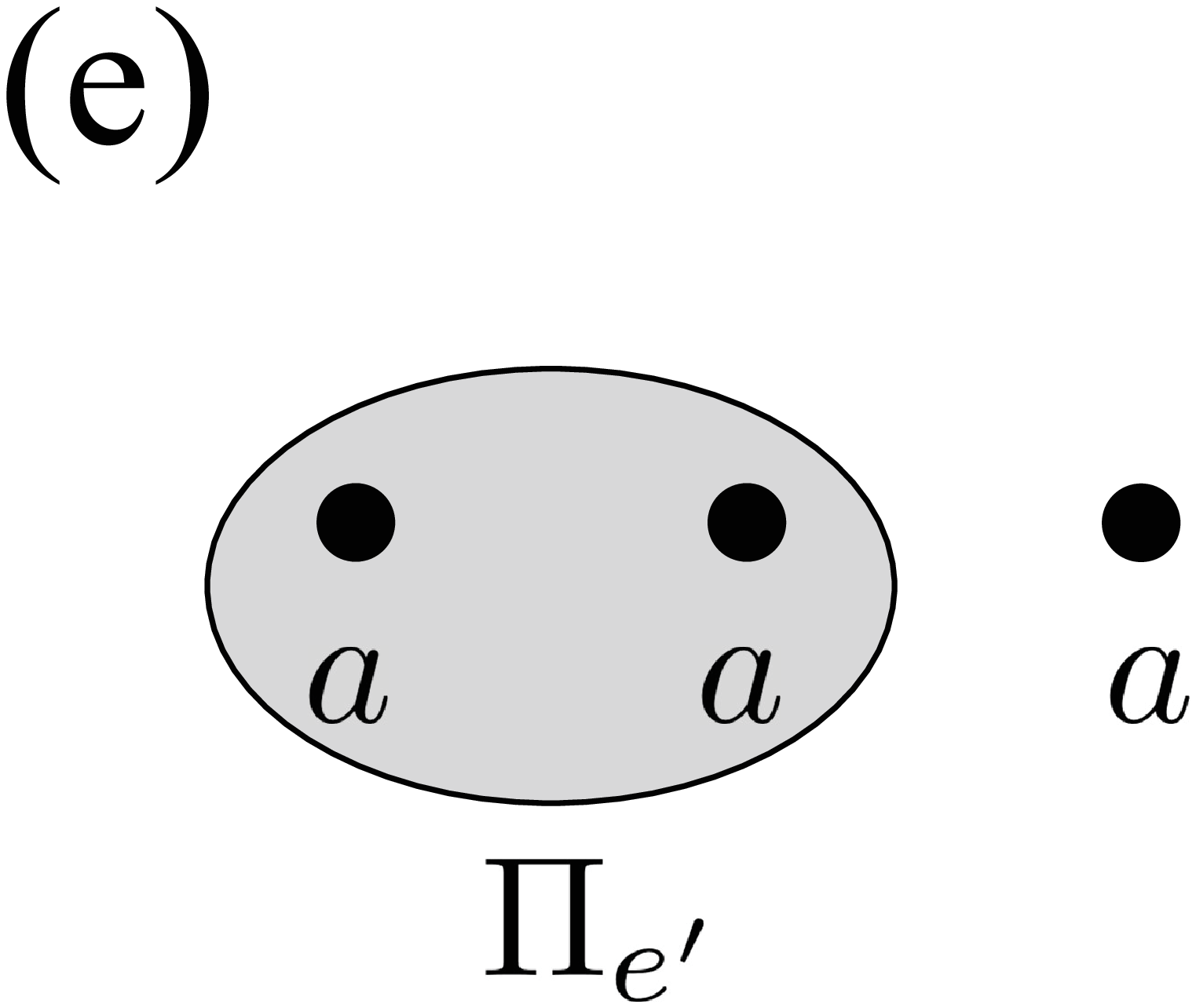}
  \caption{An experiment that can detect non-Abelian braiding. After initializing the system in the state of Fig.~\ref{fig:initialization}(a) with $a=b=c$, a commutator of braiding exchange operations $R_{12}^{-1} R_{23}^{-1} R_{12} R_{23}$ is implemented by performing the sequence of braiding exchanges shown sequentially in (a-d). (e) After performing the braiding sequence $R_{12}^{-1} R_{23}^{-1} R_{12} R_{23}$, the joint topological charge of the pair of quasiparticles at positions 1 and 2 is measured. This measured fusion channel is found to have topological charge $e'$ with probabilities $p_{aaa;d}^{(\text{com})}(e'|e)$ given in Eq.~(\ref{eq:nonAbelian_experiment}). A measurement outcome $e' \neq e$ is an indication of non-Abelian braiding statistics.}
  \label{fig:nonAbelian}
\end{center}
\end{figure*}

A property of much interest regarding anyons is whether their braiding is Abelian or non-Abelian in nature. More specifically, the question is whether the product of two distinct braiding operations is independent of the order they are applied, when the final configurations resulting from different orders of application are the same. A general experiment to test this property may be carried out for three identical quasiparticles carrying topological charge $a$, whose collective fusion channel is $d$ (and there will be a fourth quasiparticle of charge $\bar{d}$ that compensates for their collective charge). Labeling the positions (of the pinning potentials) where the three $a$ quasiparticles can be localized before and after braiding, I consider the braiding exchanges $R_{12}$ interchanging the quasiparticles at positions 1 and 2 in a counterclockwise fashion, and $R_{23}$ interchanging the quasiparticles at positions 2 and 3 in counterclockwise fashion. Then the question is whether $R_{12} R_{23}$ is equal to $R_{23} R_{12}$? The simplest way to probe this is to implement the commutator $R_{12}^{-1} R_{23}^{-1} R_{12} R_{23}$ and inspecting whether the resulting operation is identity. The initial setup for this experiment can start in the same manner as the associativity experiment by creating four quasiparticles from vacuum that respectively carry topological charge $a$, $a$, $a$, and $\bar{d}$, following the steps:
\begin{enumerate}
  \item Pair-create quasiparticles carrying topological charges $d$ and $\bar{d}$  from vacuum, and move them apart.
  \item Split quasiparticle $d$ into quasiparticles carrying topological charges $e$ and $a$, and move them apart to positions 2 and 3, respectively.
  \item Split quasiparticle $e$ into quasiparticles carrying topological charges $a$ and $a$, and move them apart to positions 1 and 2.
\end{enumerate}
With this initial setup, the braiding experiment can be performed by the following repeatable steps, shown schematically in Fig.~\ref{fig:nonAbelian}:
\begin{enumerate}
\setcounter{enumi}{3}
  \item Exchange the quasiparticles at positions $2$ and $3$ in the counterclockwise direction.
  \item Exchange the quasiparticles at positions $1$ and $2$ in the counterclockwise direction.
  \item Exchange the quasiparticles at positions $2$ and $3$ in the clockwise direction.
  \item Exchange the quasiparticles at positions $1$ and $2$ in the clockwise direction.
  \item Measure the collective topological charge of the $a$-$a$ pair of quasiparticles at positions 1 and 2.
  \item Go to step 4.
\end{enumerate}
If the braiding operations are Abelian, every repetition of this experiment will result in the measurement at step 8 finding the same topological charge value $e$ as the initial value.
Thus, if repetitions of this experiment yield the observation of more than one possible measurement outcome for the fusion channel of the quasiparticles at positions 1 and 2, then it is a clear indication that the commutator of braiding operations is not proportional to identity, and hence the braiding of type $a$ quasiparticles is non-Abelian.

Indeed, given that the initial fusion channel of the quasiparticles at positions 1 and 2 was $e$, the probability of the measurement in step 8 having outcome $e'$ is
\begin{eqnarray}
&& p_{aaa;d}^{\text{com}}(e'|e) \notag \\
&&= \left| \Pi_{e'}^{(12)}  R^{-1}_{12} R^{-1}_{23} R_{12} R_{23} \left| a ,a; e \right\rangle \left| e, a ; d \right\rangle \left| d ,\bar{d} ; 0 \right\rangle \right|^{2}
\notag \\
&&= \left| \Pi_{e'}^{(12)}  R_{23} R^{-1}_{12} \left| a ,a; e \right\rangle \left| e, a ; d \right\rangle \left| d ,\bar{d} ; 0 \right\rangle \right|^{2}
\notag \\
&&= \left| \left[B_{d}^{a a a}\right]_{ee'} \right|^{2} = \left| \left[F_{d}^{a a a}\right]_{e e'} \right|^{2}
.
\label{eq:nonAbelian_experiment}
\end{eqnarray}
The first equivalence in the above equation is obtained by application of the Yang-Baxter relation for braids $R_{12} R_{23} R_{12} = R_{23} R_{12} R_{23}$; the second equivalence is obtained by evaluating the operations; and the final equivalence is an application of Eq.~(\ref{eq:magB_magF}). Thus, $p_{aaa;d}^{\text{com}}(e'|e)  = p_{a(aa);d}(e'|e)$ from Eq.~(\ref{eq:associativity_experiment}). The fact that this probability can be reduced to an expression that only involves $F$-symbols further highlights how ``non-Abelian fusion'' (i.e. having multiple fusion channels) implies the existence of non-Abelian braiding.

I note that this experiment actually distinguishes whether or not the commutator is \emph{proportional} to identity.
As such, it is insensitive to effects (universal or non-universal) that introduce overall phase factors.
This includes moving the quasiparticles on different time scales or along different, but topologically equivalent, paths, as well as when the quasiparticles carry the same topological charge values, but are not identical objects, for example having different electric charge values.
This make the experiment a robust probe of the existence of non-Abelian braiding statistics.

\subsection{Examples}

It is instructive to consider how the fusion and braiding properties of a topological phase may be extracted from the described experiments.
This is most easily examined for the case where $a$ and $b$ have two allowed fusion channels, which I write as the charges $e_1$ and $e_2$, and $b$ and $c$ have two allowed fusion channels, which I write as the charges $f_1$ and $f_2$.

\begin{table*}
\[
\begin{array}{|c|c|c|}
\hline
\mathcal{C} = \{I, \sigma, \psi\}  & \psi\times\psi=I, \quad \sigma\times\psi=\sigma, \quad \sigma\times\sigma=I+\psi &  d_I = d_\psi =1, \quad d_{\sigma} = \sqrt{2} \\
\hline
\multicolumn{2}{|c|}{ F^{\psi\sigma\psi}_\sigma=F^{\sigma\psi\sigma}_\psi=-1, \quad \left[F^{\sigma\sigma\sigma}_\sigma\right]_{ef}=
\frac{\kappa_{\sigma}}{\sqrt{2}} \left[
\begin{matrix}
1 & 1\\
1 & -1
\end{matrix}
\right]_{ef_{\phantom{g}}}^{\phantom{T}} }          &  \kappa_{\sigma} = (-1)^{\frac{\nu^{2} -1 }{8}}  \\
\hline
\multicolumn{2}{|c|}{  R^{\psi\sigma}_\sigma=R^{\sigma\psi}_\sigma=(-i)^{\nu}, \quad R^{\sigma\sigma}_{I}=\kappa_{\sigma} e^{-i\frac{\pi}{8}\nu}, \quad R^{\sigma\sigma}_\psi=\kappa_{\sigma} e^{i\frac{3\pi}{8}\nu}} & \theta_I =1, \theta_\psi =-1, \theta_\sigma = e^{i\frac{\pi}{8}\nu }
\\
\hline
\multicolumn{2}{|c|}{
S = \frac{1}{2} \left[
\begin{array}{ccc}
1 & \sqrt{2} & 1 \\
\sqrt{2} & 0 & -\sqrt{2} \\
1 & -\sqrt{2} & 1%
\end{array}%
\right]^{\phantom{T}}_{\phantom{g}}
, \quad
S^{(\psi)} = e^{-i \frac{\pi}{4} \nu}
}  &  c_- \text{ mod }8 = \frac{\nu}{2}
\\
\hline
\hline
p_{\sigma \sigma}(I) = p_{\sigma \sigma}(\psi) = \frac{1}{2} &  p_{\sigma (\sigma \sigma); \sigma}(f|e) = p_{(\sigma \sigma) \sigma; \sigma}(e|f) = \frac{1}{2} & p_{\sigma \sigma}^{(2)}(I) = 0 , \quad p_{\sigma \sigma}^{(2)}(\psi) = 1
\\
\hline
\multicolumn{2}{|c|}{
p_{\sigma \sigma \sigma ;\sigma}^{(n)}(e'|e) = \left\{
\begin{array}{lll}
\delta_{e,e'} & \quad & \text{ for $n=0 \text{ mod }4$}   \\
\frac{1}{2} & \quad & \text{ for $n$ odd} \\
1- \delta_{e,e'} & \quad & \text{ for $n=2 \text{ mod }4$}
\end{array}
\right.
}
&  p_{\sigma \sigma \sigma ;\sigma}^{\text{com}}(e'|e) = \frac{1}{2}
\\
\hline
\end{array}
\]
\caption{The basic data of Ising$^{(\nu)}$ UMTCs, along with some of the topological invariants and the predicted probability outcomes for the experiments described in this section [Eqs.~(\ref{eq:fusion_experiment}), (\ref{eq:associativity_experiment}), (\ref{eq:purebraid_experiment}), (\ref{eq:braidexchange_experiment}), and (\ref{eq:nonAbelian_experiment})]. Here, $\nu \in \{1 , 3, \ldots , 15 \}$ is an odd integer mod 16 that characterizes the eight different UMTCs with these fusion rules, $I$ is the vacuum charge, and $e,e',f \in \{ I, \psi \}$. Fusion with vacuum, and $F$-symbols and $R$-symbols that equal 1 are not shown. These are all the possible UBTCs with these fusion rules.}
\label{Table:Ising}
\end{table*}

The fusion rules experiments will determine these to be the allowed fusion channels, and will reveal the corresponding quantum dimensions of these fusion channels.
The associativity experiments may be used to determine the magnitudes of all the $F$-symbol components.
The relevant $F$-symbol in this case is a $2 \times 2$ unitary matrix
\begin{equation}
F^{a b c}_{d} = \left[
\begin{matrix}
  F_{11} & F_{12} \\
  F_{21} & F_{22}
\end{matrix}
\right]
.
\end{equation}
The condition of unitarity requires $|F_{22}| = |F_{11}|$, $|F_{12}|=|F_{21}|=\sqrt{1 - |F_{11}|^{2}}$, and $F_{12} F_{22}^{\ast} = - F_{11}F_{21}^{\ast}$, so all the magnitudes are related. The phases of the $F$-symbol components cannot be determined by the associativity experiments, though it may be possible to derive some information about them from the algebraic conditions required to be satisfied by UMTCs.

The measurement outcome probabilities for the pure braid experiments in this case are
\begin{eqnarray}
p_{abc;d}^{(2m)}(e_{1}|e_{1}) &=& p_{abc;d}^{(2m)}(e_{2}|e_{2}) \notag \\
&=& \left|F_{11} \right|^4 +\left|F_{12} \right|^4 +2 \left|F_{11} F_{12} \right|^2 \cos(m \varphi) \notag \\
&=&  1- 2\left| F_{11} F_{12} \right|^2 \left[1 - \cos( m \varphi ) \right] , \\
p_{abc;d}^{(2m)}(e_{1}|e_{2}) &=& p_{abc;d}^{(2m)}(e_{2}|e_{1}) \notag \\
&=&  2\left| F_{11} F_{12} \right|^2  \left[1 - \cos( m \varphi ) \right]
,
\end{eqnarray}
where $e^{i \varphi} = \frac{\theta_{f_2}}{\theta_{f_1}}$ is the relative phase between the twist factors of $f_1$ and $f_2$.
Having first determined the magnitudes of the $F$-symbols from the associativity experiments, the pure braid experiments provides the value of $\cos(\varphi) = \mathrm{Re} \left(\frac{\theta_{f_2}}{\theta_{f_1}} \right)$. In this case, only the $m=1$ experiment is needed to extract the relevant braiding data, i.e. $\cos(\varphi)$; however, performing the $m\neq 1$ experiments may provide useful corroboration of the extracted information. If the topological charges are chosen such that $c=\bar{b}$, so that $f_1 =0$, this provides the real part of the twist factor of $f_2$.

Using the same assumptions, but letting $c=b$, the resulting measurement outcome probabilities for the exchange braiding experiments are similarly obtained to be
\begin{eqnarray}
p_{abb;d}^{(n)}(e_{1}|e_{1}) &=& p_{abb;d}^{(n)}(e_{2}|e_{2}) \notag \\
&=&  1- 2\left| F_{11} F_{12} \right|^2 \left[1 - \cos( n \lambda ) \right] ,
\\
p_{abb;d}^{(n)}(e_{1}|e_{2}) &=& p_{abb;d}^{(n)}(e_{2}|e_{1}) \notag \\
&=&  2\left| F_{11} F_{12} \right|^2  \left[1 - \cos( n \lambda ) \right]
,
\end{eqnarray}
where $e^{i \lambda} = \frac{R^{bb}_{f_2}}{R^{bb}_{f_1}} = \frac{\sqrt{\theta_{f_2}} \Lambda^{bb}_{f_2}}{ \sqrt{\theta_{f_1}}\Lambda^{bb}_{f_1}}$. Recall from Eq.~(\ref{eq:n1exchange_relation}) that the $n=1$ experiment only yields information about the topological order that can already be obtained from associativity experiments. Hence, this is also true for all $n$, in this situation. Moreover, setting $a = \bar{b}$ and using Eq.~(\ref{eq:magB_magF}) yields the relation
\begin{equation}
\text{Re} \left( \frac{R^{bb}_{f_2}}{R^{bb}_{f_1}} \right)  = -\frac{d_{f_1} (d_{f_1}-1) + d_{f_2} (d_{f_2}-1)}{2 d_{f_1} d_{f_2}}
,
\end{equation}
when $b\times b = f_1 + f_2$, which explicitly demonstrates this fact. However, it is nonetheless useful to perform these exchange experiments to corroborate the data and demonstrate its extraction through braiding operations.

\begin{table*}
\[
\begin{array}{|l|c|c|}
\hline
\mathcal{C} = \{I, \varepsilon \}  & \varepsilon \times \varepsilon =I + \varepsilon &  d_I = 1, \quad d_{\varepsilon} = \phi \\
\hline
\multicolumn{2}{|l|}{
\left[F^{\varepsilon \varepsilon \varepsilon}_\varepsilon \right]_{ef}=
\left[
\begin{matrix}
\phi^{-1} & \phi^{-1/2}\\
\phi^{-1/2} & -\phi^{-1}
\end{matrix}
\right]_{ef_{\phantom{g}}}^{\phantom{T}}
}          &  \kappa_{\varepsilon} = 1  \\
\hline
\multicolumn{2}{|l|}{  R^{\varepsilon \varepsilon}_{I}= e^{-i s\frac{4 \pi}{5}}, \quad R^{\varepsilon \varepsilon }_\varepsilon = e^{i s \frac{3\pi}{5}}
} & \theta_I =1, \quad \theta_\varepsilon = e^{i s\frac{4\pi}{5}}
\\
\hline
\multicolumn{2}{|l|}{
S = \frac{1}{\sqrt{\phi^{2}+1}} \left[
\begin{array}{cc}
1 & \phi \\
\phi & -1
\end{array}%
\right]^{\phantom{T}}_{\phantom{g}}
, \quad
S^{(\varepsilon)} = e^{ i s\frac{3\pi}{10}}
}  &  c_- \text{ mod }8 = s \frac{14}{5}
\\
\hline
\hline
p_{\varepsilon \varepsilon}(I) = 1- p_{\varepsilon \varepsilon}(\varepsilon) = \phi^{-2} &
\multicolumn{2}{|l|}{ p_{\varepsilon (\varepsilon \varepsilon); \varepsilon}(I|I) = p_{\varepsilon (\varepsilon \varepsilon); \varepsilon}(\varepsilon|\varepsilon) =
1- p_{\varepsilon (\varepsilon \varepsilon); \varepsilon}(I| \varepsilon) = 1- p_{\varepsilon (\varepsilon \varepsilon); \varepsilon}(\varepsilon|I)
= \phi^{-2}
}
\\
\hline
p_{ \varepsilon  \varepsilon}^{(2)}(I) =1 -p_{ \varepsilon  \varepsilon}^{(2)}(\varepsilon) = \phi^{-4}
& \multicolumn{2}{|l|}{
p_{\varepsilon \varepsilon \varepsilon; \varepsilon}^{\text{com}}(I|I) = p_{\varepsilon \varepsilon \varepsilon; \varepsilon}^{\text{com}}(\varepsilon|\varepsilon) =
1- p_{\varepsilon \varepsilon \varepsilon; \varepsilon}^{\text{com}}(I| \varepsilon) = 1- p_{\varepsilon \varepsilon \varepsilon; \varepsilon}^{\text{com}}(\varepsilon|I)
= \phi^{-2}
}
\\
\hline
\multicolumn{3}{|l|}{
p_{\varepsilon \varepsilon \varepsilon; \varepsilon}^{(n)}(I|I) =  1- p_{\varepsilon \varepsilon \varepsilon; \varepsilon}^{(n)}(I| \varepsilon) = 1- p_{\varepsilon \varepsilon \varepsilon; \varepsilon}^{(n)}(\varepsilon|I) = p_{\varepsilon \varepsilon \varepsilon; \varepsilon}^{(n)}(\varepsilon|\varepsilon) = \left\{
\begin{array}{lll}
1 & \quad & \text{ for $n=0 \text{ mod }10$}  \\
\phi^{-2} & \quad & \text{ for $n=1,9 \text{ mod }10$} \\
\phi^{-4} & \quad & \text{ for $n=2,8 \text{ mod }10$} \\
\phi^{-4} + 2 \phi^{-2} & \quad & \text{ for $n=3,7 \text{ mod }10$} \\
2 \phi^{-4} + \phi^{-2} & \quad & \text{ for $n=4,6 \text{ mod }10$} \\
\phi^{-6}  & \quad & \text{ for $n=5 \text{ mod }10$}
\end{array}
\right.
}
\\
\hline
\end{array}
\]
\caption{The basic data of Fibonacci UMTCs, along with some of the topological invariants and the predicted probability outcomes for the experiments described in this section [Eqs.~(\ref{eq:fusion_experiment}), (\ref{eq:associativity_experiment}), (\ref{eq:purebraid_experiment}), (\ref{eq:braidexchange_experiment}), and (\ref{eq:nonAbelian_experiment})]. Here, $s = \pm 1$ indicates the chirality of the two UMTCs with these fusion rules, $\phi = \frac{1 + \sqrt{5}}{2}$ is the Golden ratio, and $I$ is the vacuum charge. Fusion with vacuum, and $F$-symbols and $R$-symbols that equal 1 are not shown. These are all the possible UBTCs with these fusion rules.}
\label{Table:Fib}
\end{table*}

Now considering specific UMTCs, I first examine the Ising$^{(\nu)}$ topological phases, the basic data of which are given in Table~\ref{Table:Ising}. Here, $\nu$ is an odd integer mod~16 that distinguishes the eight different theories that share the same fusion rules. Table~\ref{Table:Ising} also lists some useful topological invariants and the predicted measurement probabilities for the bulk quasiparticle experiments described in this paper. The $\sigma$ topological charge is the only one that is non-Abelian, so these quasiparticle types are the only ones that will generate nontrivial information from the bulk quasiparticle experiments. For these topological orders, all of the fusion rules, quantum dimensions, and magnitudes of the $F$-symbols can be determined using the fusion and associativity experiments. The braiding experiments (using $\sigma$ quasiparticles) can determine the value of $\theta_{\psi}$. However, the values of $\kappa_{\sigma}$ and $\theta_{\sigma}$ cannot be obtained from such experiments. Thus, these experiments can determine the topological order to be of the Ising$^{(\nu)}$ type, but cannot distinguish between different $\nu$.

Next, I examine the Fibonacci topological phases, whose basic data are given in Table~\ref{Table:Fib}. In this case, $s =\pm 1$ labels the two theories with the same fusion rules, which are simply distinguished by their chirality. Table~\ref{Table:Fib} also lists some useful topological invariants and the predicted measurement probabilities for the bulk quasiparticle experiments described in this paper. In this case, the $\varepsilon$ topological charge is the only nontrivial one and it is non-Abelian. For these topological orders, all of the fusion rules, quantum dimensions, and magnitudes of the $F$-symbols can be determined using the fusion and associativity experiments. The braiding experiments can extract the value of $\mathrm{Re}(\theta_{\varepsilon})$, but cannot determine which chirality $s =\pm 1$ is present.

The Ising$^{(\nu)}$ and Fibonacci topological phases are examples for which the bulk quasiparticle braiding experiments described do not provide additional information about the topological order, beyond what may be obtained from the fusion and associativity experiments (together with consistency conditions). However, the braiding experiments provide additional verification of the information learned from the associativity experiments, as well as direct probes of the braiding properties.

\section{Additional Experiments}
\label{sec:Additional}

Even though bulk quasiparticle experiments of the type discussed in Sec.~\ref{sec:Bulk} can provide a large amount of information about the topological order of a phase, they generally will not provide complete information, as was demonstrated for the Ising$^{(\nu)}$ and Fibonacci topological phases.
The primary limitations of those experiments stem from the fact that the position and topological charge of a localized quasiparticle behave as classically controlled parameters that take definite values.
As such, those experiments are unable to utilize superpositions of different quasiparticle trajectories and can only utilize coherent superpositions of different topological charges that occur as common fusion channels of the same sets of localized quasiparticles (which requires the localized quasiparticles to be non-Abelian).
The impact of these limitations is most extreme for Abelian topological phases, for which no nontrivial information about associativity or braiding can be obtained using these bulk quasiparticle experiments; they can only access the fusion rules.
Moreover, a general limitation due to the intrinsic nature of the bulk quasiparticle experiments is that they are unable to probe the chirality of the topological phase, as doing so requires some form of external reference.

In light of these limitations, additional types of experiments should be considered in order to access additional information about the topological order.
These additional experiments come with their own implementation difficulties and obstacles, so it is useful to weigh these against what information they may provide.
In this section, I consider several classes of additional experiments, determine the invariant information they access, and discuss some of the difficulties and challenges with their implementations.
This is not meant to be an exhaustive list, but rather is intended to convey that there are numerous approaches one may pursue to provide additional information in certain circumstances.

\subsection{Edge Modes}
\label{sec:EdgeModes}

Many experiments have been proposed to probe the topological order of topological phases with gapless edge modes, which most notably include fractional quantum Hall states.
The gapless edge modes can be used to reveal information about the bulk topological order, because the bulk-edge correspondence allows the bulk quasiparticles' topological quantum numbers to also manifest on the edge excitations.
This is well-understood, in particular, when the effective theory of the bulk is described by a Chern-Simons quantum field theory, in which case the edge manifests a corresponding conformal field theory (CFT).
Such a Chern-Simons and CFT description is typically applicable for the long-distance effective theory of the bulk and gapless edge modes of topological phases, such as fractional quantum Hall states with sharp edges, so I will view the gapless edge modes from this perspective.
In this section, I specifically focus on three types of experiments that can provide the most direct access to missing information about the topological order: point contact tunneling, interferometry, and thermal Hall transport.

\subsubsection{Point Contact Tunneling}

The first type of edge mode experiments may be performed using a quantum point contact which allows controllable tunneling to occur between two different locations along the edge.
A point contact may be created by deforming the path of edge modes, e.g. using electrostatic gates, so that distinct locations along the edge are brought closer to each other in terms of distance through the bulk.
In the weak backscattering limit, the region within the point contact constriction is to remain in the same topological phase as the bulk to either side of the point contact.
In this case, the point contact induces tunneling of quasiparticles through the bulk between the edge modes.
This is in contrast to the strong backscattering limit, where the point contact essentially pinches off the topological phase, in which case there is no tunneling of quasiparticles through the bulk, but one can instead probe tunneling of constituent particles, e.g. electrons, across non-topological regions.
Hence, I will focus only on the weak backscattering limit, as it is the regime that allows one to probe properties of the quasiparticles.

In the case of quantum Hall states, and other topological phases whose edge modes transport electrical charge, one can probe the tunneling current as a function of temperature and the voltage applied across the point contact.
The expected tunneling current associated with an edge excitation, in the appropriate limits, takes the form~\cite{Wen92b,Wen95,Fendley06a,Fendley07a,Lee07,Levin07}
\begin{equation}
\label{eqn:power-laws}
I_{\text{tun}} \propto \left\{
\begin{array}{ll}
T^{2g -2}\,V  & \text{for small } eV \ll {k_B}T \\
V^{2g - 1}    & \text{for small } eV \gg {k_B}T
\end{array}
\right.
,
\end{equation}
where $V$ is the voltage difference across the point contact, $T$ is temperature, and $g$ is the tunneling exponent of the tunneling quasiparticles.

In principle, all possible types of quasiparticle excitations of the topological order may contribute to the tunneling current, so the expression for tunneling current will actually involve a sum over all possible quasiparticle types, with different prefactors and values of $g$ for each quasiparticle type.
However, in the suitable limit, the tunneling is expected to be dominated by the excitations whose tunneling operators are the most relevant (in the renormalization group sense), i.e. those with the smallest values of $g$.

When the system has a single edge mode described by CFT, the tunneling exponent of a quasiparticle carrying topological charge $a$ is given by $g_{a} = 2h_{a}$, twice the conformal scaling dimension of the corresponding CFT operator associated with $a$.
The bulk-edge correspondence provides a relation between the topological twist factors of the quasiparticle and the conformal scaling dimensions of the edge excitations
\begin{equation}
\theta_{a} = e^{i 2\pi h_a}
.
\end{equation}
This provides an extraction of the topological twist factors of tunneling quasiparticles from measurements of the tunneling current.

When the system has multiple edge modes that all propagate in the same direction, a similar statement holds, though the topological charge can be written as a multi-component quantity $a = (a_1 , \ldots , a_n)$, with each component corresponding to a component of the edge modes, and $h_{a} = \sum_{j} h_{a_j}$.
In these cases, the tunneling exponent of a quasiparticle is universal and independent of the edge potentials and interactions.

When there are multiple edge modes that do not all propagate in the same direction, the situation is more complicated and will depend on the interactions and edge potentials.
For quantum Hall states, long-range Coulomb interactions and impurity scattering at the edge are expected to cause the edge modes to equilibrate into a single electrically charged mode and the neutral sector modes.
In this case, the conformal scaling dimensions can be written as $h_a = h_{a,\text{c}} + h_{a,\text{n}}$, where $h_{a,\text{c}}$ and $h_{a,\text{n}}$ are projections of the conformal dimensions onto the charge and neutral components, respectively.
In this way, the tunneling exponent of a quasiparticle can be written in terms of the charge and neutral sectors as
\begin{equation}
g_{a} = 2\left(h_{a,\text{c}} + \left|h_{a,\text{n}} \right|\right)
.
\end{equation}
When the edge modes all propagate in the same direction, $h_{a_j}$ all have the same sign, so $h_{a,\text{n}} \geq 0$, and this returns the previously stated result.
When there are counter-propagating edge modes, $h_{a_j}$ may have different signs, and $h_{a,\text{n}}$ can be negative.
In this case, more effort is required to establish the relation to the topological twist factors.
For a fractional quantum Hall state, it is always the case that
\begin{equation}
h_{a,\text{c}} = \frac{Q_{a}^{2}}{2 \tilde{\nu}}
,
\end{equation}
where $\tilde{\nu} = \nu \mod 1$ is the fractional part of the filling, and $Q_{a}$ is the electrical charge (in units where the electron has $Q_{e^{-}} =-1$) of the quasiparticle of type $a$.
The electrical charge of tunneling quasiparticles can be extracted from the tunneling shot noise in the same point contact experimental setup~\cite{Kane94,Chamon95,Fendley96}.
Thus, one can piece together the relation to infer the topological twist by measuring the electric charge and tunneling exponent of the tunneling quasiparticles in a quantum Hall state of known filling.
Even though the extraction of the tunneling exponent does not constitute a direct probe of the braiding statistics, the ability to yield the values of topological twist factors potentially makes it a powerful method for identifying the UMTC.

As an aside, I mention that even though the fractional electric charge values carried by quasiparticles do not provide as direct information about the topological order as do the topological twist factors, they do provide some important information.
In particular, fractional electric charge is a property associated with the symmetry fractionalization class of a topological phase with U$(1)$ charge conservation symmetry. The manner in which symmetry charges can fractionalize for quasiparticles are determined by their braiding statistics with the Abelian quasiparticles of the topological phase~\cite{Barkeshli19}.
When the quasiparticles that carry topological charge $a$ also carry fractional electric charge $Q_{a}$, it implies that the UMTC includes an Abelian topological charge $q$ (ascribed to the Laughlin quasihole in quantum Hall states) which has braiding $R^{aq}R^{qa} = \exp(i 2 \pi Q_{a} )$ for any $a$.
It also follows that the Hall conductance $\sigma_{H}$ (in units of $e^2 / h$), which equals the filling $\nu$ for quantum Hall states, satisfies $e^{i 2 \pi \sigma_{H} } = e^{i 2 \pi Q_{q}}$, so $\theta_{q} = \pm e^{i \pi \sigma_{H}}$~\cite{Meng16}.

The fractional electric charge of tunneling quasiparticles has been measured through tunneling noise experiments for quantum Hall states at various filling fractions~\cite{Saminadayar97,Picciotto97,Chung03,Dolev08,Radu08,Lin12,Baer14}, with results that fairly accurately match predictions for the expected states.
On the other hand, measurement of the tunneling exponents appears to be more challenging, and has only been reported for the states at $\nu = \frac{7}{3}$, $\frac{5}{2}$, and $\frac{8}{3}$~\cite{Radu08,Lin12,Baer14}, for which the match to predictions for candidate states is not definitive.

Perhaps the most significant challenges associated with the edge mode tunneling experiments have to do with ensuring that system is exhibiting the desired tunneling behavior.
One aspect of this is constructing the point contact so that the region within it supports the same topological phase as the rest of the bulk, as is necessary for the excitations tunneling to correspond to quasiparticles of the bulk topological order being probed.
Another is aspect is ensuring the edge modes are fully equilibrated, as otherwise determining the relation between measured quantities and UMTC information could be made intractable by the non-universal physics.
Some of these issues may be mitigated by improved sample designs, such as screening well layers that could produce more sharply defined edges~\cite{Willett19,Nakamura20}.

The tunneling behavior may change significantly as system parameters, such as $T$ and $V$, are varied~\cite{Radu08,Lin12,Baer14}, so it is important to verify that these are in an appropriate range.
Indeed, Ref.~\onlinecite{Chung03} observed a transition of the tunneling quasiparticles' charge for hierarchy filling fractions ($\nu = \frac{2}{5}$ and $\frac{3}{7}$) from what can be interpreted as minimal charge quasiparticles at higher temperatures to minimal scaling dimension quasiparticles at lower temperatures, though the scaling exponent was not extracted and it is not clear the regimes of Eq.~(\ref{eqn:power-laws}) were reached.

Even when the tunneling is behaving as desired, these experiments will have limited ability to control which quasiparticle types one can extract information about, as working in the desired regime will likely give tunneling that is dominated by one type that has the most relevant tunneling (smallest tunneling exponent), or maybe a very few number of the most relevant types.
If the most relevant tunneler does not happen to also be a generating quasiparticle (e.g. the fundamental quasiparticle), then the information provided by the extracted twist factor may not be very distinguishing.

In the case of topological phases whose edge modes do not transport electric charge, the tunneling experiments can still be performed, but require using non-electric transport methods, such as thermal transport, or additional structures that are able to interface electrical current with the topological phases' neutral edge modes, such as those considered in Ref.~\onlinecite{Aasen20a}.
These implementations can be expected to be more challenging, as electrical transport is relatively straightforward to control and measure.

\subsubsection{Interferometry}

\begin{figure}[t!]
\begin{center}
  \includegraphics[scale=0.25]{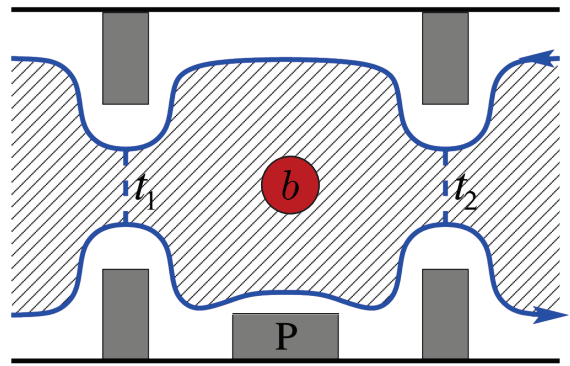}
  \caption{A two point-contact interferometer may be used for learning information about quasiparticles' braiding statistics in a topological phase with gapless edge modes. The shaded region is the topological phase and the blue lines indicate the gapless edge modes.
  The point-contacts induce tunneling of excitations between the bottom and top edges with amplitudes $t_1$ and $t_2$.
  This establishes an interference loop around the central region between the two point-contacts.
  The quasiparticle content in the central region can be controlled using a gate (red dot) that creates a pinning potential well.
  The braiding of the edge excitations with the collective topological charge $b$ of bulk quasiparticles in the central interference region enters the interference term of the edge current flowing from the bottom to the top edge.
  A plunger gate $P$ can be utilized to modulate the total magnetic flux passing through the interference loop by changing the area of the central region, which generates Aharonov-Bohm oscillations in the interference term.}
  \label{fig:FQHint}
\end{center}
\end{figure}

The second type of experiments utilizing edge modes that I focus on is interferometry.
Gapless edge modes provide the capability of establishing coherent superpositions of different paths taken by a quasiparticle.
When the different paths involve different braiding topologies, e.g. going around the left or right side of another stationary quasiparticle, it can directly probe the braiding statistics between the quasiparticles involved.
In particular, this scenario can be realized by a Fabry-P\'erot type interferometer formed using two point contacts, as indicated in Fig.~\ref{fig:FQHint}.
This configuration generates interference between the two tunneling paths edge excitations may travel around a central region.
For quantum Hall states, since the edge excitations that encircle the central interference region carry electric charge, it is useful to have a mechanism for modulating the magnetic flux through the central region, in order to generate Aharonov-Bohm oscillations in the interference signal.
This may be done by varying the magnetic field or the area of the central region, e.g. through the use of a plunger gate.

Such interferometers were first proposed and analyzed for Abelian quantum Hall states in Ref.~\onlinecite{Chamon97}, for the non-Abelian Pfaffian state expected at $\nu=\frac{5}{2}$ in Refs.~\onlinecite{Fradkin98,DasSarma05,Stern06a,Bonderson06a}, and for general quantum Hall states in Refs.~\onlinecite{Bonderson06b,Bonderson07c,Bishara09}.

For the Fabry-P\'erot two point-contact interferometer in quantum Hall states, the leading order contribution to the current across the Hall bar due to tunneling of quasiparticles of type $a$ is
\begin{equation}
I^{(a)} \approx I_{1}^{(a)} + I_{2}^{(a)} + I_{12}^{(a)}
.
\end{equation}
Here, $I_{1}^{(a)}$ and $I_{2}^{(a)}$ are the tunneling currents that point-contacts 1 and 2 would individually contribute in the absence of quantum interference, so they would be expected to have the same behavior of single point-contact tunneling, as in Eq.~(\ref{eqn:power-laws}).
$I_{12}^{(a)}$ is the (leading order) interference term between the two tunneling paths, which takes the form
\begin{equation}
I_{12}^{(a)} \propto \text{Re}\left[ e^{i Q_{a} \Phi} M_{ab}^{\ast} F_{a}(T,V,L) \right]
.
\end{equation}
The first factor is the Aharonov-Bohm phase associated with the edge excitation $a$ encircling the flux $\Phi$ that passes through the central interferometry region.
The last factor $F_{a}(T,V,L)$, which can be computed from the CFT, is a function of the temperature $T$, the voltage difference $V$ between the top and bottom edges, and the length $L$ of the perimeter of the central region.
The factor
\begin{equation}
M_{ab} = \frac{S_{ab}^{\ast} S_{00}}{ S_{0a} S_{0b}}
\end{equation}
is the monodromy scalar element, which is due to the braiding of the edge excitation $a$ around the total collective topological charge $b$ of the bulk quasiparticles contained in the central interference region.
In principle, the quasiparticle content in the central region, and hence the topological charge $b$ and the factor $M_{ab}$ in the interference term, may be changed independent of the other variables.
The Aharonov-Bohm term may also be modulated, e.g. using a plunger gate, nearly independently of the other variables.
With control over the topological charge $b$ in the central region and the Aharonov-Bohm phase of the probe quasiparticles around the central, the braiding statistics encoded in $M_{ab}$ can be extracted from the interference signal in the tunneling current.
Such experiments provide information beyond that of the bulk quasiparticle braiding $S$-matrix experiments because here the phases of the elements of $M$ can, in principle, be extracted.

Such interferometers have been implemented in Refs.~\onlinecite{Camino05a,Willett09a,Ofek10,McClure12,An11,Willett2013,Nakamura19,Willett19,Nakamura20}, with varying degrees of success in the experimental results. In particular, improved sample design, e.g. involving screening well layers that screen impurities, suppress charging energies, and yield more cleanly defined edges have allowed for better behaved experiments~\cite{Willett19,Nakamura20}.

The interferometry experiments face many of the same implementation issues that negatively affect the single point contact tunneling experiments, as they are based on the same point contact tunneling technology.
Furthermore, the interferometer must be designed so as to maintain a uniform topological phase throughout the bulk, inside and outside the central interference region, as well as within the point contact constrictions, and ensure the interferometer is operating in the Aharonov-Bohm regime, not the Coulomb-dominated regime~\cite{Halperin11}, .
This must also be balanced by need to keep $L$ short enough that the interference remains coherent.

As in the case of the single point contact tunneling experiment, there are some limitations in the accessible information, even when the device is functioning as desired.
Since the tunneling will be dominated by one or very few of the most relevant types of quasiparticles, with little control over which these are, one may expect to only determine a single row or very few rows of the braiding matrix $M$.
Also, calibration of the device signal to recognize $b=0$ may be challenging, in which case it may only be possible to assess relative differences in braiding phases, with no ``zero'' reference point.

While the monodromy matrix $M$ does not provide as much additional information as the topological twist factors, the interferometry experiments have certain advantages over single point contact tunneling experiments.
One advantage is that the probe of braiding statistics is more direct, in that the dependence on $M$ in the interferometry experiments are due to the edge quasiparticles actually moving around the bulk quasiparticles.
Another advantage is that the data may be extracted at a fixed temperature and potential difference between the edges, rather than being extracted through a scaling relation.
Related to this, non-universal edge physics likely only reduces the visibility of the interference term for interferometry, whereas it can render the relation of tunneling scaling to UMTC data irretrievable.

\subsubsection{Thermal Hall Transport}

The third type of edge mode experiments involve measuring the thermal Hall conductivity which relates the temperature difference between edges to the perpendicular thermal current carried by the edge modes.
As mentioned in Sec.~\ref{sec:Intro}, the thermal Hall conductivity of a $(2+1)$D topological phase is~\cite{Kane97,Read00,Cappelli02}
\begin{equation}
\kappa_{H} = \frac{\pi}{6}T c_{-}
,
\end{equation}
where the chiral central charge $c_{-} = c_{L} - c_{R}$ is the difference of the positive-valued central charges $c_L$ and $c_R$ of the left-moving edge modes and the right-moving edge modes, respectively.
Thus, such thermal transport edge mode experiments can provide the chiral central charge.
However, this prediction only applies in the regime where the edge mode are fully thermally equilibrated (or rather the length scales involved are long compared to the thermal equilibration length).
In the opposite limit, where the edge modes are completely non-equilibrated, one might expect each edge mode to contribute positively to the thermal transport, regardless of its chirality, that is $\kappa_{H}$ would equal $\frac{\pi}{6}T \left(c_{L} + c_{R}\right)$.
However, in this case spurious contributions to $\kappa_{H}$, such as from edge reconstruction, may not be assumed to cancel out, and the thermal Hall conductivity would be expected to depend on detailed, non-universal edge physics.
Thus, when the edge modes are not in the fully thermally equilibrated regime, one could expect any (not necessarily quantized) value $\kappa_{H} \geq \frac{\pi}{6}T c_{-}$, which would not indicate much about the topological order.

Ref.~\onlinecite{Jezouin13} measured $|\kappa_{H}|$ for integer quantum Hall states in GaAs heterostructures using methods involving quantum point contacts designed to separate the heat flow across edge mode channels from the heat flow to bulk phonons.
Building on these methods, Refs.~\onlinecite{Banerjee17} and \onlinecite{Banerjee18} performed similar measurements for fractional quantum Hall states in GaAs heterostructures at $\nu=\frac{1}{3}$, $\frac{2}{3}$, $\frac{3}{5}$, and $\frac{4}{7}$, and $\nu = \frac{7}{3}$, $\frac{5}{2}$, and $\frac{8}{3}$, respectively.
The reported values of $\kappa_{H}$ for $\nu = \frac{1}{3}$, $\frac{3}{5}$, $\frac{4}{7}$, and $\frac{7}{3}$ appear to be in sharp agreement with the predictions, $c_{-} = 1$, $-1$, $-2$, and $3$ for the corresponding Laughlin~\cite{Laughlin83} and Haldane-Halperin hierarchy~\cite{Haldane83,Halperin84} states.
The reported values of $\kappa_{H}$ for $\nu = \frac{2}{3}$ and $\frac{8}{3}$ appear to be near, but significantly deviated from the predictions $c_{-} = 0$ and $2$, for these being particle-hole conjugated Laughlin states.
For $\nu = \frac{2}{3}$, this is attributed to the diffusive thermal transport expected to occur when $c_{-} =0$.
The reported value of $\kappa_{H}$ for $\nu = \frac{5}{2}$ is far removed from any of the previously expected candidate states, and argued to be in agreement with the prediction, $c_{-} = 2.5$, for the (unexpected) particle-hole symmetric Pfaffian state, despite theoretical issues with this candidate state.
However, various arguments have been put forth to explain the mysterious occurrence of this $c_{-} = 2.5$ value~\cite{Mross17,Wang17,Simon18,Feldman_Comment,Simon_Reply,Ma18,Simon19,Asasi20} in terms of the previously expected candidate states, which suggest significant dependence on non-universal physics, such as partially equilibrated edge modes or disorder induced nucleation of domains of different bulk topological order.
It is worth noting that the experimentally extracted value of $\kappa_{H}$ for $\nu = \frac{5}{2}$ comes near the $c_{-} = 2.5$ value in the temperature range $18-25$mK (only within error bars at $22$mK), and diverges significantly from it, increasing as temperature continues to be lowered.
This behavior is posited to result from the equilibration length increasing as temperature decreases~\cite{Banerjee18} and appear similar to the temperature dependence behavior observed for the $\nu = \frac{2}{3}$ state~\cite{HeiblumPC}.
(The temperature dependence of $\kappa_{H}$ for other $\nu$ is not presented.)

In contrast, Ref.~\onlinecite{Srivastav20} measured $|\kappa_{H}|$ for fractional quantum Hall states in bilayer graphene at $\nu = \frac{4}{3}$, $\frac{5}{3}$, $\frac{7}{3}$, and $\frac{8}{3}$ using two-terminal measurements similar to Ref.~\onlinecite{Jezouin13}, though without the use of quantum point contacts to isolate the edge mode heat flow from that of bulk phonons.
Nonetheless, the heat flow due to coupling to phonons was observed to become negligible at sufficiently small temperatures ($T \lesssim 60$mK).
The reported values of $\kappa_{H}$ appear to be in sharp agreement with the predictions expected for \emph{completely non-equilibrated} edge modes that are entirely free of spurious contributions, that is $c{L} + c_{R} = 2$, $3$, $3$, and $4$, respectively.
The authors posit that these results can be attributed to a sharp confining potential generating strong interactions and very long thermal equilibration length.

Refs.~\onlinecite{Kasahara18,Yamashita20} measured $\kappa_{H}$ for $\alpha$-RuCl$_{3}$ using a straightforward approach.
These experiments reported the observation of $c_{-} = 0.5$ in certain parameter ranges.
However, they also found the diagonal thermal conductivity $\kappa_{xx}$ to be much larger than the thermal Hall conductivity, whereas one na\"ively expects it to be much smaller in order to attain quantization of $\kappa_{H}$.
Attributing the $\kappa_{xx}$ thermal conductivity to bulk phonons, it was argued that thermalization of the edge modes interacting with the bulk phonons can lead to approximate quantization of an effective thermal Hall conductivity~\cite{Ye18,Vinkler18}.
The experiments also found to have sample dependence, where those with smaller $\kappa_{xx}$ failed to exhibit thermal Hall conductivity indicative of $c_{-} = 0.5$ ~\cite{Yamashita20}.

It is not unexpected that thermal Hall conductivity experiments would be more difficult to perform and interpret than corresponding electrical transport measurements of quantum Hall states.
Indeed, a well-developed quantum Hall plateau with vanishing diagonal conductivity does not guarantee a similarly well-quantized thermal Hall conductivity.
There appear to be numerous questions left unanswered regarding when the systems are operating in a regime where the experiments are actually providing invariant universal data.
In addition to the challenges of isolating the thermal transport of the edge modes from bulk phonon thermal transport, or ensuring the interaction and thermalization of the edge modes and bulk phonons are in a suitable regime, one must ensure that multiple edge modes are fully equilibrated, that the temperature is sufficiently low (but not too low), and that the edge length of the system is much larger than the equilibration length.
Different sample quality and design, such as those with screening wells, may also help mitigate some of these issues.
It is important to also recognize that different materials and samples may actually favor different topological phases, even at the same filling fraction, or different edge physics.

Assuming the value of $c_{-}$ is accurately extracted experimentally, it is not, by itself, a very strong indicator of the topological order.
As mentioned in Sec.~\ref{sec:Intro}, a given value of $c_{-}$ can be associated with infinitely many different UMTCs; for example, all Drinfeld doubles (an infinite set) have $c_{-} =0$.
Perhaps the most definitive information the chiral central charge can provide by itself is that a non-integer value is indicative of a UMTC with non-Abelian quasiparticles.
On the other hand, $c_{-}$ can be much more useful in conjunction with other, more constricting data, such as the fusion rules.

\subsection{Topological Defects}
\label{sec:Defects}

Topological defects in the system, such as symmetry defects (fluxes) or domain walls, may provide further means of obtaining additional information about the topological order.
In contrast to quasiparticles, such defects are extrinsic objects.
They can only be introduced in the system by altering the Hamiltonian along a string-like region, not point-like pinning potentials.
As such, they may be viewed as confined objects in the unmodified Hamiltonian, whose energy depends on the length of the string connecting defects.
However, they have fusion and braiding behavior that is similar to that of quasiparticles, though with additional structure.

In the case where a topologically ordered system also has unbroken global symmetry, the interplay between topology and symmetry give rise to symmetry enriched topological phases, which can potentially exhibit different classes of symmetry fractionalization and defects.
The algebraic structure of $(2+1)$D symmetry enriched topological phases (with orientation preserving symmetry) is captured by $G$-crossed UMTCs, which is a generalization of UMTCs that further incorporates the symmetry action, fractionalization, and defects~\cite{Barkeshli19}.

The action of symmetry on the emergent UMTC degrees of freedom is realized through the topological symmetries, which are the maps from the UMTC back to itself that can permute topological charges, but must preserve the basic data, up to gauge transformations.
In other words, the gauge invariant quantities associated with topological charges related by topological symmetries are identical.
For example, when $\varphi$ is a topological symmetry, $\theta_{\varphi(a)} = \theta_{a}$ and $S_{\varphi(a)\varphi(b)} = S_{ab}$.

Symmetry fractionalization arises from the global symmetry acting on quasiparticles, which manifests in the physical Hilbert space as a combination of the topological symmetry action (on the physical manifestation of the UMTC degrees of freedom) and localized symmetry operators acting in small regions containing a quasiparticle.
The fractionalization patterns are constrained by the symmetry group and action, as well as the fusion and braiding properties of the UMTC.
Conversely, observation of certain fractionalization patterns allows information to be inferred about the UMTC, as was mentioned in Sec.~\ref{sec:EdgeModes} for U$(1)$ symmetry.
The symmetry factionalization class may generally be measured through some extrinsic quantities that are viewed as fractionalized symmetry charges or global invariants, e.g. fractional electric charges carried by quasiparticles or Hall conductance.
My focus in this section is the more direct information about the UMTC that can be gained from defects, so I will not go into further detail on the information determined by symmetry fractionalization.

The fusion and associativity of defects described by a $G$-crossed theory are that of a fusion tensor category, as is the case for UMTCs only describing quasiparticles.
The only differences are that the fusion rules are required to be $G$-graded, and they are not required to be commutative, but rather must be compatible with $G$-crossed braiding.
This does not disrupt the general fusion category properties, but merely introduces and modifies some additional constraints.
More specifically, the topological charges associated with symmetry defects (which are not simply those of quasiparticles) are ascribed the corresponding labels ${\bf g} \in G$ of the global symmetry group and their fusion must respect the group multiplication structure, i.e. $a_{\bf g} \times b_{\bf h} = \sum\limits_{c_{\bf gh}} N_{ab}^{c} c_{\bf gh}$.
In this notation, the quasiparticles are ascribed the trivial group element $\bf 0$; in other words, the original UMTC that describes the quasiparticles constitutes the $\bf 0$-sector of the $G$-crossed UMTC.
When a ${\bf g}$-symmetry action permutes topological charge types, ${\bf g}$-symmetry defects will necessarily be non-Abelian.
Such non-Abelian defects may be utilized for bulk experiments in a very similar manner to those of Sec.~\ref{sec:Bulk}.
Indeed, the fusion and associativity experiments are essentially identical to those described in Secs.~\ref{sec:FusionExp} and \ref{sec:AssociativityExp}.
The only differences are that the defects cannot be localized by point-like pinning potentials, but instead require string-like modifications of the Hamiltonian, and the topological charges carry group element labels corresponding to the types of defects, e.g. $a_{\bf g}$, $b_{\bf h}$, $c_{\bf k}$, $e_{\bf gh}$, $f_{\bf hk}$, and $d_{\bf ghk}$ for the associativity experiment.

On the other hand, the braiding of symmetry defects in $G$-crossed UMTCs is generalized from that of UMTCs.
In particular, defect braiding incorporates the symmetry action and fractionalization of the corresponding symmetry group label, giving ``$G$-crossed braiding.''
For example, transporting a quasiparticle of topological charge $a_{\bf 0}$ around a ${\bf g}$-defect in the counterclockwise sense transforms its topological charge to $\rho_{\bf g}(a_{\bf 0})$, where $\rho : G \times \mathcal{C} \rightarrow \mathcal{C}$ is the group action that specifies how the global symmetries act on the UMTC degrees of freedom.

\begin{figure}[t!]
\begin{center}
  \includegraphics[scale=0.15]{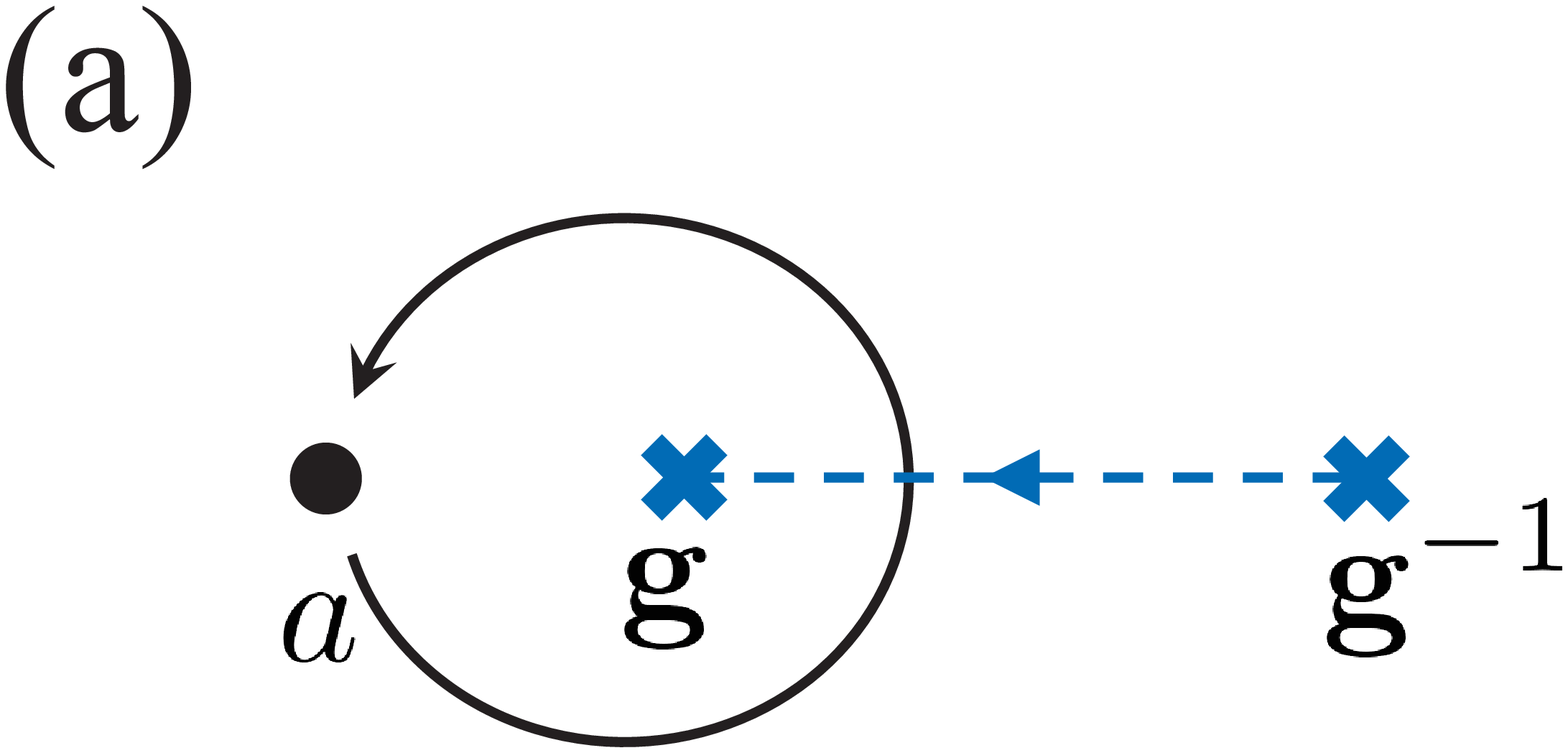} \quad
  \includegraphics[scale=0.15]{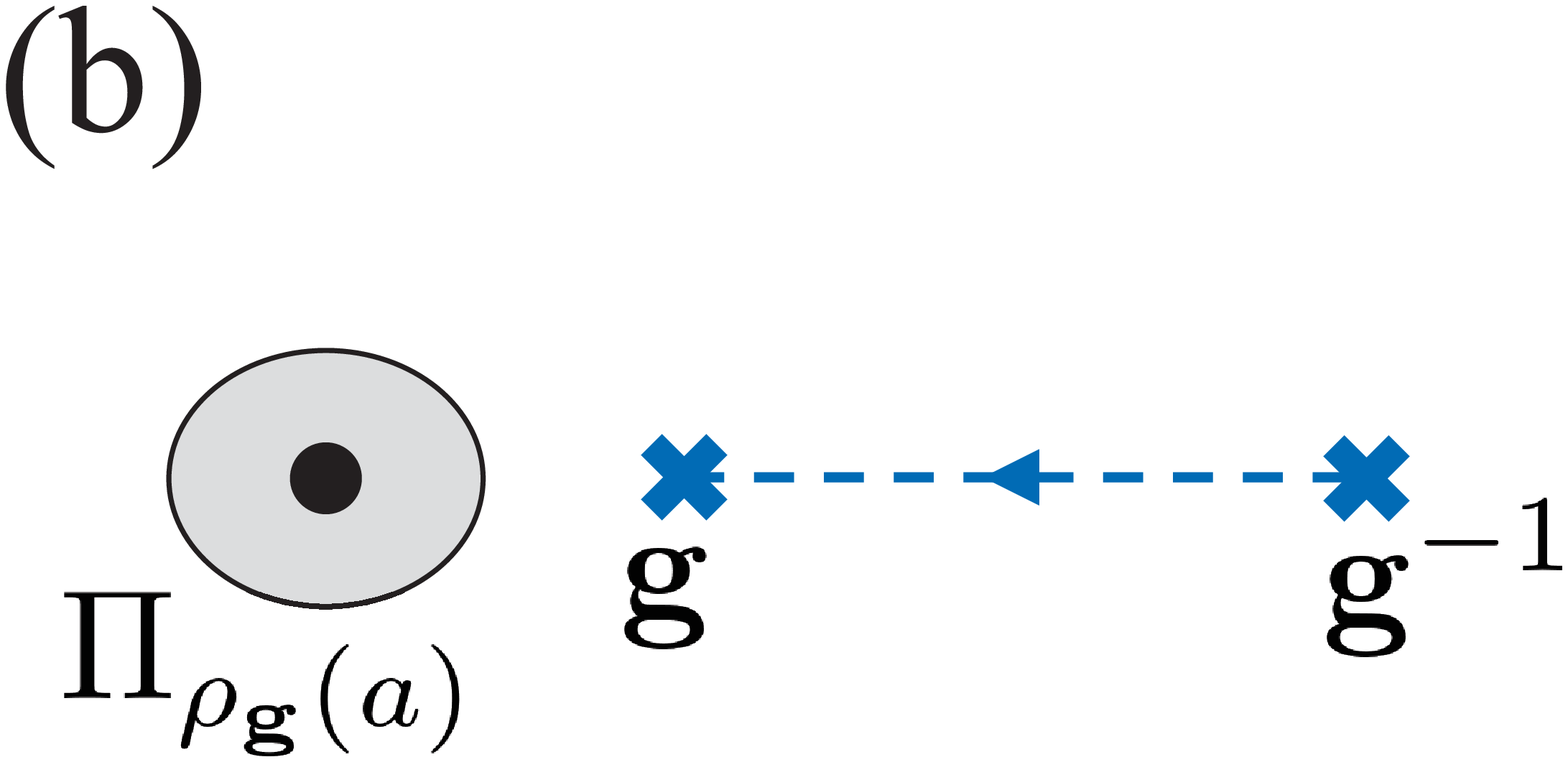}
  \caption{An experiment that can determine the symmetry action on topological charges. (a) After initializing the system in a state that includes a ${\bf g}$-defect and a quasiparticle with topological charge $a$, the quasiparticle is transported once around the defect. (b) After the braiding operation the topological charge of the quasiparticle is measured. The resulting topological charge is $\rho_{\bf g}(a)$.}
  \label{fig:defectbraiding}
\end{center}
\end{figure}

This suggests a braiding experiment to determine the symmetry action through the following the steps, shown schematically in Fig.~\ref{fig:defectbraiding}:
\begin{enumerate}
  \item Create the system with a ${\bf g}$-defect branch line, which has ${\bf g}$ and ${\bf g}^{-1}$ defects at the endpoints.
  \item Pair-create quasiparticles carrying topological charges $\bar{a}_{\bf 0}$ and $a_{\bf 0}$  from vacuum, and move them apart.
  \item Move quasiparticle $a_{\bf 0}$ around the ${\bf g}$-defect once in the counterclockwise direction.
  \item Measure the topological charge of the quasiparticle, which now carries topological charge $\rho_{\bf g}(a_{\bf 0})$.
\end{enumerate}
This experiment can also be carried out using defects instead of the quasiparticle, but doing so will provide information about symmetry action on defects, which is not as direct information about the UMTC describing the topological order.

The braiding experiments of Secs.~\ref{sec:exp_braiding} and \ref{sec:exp_exchange} can be carried out with defects instead of quasiparticles.
In this case, the topological charges will also carry group element labels, depending on what type of defects they are.
The most useful braiding experiments for determining additional information about the topological order (specifically, about the UMTC describing the quasiparticles), are the pure braiding experiments with defects $a_{\bf g}$, $b_{\bf h}$, $c_{{\bf h}^{-1}}$, and $d_{\bf g}$.
In this case, the fusion channels of $b$ and $c$ correspond to quasiparticles with topological charges $f_{\bf 0}$.
Additionally, the pertinent pure braid operation takes a simple form, since ${\bf h}$ and ${\bf h}^{-1}$ defects act trivially on each others' topological charges, i.e.  $\rho_{{\bf h}^{-1}}(b_{\bf h})= b_{\bf h}$, and the $G$-crossed version of the ribbon property gives~\cite{Barkeshli19}
\begin{equation}
R^{ b c}_{f} R^{c b}_{f} = \frac{\theta_{f}}{\theta_{b}  \theta_{c} \eta_{b}({\bf h}, {\bf h}^{-1}) \eta_{c}({\bf h}^{-1}, {\bf h}) }
,
\end{equation}
where the group labels of charges are left implicit, and the $\eta$-symbols are phase factors associated with symmetry factionalization.
As such, when the $a_{\bf g}$-$b_{\bf h}$ pair initially had collective charge $e_{\bf gh}$, the probability of the measurement of this pair after the braiding having outcome $e'_{\bf gh}$ will again be given by Eq.~(\ref{eq:purebraid_experiment}).
However, the important distinction is that the defects can support sums over fusion channels $f_{\bf 0}$ that may not be possible when the experiment only involves quasiparticles.
Different choices of $a_{\bf g}$ may help give more information about the quasiparticles by allowing the to be probed with different $F$-symbols and superpositions of $f_{\bf 0}$ fusion channels that enter these expressions, though it may be most convenient to choose ${\bf g} = {\bf h}^{-1}$, so that $e_{\bf 0}$ and $e'_{\bf 0}$ are also in the quasiparticle sector.

Clear examples of defects providing superpositions of quasiparticle fusion channels that cannot occur with quasiparticles alone are Abelian topological orders with symmetries that permute the quasiparticle types.
A simple example of this is the toric code D$(\mathbb{Z}_{2})$ topological order with $G=\mathbb{Z}_{2}$ symmetry that interchanges the $e$ and $m$ topological charges (electromagnetic duality).
In this case, there are two types of symmetry defects, denoted $\sigma_{\bf 1}^{+}$ and $\sigma_{\bf 1}^{-}$, which are related to each other by fusion with a $e$ or $m$ charge.
These defects have fusion rules $\sigma_{\bf 1}^{\pm} \times \sigma_{\bf 1}^{\pm} = I + \psi$ and $\sigma_{\bf 1}^{\pm} \times \sigma_{\bf 1}^{\mp} = e + m$.
(See Ref.~\onlinecite{Barkeshli19} for the complete $G$-crossed data.)
In this example, the pure braid experiment involving defects can reveal that $\theta_{\psi}=-1$ and $\theta_{e} = \theta_{m}$ (the latter is also revealed by the symmetry action experiment described in this section), whereas no information beyond the fusion rules can be gained from the experiments in Sec.~\ref{sec:Bulk} involving only quasiparticles.

More generally, topological defects do not require a global symmetry of the Hamiltonian, but may, nonetheless, generate action of the topological symmetries (i.e. symmetries of the UMTC, not of the Hamiltonian) on quasiparticles, in a similar manner.
Examples of this include twist defects (lattice dislocations that effect $e-m$ interchange in the toric code)~\cite{Bombin2010}, genons (layer interchanging structures of multi-layer topological phases)~\cite{Barkeshli12a}, and Parafendleyon wires (topological charge conjugating domain walls, e.g. formed in fractional quantum Hall gapped by a superconducting proximity effect)~\cite{Clarke2012,Lindner2012,Cheng2012}.
As such, the $G$-crossed formalism provides a useful means of describing the algebraic structure of the defects' fusion and braiding operations, even when there is no global symmetry.

Creating and manipulating the topological defects in the controlled manner that is needed for these experiments will likely prove significantly more difficult than creating and manipulating quasiparticles, since these are extrinsic objects that require string-like modifications of the Hamiltonian that are sometimes highly nontrivial.
The physical realization of such defects that have been proposed have challenging obstacles, such as generating specific interlayer interactions for genons or interfacing superconductors with quantum Hall systems for Parafendleyon wires.
The prospect of furthermore implementing physical transportation of the defects for the purposes of braiding experiments may seem daunting.
In this case, it may be beneficial to resort to measurement-only methods~\cite{Bonderson08a,Bonderson08b,Bonderson12a} to effect the braiding transformations of the defects without actually moving them.
However, the limitations described in Appendix~\ref{sec:MObraiding} will also apply to defects, potentially restricting the information that can be accessed.

I end this section by remarking that if one is able to create and manipulate topological defects, then one would likely also be interested in learning information about the defects themselves and characterizing the symmetry enriched topological phase, not just the UMTC describing the quasiparticles.
In this regard, all the experiments of Sec.~\ref{sec:Bulk} can be performed, using all possible configurations of defects.
For this, the expressions for the predicted probabilities need to be modified to take into account the more generalize $G$-crossed braiding.

\subsection{Nontrivial Topology and Mapping Class Group Transformations}
\label{sec:Topology}

Another way to access additional information about the UMTC of a topological phase is by making use of its full topological nature as described by topological quantum field theory (TQFT).
A TQFT specifies the long-distance, universal properties of a topological phase when it is manifested on manifolds with nontrivial topology.
For a $(2+1)$D topological phase, the TQFT can be defined in terms of the corresponding UMTC.
On a surface of nontrivial genus, e.g. a torus, a topological phase will generally have a degenerate ground state space, even without non-Abelian quasiparticles.
These ground state degeneracies are examples of topological invariants associated with higher genus surfaces, though these can be expressed in terms of the fusion rules.
The TQFT structure specifies, among other things, these degeneracies and topological operations on the state space, such as the action of the mapping class group, i.e. the isotopy equivalence classes of automorphisms of the surface.
The mapping class group transformations can be viewed as ``active,'' being attained through adiabatic modification of the microscopic Hamiltonian, or ``passive,'' arising as changes of basis which correspond to different choices of measurements.
I focus primarily on the passive approach, since active mapping class group transformations seem less practical to implement.

\begin{figure}[t!]
\begin{center}
  \includegraphics[scale=0.4]{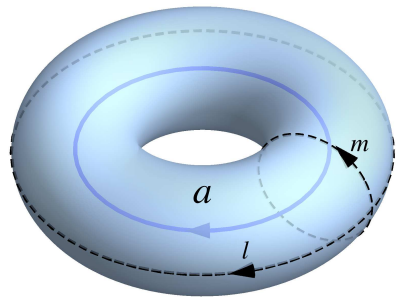}
  \caption{A basis for the degenerate ground states of a topological phase on a torus may be specified in terms of the topological charges $a \in \mathcal{C}$ together with a generating pair of cycles $(l,m)$. The topological flux line $a$ for the basis state $\left| {\Phi_{a}}\right\rangle_{(l,m)}$ is understood to wind once around the $l$ cycle, with no winding (twisting) around the $m$ cycle.}
  \label{fig:torus}
\end{center}
\end{figure}

In the case of the torus, a basis for the ground state space can be defined in terms of the topological charge values $a \in \mathcal{C}$ together with an ordered pair $(l,m)$ of generating cycles on the surface, i.e. two homotopy classes of non-contractible loops whose algebraic intersection number is $+1$, as shown in Fig.~\ref{fig:torus}.
The ground state degeneracy is $|\mathcal{C}|$, the number of distinct topological charge types, and the basis states are denoted as $\left| \Phi_{a} \right\rangle_{(l,m)}$.
The state $\left| \Phi_{a} \right\rangle_{(l,m)}$ is defined by the properties that a topological charge measurement around the cycle $m$ is found to have value $a$ (i.e. there is a topological flux $a$ threading the cycle $m$), and it can be obtained from $\left| \Phi_{0} \right\rangle_{(l,m)}$ by pair-creating quasiparticles of charge $a$ and $\bar{a}$, transporting $a$ around $l$, and then pair-annihilating the quasiparticles (which threads the topological flux $a$ along the path $l$).

The mapping class group of the torus, also known as the modular group $\text{SL}(2,\mathbb{Z})$, relates any two choices $(l,m)$ and $(l',m')$ of generating pairs of cycles of the torus as
\begin{equation}
(l,m) = \mathfrak{q}(l',m') = (\alpha l' + \beta m' , \gamma l' + \delta m' )
\end{equation}
where $\alpha, \beta, \gamma, \delta \in \mathbb{Z}$ and $\alpha \delta - \beta \gamma=1$.
With the matrix presentation
\begin{equation}
\mathfrak{q} \cong
\left[
\begin{matrix}
\alpha & \beta \\
\gamma & \delta
\end{matrix}
\right]
,
\end{equation}
of such transformations, the modular group can be generated by the two elements
\begin{equation}
\mathfrak{s} \cong \left[ \begin{matrix}0&-1\\1&0\end{matrix} \right], \qquad
\mathfrak{t} \cong \left[ \begin{matrix}1&1\\0&1\end{matrix} \right],
\end{equation}
which satisfy the group relations $(\mathfrak{s}\mathfrak{t})^2 = \mathfrak{s}^2$ and $\mathfrak{s}^4 = \openone$.

Alternatively, the mapping class groups can always be generated by ``Dehn twists'' around a complete set of generating cycles. For the torus, the modular group can be generated by the Dehn twists around cycles $l$ and $m$, corresponding to
\begin{equation}
\mathfrak{t}_{l} \cong \left[ \begin{matrix}1&0\\-1&1\end{matrix} \right], \qquad
\mathfrak{t}_{m} \cong \left[ \begin{matrix}1&1\\0&1\end{matrix} \right],
\end{equation}
which satisfy the group relations $\mathfrak{t}_{m}\mathfrak{t}_{l}\mathfrak{t}_{m} = \mathfrak{t}_{l}\mathfrak{t}_{m}\mathfrak{t}_{l}$ and $(\mathfrak{t}_{m}\mathfrak{t}_{l})^{6}=\openone$.
Clearly, $\mathfrak{t}_{m} =\mathfrak{t}$ and $\mathfrak{t}_{l} = \mathfrak{s}\mathfrak{t}\mathfrak{s}^{-1}$.

The topological $S$ and $T$ matrices of the UMTC provide projective representations of the corresponding generating modular transformations, and so the corresponding basis transformations are
\begin{eqnarray}
\left| {\Phi_{a}}\right\rangle_{(l,m)} &=& \sum_b S_{ab} \left| {\Phi_{b}}\right\rangle_{(m,-l)}
,
\\
&=& \sum_b T_{ab} \left|{\Phi_{b}}\right\rangle_{(l-m,m)}
.
\end{eqnarray}

For a general modular transformation $\mathfrak{q}$ that maps $(l',m')$ to $(l,m) = \mathfrak{q}(l',m')$, the projective representation of its action is given by
\begin{equation}
\left| {\Phi_{a}}\right\rangle_{(l,m)} = \sum_b Q_{ab} \left| {\Phi_{b}}\right\rangle_{(l',m')}
,
\end{equation}
where $Q$ can be expressed in terms of the $S$ and $T$ matrices in the same manner that $\mathfrak{q}$ is generated from $\mathfrak{s}$ and $\mathfrak{t}$.
An experiment that can infer the magnitude of the components of $Q$ is given by the following steps:
\begin{enumerate}
  \item Measure the topological charge around the cycle $m$ of the torus.
  \item Measure the topological charge around the cycle $m'$ of the torus.
  \item Go to step 1.
\end{enumerate}
The probability that the measurement around the cycle $m'$ will have outcome $b$, given that measurement around the cycle $m$ had value $a$ is
\begin{equation}
\label{eq:MCG_experiment}
p_{\mathfrak{q}}(b|a)  = \left| Q_{ab} \right|^{2}
.
\end{equation}

If one were able to use active implementations of mapping class group transformations, i.e. operations that implement $U_{\mathfrak{q}} = Q$ (up to an overall phase) on the states for a fixed generating pair basis, the experiment could be perform in the following steps:
\begin{enumerate}
  \item Measure the topological charge around the cycle $m$ of the torus.
  \item Apply the operations realizing the mapping class group element $\mathfrak{q}$.
  \item Measure the topological charge around the cycle $m$ of the torus.
  \item Go to step 2.
\end{enumerate}
Eq.~(\ref{eq:MCG_experiment}) similarly describes the conditional probabilities of the measurement outcome $b$ after the application of $\mathfrak{q}$ in a given round when $a$ was the outcome of the measurement before applying $\mathfrak{q}$.

For example, the modular transformation $\mathfrak{s}$ gives
\begin{equation}
p_{\mathfrak{s}}(b|a)  = \left| S_{ab} \right|^{2}
.
\end{equation}
In this case, the measurements are performed around the cycles $m$ and $m'=-l$, respectively (using $m'=l$ will yield the same probabilities).
An important property of this experiment is that, assuming the implementation of the system and measurements on the torus are not faulty, it will reveal all the topological charge types of the UMTC, since $S_{a 0 } =S_{0 a}= \frac{d_{a}}{\mathcal{D}}$ for all $a\in \mathcal{C}$.
In other words, there ought to be no inadvertent restrictions to a subtheory of the full UMTC.

A useful set of modular transformations are $\mathfrak{s} \mathfrak{t}^{n} \mathfrak{s}^{-1}$, which give
\begin{equation}
p_{\mathfrak{s} \mathfrak{t}^{n} \mathfrak{s}^{-1}}(b|a)  = \left| \sum_{c} S_{ac} \theta_{c}^{n} S_{bc}^{\ast} \right|^{2}
.
\end{equation}
In this case, the measurement cycles are $m$ and $m' = m + nl$.
Combined with knowledge of the $S$-matrix, e.g. obtained from the $\mathfrak{s}$ experiment and unitarity, this set of experiments can provide information about the topological twist factors.
The fact that all topological charges can enter the expression nontrivially, i.e. the torus enables superpositions of all topological charges, makes this a powerful probe of the twist values.
As was the case for bulk quasiparticle experiments, there will be redundancies between different experiments in regard to the information about the UMTC that can be gained, for example,  $p_{\mathfrak{s} \mathfrak{t} \mathfrak{s}^{-1}}(b|a)  = p_{\mathfrak{s}}(b|a)$.

Experiments utilizing the mapping class group transformations on higher genus surfaces can be similarly analyzed using higher genus generalizations, where the surface can be decomposed into punctured torii and three-punctured spheres for the purposes of representing states and mapping class group transformations.
I have no concrete examples of this providing information about the UMTC that could not be obtained from the torus or bulk quasiparticles, so I will not discuss this generalization in detail.

The obstacles to implementing these experiments may seem a bit daunting, though some of them may be mitigated.
The first challenge is simply realizing the topological phase on a system with nontrivial genus.
This not only involves implementing phase on a system with enough curvature to achieve the nontrivial topology (embedded in 3D real space), but also ensuring the lengths of all cycles are much larger than the correlation length, to maintain the topologically protected degeneracies.

Implementing the topological charge measurements along nontrivial cycles poses another significant challenge.
In general, these involve measurements of nonlocal operators, whose support is a ribbon along the length of the cycle being measured.
One might envision measuring these using some interferometric measurement or by modifying the Hamiltonian along the cycle in some manner, e.g. lowering the gap or creating boundaries, that allows the topological charge value of interest to couple to a local observable~\cite{Bonderson10}.
The difficulty of performing such measurements along cycles with multiple windings around the longitudinal or meridional cycles of the embedded torus are likely to scale poorly with the complexity of the cycle.

Active modular transformations that involve adiabatically modifying the Hamiltonian to perform Dehn twists around cycles likely require a greater degree of knowledge and control over the microscopic Hamiltonian than is realistic.
Even if possible, each Dehn twist increases the distance over which interactions between the microscopic degrees of freedom must be able to interact (in a controlled manner).
Spreading this out evenly across the system can minimize this to one lattice step increase in range for each site, but even in this way, the number of consecutive Dehn twists that can be performed is likely very small.

Some of the measurement and active mapping class group transformation difficulties may be mitigated by using measurement-only type methods to implement mapping class group transformations.
This strategy utilizes additional genus as ancillary degrees of freedom, and then employs measurements or tunable interactions along a minimal set of cycles to generate arbitrary mapping class group transformations on the subsystem~\cite{Barkeshli16a}.
This removes the need for active implementations of Dehn twists or measurements around very complicated cycles, at the cost of requiring at least one higher genus and utilizing a longer sequence of operations (measurements or tuned interactions) to effectively generate the Dehn twist transformations.

Another strategy that may potentially mitigate many of the difficulties with implementing these experiments is to utilize ``genons.''
As briefly mentioned in Sec.~\ref{sec:Defects}, genons are layer interchanging defects in a system with two layers realizing the same topological order~\cite{Barkeshli12a}.
In more detail, the bilayer topological phase is assumed to have the same UMTC described by $\mathcal{C}$ in each layer, so the combined UMTC describing the bilayer quasiparticles is $\mathcal{C} \boxtimes \mathcal{C}$.
Layer interchange can be associated with a $\mathcal{Z}_{2}$ topological symmetry (a global symmetry of the microscopic Hamiltonian is not strictly necessary).
When a quasiparticle from one layer is transported around a genon, it ends up in the other layer.
Denoting the ``bare'' genon defect as $X$, it has the fusion rule
\begin{equation}
X \times X = \sum_{c\in \mathcal{C}} (c,\bar{c}).
\end{equation}
Up to overall phases that do not depend on the fusion channels, the bare defects' $F$-symbols are given by the $S$-matrix of $\mathcal{C}$ and their $R$-symbols are given by the topological twists of $\mathcal{C}$, that is
\begin{align}
[F^{XXX}_{X}]_{(e,\bar{e}) (f,\bar{f})} &\propto  S_{ef} \\
R^{XX}_{(c,\bar{c})} &\propto \theta_{c} .
\end{align}

\begin{table*}
\[
\begin{tabular}{|l|c|c|c|c|}
\hline
UBTC &  $c_- \text{ mod }8$ & $T$ & $S$ & $ST^{2}S^{-1}$
\\
\hline
${\rm D}(\mathbb{Z}_{2})
\phantom{\begin{array}{r}
1 \\
1 \\
1
\end{array}}
$ & $0$ & $[1,1,1,-1]$ &
\multirow{2}{*}{
$\frac{1}{2} \left[
\begin{array}{rrrr}
1 & 1 & 1 & 1 \\
1 & 1 & -1 & -1 \\
1 & -1 & 1 & -1 \\
1 & -1 & -1 & 1
\end{array}
\right]
$
}
&
\multirow{2}{*}{$
\left[
\begin{array}{rrrr}
1 & 0 & 0 & 0 \\
0 & 1 & 0 & 0 \\
0 & 0 & 1 & 0 \\
0 & 0 & 0 & 1
\end{array}
\right]^{\phantom{T}}
$ }
\\
\cline{1-3}
${\rm SO}(8)_{1}
\phantom{\begin{array}{r}
1 \\
1 \\
1
\end{array}}
$ & $4$ & $[1,-1,-1,-1]$ & &
\\
\hline
$\mathbb{Z}_{2}^{(\frac{1}{2})} \boxtimes \mathbb{Z}_{2}^{(\frac{1}{2})}
\phantom{\begin{array}{r}
1 \\
1
\end{array}}
$ & $2$ & $[1,i,i,-1]$ &
\multirow{3}{*}{
$\frac{1}{2} \left[
\begin{array}{rrrr}
1 & 1 & 1 & 1 \\
1 & -1 & 1 & -1 \\
1 & 1 & -1 & -1 \\
1 & -1 & -1 & 1
\end{array}
\right]
$
}
&
\multirow{3}{*}{$
\left[
\begin{array}{rrrr}
0 & 0 & 0 & 1 \\
0 & 0 & 1 & 0 \\
0 & 1 & 0 & 0 \\
1 & 0 & 0 & 0
\end{array}
\right]^{\phantom{T}}
$ }
\\
\cline{1-3}
$\mathbb{Z}_{2}^{(-\frac{1}{2})} \boxtimes \mathbb{Z}_{2}^{(-\frac{1}{2})}
\phantom{\begin{array}{r}
1 \\
1
\end{array}}
$ & $6$ & $[1,-i,-i,-1]$ & &
\\
\cline{1-3}
$\mathbb{Z}_{2}^{(\frac{1}{2})} \boxtimes \mathbb{Z}_{2}^{(-\frac{1}{2})}
\phantom{\begin{array}{r}
1 \\
1
\end{array}}
$ & $0$ & $[1,-i,i,1]$ & &
\\
\hline
\hline
$\mathbb{Z}_{2}^{(0)} \boxtimes \mathbb{Z}_{2}^{(0)}
\phantom{\begin{array}{r}
1 \\
1 \\
1
\end{array}}
$ & \multirow{2}{*}{$\ast$} & $[1,1,1,1]$ &
\multirow{2}{*}{
$\frac{1}{2} \left[
\begin{array}{rrrr}
1 & 1 & 1 & 1 \\
1 & 1 & 1 & 1 \\
1 & 1 & 1 & 1 \\
1 & 1 & 1 & 1
\end{array}
\right]
$
}
&
\multirow{2}{*}{$\ast$}
\\
\cline{1-1}\cline{3-3}
$\mathbb{Z}_{2}^{(0)} \boxtimes \mathbb{Z}_{2}^{(1)}
\phantom{\begin{array}{r}
1 \\
1 \\
1
\end{array}}
$ &  & $[1,-1,1,-1]$ & &
\\
\hline
$\mathbb{Z}_{2}^{(0)} \boxtimes \mathbb{Z}_{2}^{(\frac{1}{2})}
\phantom{\begin{array}{r}
1 \\
1
\end{array}}
$ & \multirow{3}{*}{$\ast$} & $[1,i,1,i]$ &
\multirow{3}{*}{
$\frac{1}{2} \left[
\begin{array}{rrrr}
1 & 1 & 1 & 1 \\
1 & -1 & 1 & -1 \\
1 & 1 & 1 & 1 \\
1 & -1 & 1 & -1
\end{array}
\right]
$
}
&
\multirow{3}{*}{$\ast$}
\\
\cline{1-1}\cline{3-3}
$\mathbb{Z}_{2}^{(0)} \boxtimes \mathbb{Z}_{2}^{(-\frac{1}{2})}
\phantom{\begin{array}{r}
1 \\
1
\end{array}}
$ & & $[1,-i,1,-i]$ & &
\\
\cline{1-1}\cline{3-3}
$\mathbb{Z}_{2}^{(1)} \boxtimes \mathbb{Z}_{2}^{(\frac{1}{2})}
\phantom{\begin{array}{r}
1 \\
1
\end{array}}
$ & & $[1,i,-1,-i]$ & &
\\
\hline
\end{tabular}
\]
\caption{All UBTCs with $\mathbb{Z}_{2} \times \mathbb{Z}_{2}$ fusion rules (see, e.g. Ref.~\onlinecite{Bonderson07b}), and their relevant topological quantities (only the diagonal elements of the $T$-matrix are listed). The first five rows are the UMTCs, also known as Toric Code, 3-Fermion Model, Semion-Semion, $\overline{\text{Semion}}\text{-}\overline{\text{Semion}}$, and Double Semion, respectively. The last five theories are not modular (as can be seen from the non-unitary $S$-matrix), but could occur as subsectors of certain other UMTCs. As such, their chiral central charges and modular transformations are only defined by the full UMTC to which they belong. For these fusion rules, the full set of topological twist factors of anyons (encoded in the $T$-matrix) completely distinguishes between all these UBTCs, while the other discussed topological invariants distinguish between certain subsets of the theories.}
\label{Table:Z2Z2}
\end{table*}

Thus, for bulk defect experiments using $a=b=c=d=X$, the associativity experiments of Secs.~\ref{sec:AssociativityExp} and \ref{sec:Defects} yield the measurement probabilities
\begin{equation}
p_{X(XX);X}((f,\bar{f})|(e,\bar{e}))  = \left| S_{ef} \right|^{2} = p_{\mathfrak{s}}(f|e)
,
\end{equation}
and the braiding exchange experiments of Secs.~\ref{sec:exp_exchange} and \ref{sec:Defects} yield the measurement probabilities
\begin{equation}
p_{XXX;X}^{(n)}((e',\bar{e}')|(e,\bar{e}))  = \left| [ST^{n}S^{-1}]_{ee'}  \right|^{2} = p_{\mathfrak{s} \mathfrak{t}^{n} \mathfrak{s}^{-1}}(e'|e)
.
\end{equation}

More generally, there is a bijection between genons in a bilayer system with $\mathcal{C} \boxtimes \mathcal{C}$ and systems on nontrivial surfaces with $\mathcal{C}$.
A bilayer system with $2g+2$ genons maps to a surface with genus $g$.
Fusion and braiding of genons correspond to the mapping class group transformations on the corresponding surface, e.g. braiding genons correspond to Dehn twist operations.
Again, braiding transformations of genons can be realized using measurement-only methods~\cite{Bonderson08a,Bonderson08b,Bonderson12a}, which also maps to those applied for surfaces with genus~\cite{Barkeshli12a}.
Though realizing genons and the desired experiments in bilayer topological phases would undoubtedly be challenging, it would likely be significantly less difficult that realizing topological phases on surfaces with genus and their mapping class group transformations.

\subsection{Examples}

In general, the experiments discussed in this section can provide additional information about the topological order that could not be accessed through the bulk quasiparticle experiments of Sec.~\ref{sec:Bulk}. In addition to revisiting the Ising$^{(\nu)}$ and Fibonacci topological orders (described in Tables~\ref{Table:Ising} and \ref{Table:Fib}, respectively), it is useful to consider some Abelian topological orders. For this, Table~\ref{Table:Z2Z2} lists all UBTCs whose topological charges have $\mathbb{Z}_{2} \times \mathbb{Z}_{2}$ fusion rules, together with their twist factors and $S$-matrix, as well as $c_- \text{ mod }8$ and $ST^{2}S^{-1}$ for the modular theories listed.

Though not as direct a probe of the bulk topological order, since they rely on the bulk-edge correspondence, the edge mode experiments of Sec.~\ref{sec:EdgeModes} can be very powerful in terms of the information they provide. In particular, the chiral central charge $c_{-}$ distinguishes between all the UMTCs within the Ising$^{(\nu)}$ family, between the two chiralities of the Fibonacci UMTC, and distinguishes all $\mathbb{Z}_{2} \times \mathbb{Z}_{2}$ UMTCs, except not between the toric code and double semion theories. The topological twist factors distinguish between all of these, as well as all the $\mathbb{Z}_{2} \times \mathbb{Z}_{2}$ UBTCs, though this is under the assumption that all twist values can be obtained for these theories, which may not be true. The interferometry experiments that measure the monodromy matrix $M$ can only distinguish within the $\mathbb{Z}_{2} \times \mathbb{Z}_{2}$ family whether a theory is modular vs non-modular, and semionic vs. non-semionic, again assuming all components of $M$ are obtainable.

The experiments using topological defects (Sec.~\ref{sec:Defects}) or nontrivial topology and mapping class group transformations (Sec.\ref{sec:Topology}) are more direct probes of the bulk topological order, though may be more limited the information they provide. For the Ising$^{(\nu)}$ topological phases, the $\mathfrak{s}\mathfrak{t}^{n} \mathfrak{s}^{-1}$ experiments will determine that $\theta_\sigma$ is a primitive 16th root of unity, as well as extract the value of $\rm{Re} [\theta_{\sigma}^{2}]$. This will identify whether $\nu \in \{1,7,9,15 \}$ or $\{ 3,5,11,13 \}$, which also determines the value of $\kappa_\sigma$. For the $\mathbb{Z}_{2} \times \mathbb{Z}_{2}$ UMTCs, the $\mathfrak{s}\mathfrak{t}^{n} \mathfrak{s}^{-1}$ experiments will distinguish between the semionic and non-semionic theories. A nontrivial symmetry action permuting quasiparticle types will distinguish toric code from the double semion theory. As previously mentioned, the $\mathfrak{s}$ experiment will also reveal all the topological charge values, so it will distinguish between the modular and non-modular $\mathbb{Z}_{2} \times \mathbb{Z}_{2}$ theories.

\section{Discussion}

There is no single experiment that serves as a conclusive ``smoking gun'' for determining the topological order of a topological phase.
Any given experiment will provide partial information that will narrow the scope of possible topological orders of the system by a certain amount, so various experiments must be conducted to piece together a more complete picture.
Moreover, it is useful to conduct different experiments that provide redundant information obtained in different manners, as different ways of probing the same information verifies the consistency of the results and the theory.
Pragmatism demands balancing this voracity for experimental tests with the need to efficiently utilize limited resources.
In this regard, the importance of bolstering and guiding experimental investigation through theoretical arguments and numerical simulations should be recognized.

In terms of efficient use of experimental resources, the bulk quasiparticle fusion rules and associativity experiments of Secs.~\ref{sec:FusionExp} and \ref{sec:AssociativityExp} provide very strong information about the topological order, while remaining relatively simple.
Indeed, UMTCs that share fusion rules and magnitudes of $F$-symbols can generally be viewed as belonging to a closely related family, which have relatively minor differences in their $F$-symbols and braiding.
From the perspective of quantum computational operations obtained from a topological phase, the computational gates obtained from a given physical operation appear to be isomorphic for UMTCs within such a family. (One may conjecture this to be generally true.)
Along these lines, there is additional incentive to pursuing the bulk quasiparticle experiments of Sec.~\ref{sec:Bulk}, since the architectures and operations used for these experiments are the same ones that would be developed to perform topological quantum computation.

\acknowledgements
I thank C.~Nayak, K. Shtengel, Z.~Wang, and M.~M.~Zaletel for useful discussions. I acknowledge the hospitality and support of the Aspen Center for Physics, which is supported by National Science Foundation grant PHY-1607611.

\appendix

\section{Splitting and Fusing Localized Anyons}
\label{sec:Splitting_Fusing}

In this section, I consider a simple idealized model for the splitting and fusion operations. The simplest scenario involves two sites with tunable pinning potential and tunable interactions between the two sites. The tunable pinning potential at a site $j$ is modeled by an on-site operator
\begin{equation}
{\bf V}^{(j)} = \sum_{x \in \mathcal{C}} E_{x}^{(j)} {\bf P}_{x}^{(j)}
,
\end{equation}
where ${\bf P}_{x}^{(j)}$ is a local projection operator that takes value $1$ for any state localizing a quasiparticle carrying topological charge $x$ at site $j$, and the value $0$ otherwise. The on-site property for these projectors implies ${\bf P}_{x}^{(j)}$ and ${\bf P}_{y}^{(k)}$ commute for $j \neq k$. (In more realistic models, this property would be true up to exponentially suppressed corrections.) The projectors can be written as ${\bf P}_{x}^{(j)} = \left| x^{(j)} \right\rangle \left\langle x^{(j)} \right|$, with the understanding that it requires the appropriate anyonic state interpretation, since a single nontrivial topological charge does not correspond to a local operator. More specifically, the nonlocal nature of anyonic states requires
\begin{equation}
{\bf P}_{x}^{(j)} \otimes {\bf P}_{y}^{(k)} = \sum_{z} \left| x^{(j)} , y^{(k)} ; z^{(jk)}  \right\rangle \left\langle x^{(j)} , y^{(k)} ; z^{(jk)} \right|,
\end{equation}
and
\begin{align}
{\bf P}_{x}^{(j)} + {\bf P}_{y}^{(k)} &= 2 \sum_{z} \left| x^{(j)} , y^{(k)} ; z^{(jk)}  \right\rangle \left\langle x^{(j)} , y^{(k)} ; z^{(jk)} \right| \notag \\
& + \sum_{\substack{q \neq y \\ z}} \left| x^{(j)} , q^{(k)} ; z^{(jk)}  \right\rangle \left\langle x^{(j)} , q^{(k)} ; z^{(jk)} \right| \notag \\
& + \sum_{\substack{q \neq x \\ z}} \left| q^{(j)} , y^{(k)} ; z^{(jk)}  \right\rangle \left\langle q^{(j)} , y^{(k)} ; z^{(jk)} \right|,
\end{align}
with further appropriate interpretation as operators on additional sites are included.

The $E_{x}^{(j)}$ are treated as tunable parameters to control which quasiparticle type is energetically preferred at the $j$th site. The terms explicitly written in these pinning potentials are the only ones that will be allowed to exist below the energy gap $\Delta$ of the system. Any additional terms that may exist (and make the model more physically realistic) will be assumed to occur at energies above the gap, and so can be neglected for the purposes considered here. In this idealized model, superpositions of different localized topological charge values on a given site are not prohibited; they simply occur at different energies due to the pinning potential. However, in physically situations, superpositions of different localized topological charge values will rapidly decohere due to local noise.

A representative configuration realizing the ``vacuum state'' is obtained by taking $E_{0}^{(j)} =0$ and $E_{x}^{(j)} \geq \Delta$ for $x \neq 0$ for all $j$. Chosen in this way, the pinning potentials locally maintain an energy gap $\Delta$.

A representative configuration localizing a quasiparticle with charge $a$ at site 1 and a quasiparticle with charge $\bar{a}$ at site 2, while similarly maintaining an energy gap of $\Delta$ at the sites, is obtained by taking $E_{a}^{(1)} =0$ and $E_{x}^{(1)} \geq \Delta$ for $x \neq a$ for site 1, and $E_{\bar{a}}^{(2)} =0$ and $E_{x}^{(2)} \geq \Delta$  for $x \neq \bar{a}$ for site 2.

A localized quasiparticle pair-creation process at these two sites may be implemented by adiabatically tuning from the vacuum configuration to the $a$-$\bar{a}$ configuration. Focusing on the corresponding low-energy basis states of the initial and final configurations, $\left| 0, 0; 0 \right\rangle$ and $\left| a, \bar{a}; 0 \right\rangle$, respectively, the effective Hamiltonian during this process takes the form
\begin{equation}
{\bf H}_{0} = \left[
\begin{array}{cc}
E_{0}^{(1)}+ E_{0}^{(2)} & \Gamma_{\bar{a}}^{\ast} \\
\Gamma_{\bar{a}} & E_{a}^{(1)} + E_{\bar{a}}^{(2)}
\end{array}
\right]
,
\end{equation}
with initial and final configurations
\begin{equation}
{\bf H}_{0}(t_{\rm i}) = \left[
\begin{array}{cc}
0  & 0 \\
0 & 2 \Delta
\end{array}
\right]
, \quad
{\bf H}_{0}(t_{\rm f}) = \left[
\begin{array}{cc}
2 \Delta  & 0 \\
0 & 0
\end{array}
\right]
.
\end{equation}
Here, $\Gamma_{\bar{a}}$ is the amplitude for a charge $\bar{a}$ to hop from site 1 to 2, or for a charge $a$ to hop from site 2 to 1. (The complex conjugate corresponds to the amplitude of the reverse process.) Such hopping amplitudes arise from topological interactions between sites 1 and 2, which may be induced by bringing the two pinning potential close to each other and turning on a local interaction, or perhaps via some nonlocal effect. Here, I simply assume the ability to turn it on in a controlled manner.

Presented in this way, the localized quasiparticle pair-creation process is seen to be locally equivalent to a tuning process for a two-level system, which is well understood. As such, it is clear that nonzero $\Gamma_{\bar{a}}$ is necessary to enable the mixing of states required to evolve between the two basis states. Moreover, the hopping amplitude provides an avoided level-crossing, with gap $2 |\Gamma_{\bar{a}} |$ when $E_{0}^{(1)}+ E_{0}^{(2)} = E_{a}^{(1)} + E_{\bar{a}}^{(2)}$. Since adiabaticity is defined with respect to the instantaneous energy gap, the avoided crossing must be large compared to the rate at which the system parameters are varied in order for the state to persist in the instantaneous ground state, as desired. Satisfying these conditions on tuning the system parameters, one has a pair-creation process, effecting evolution from the vacuum state $\left| 0, 0; 0 \right\rangle$ to the state $\left| a, \bar{a}; 0 \right\rangle$ with localized quasiparticles carrying charge $a$ and $\bar{a}$.

If $a$ is a non-Abelian topological charge, the joint fusion channel of these two quasiparticles may potentially take any value $z$ with $N_{a \bar{a}}^{z} \neq 0$. In this case, the low-energy theory should also take into consideration the states $\left| a, \bar{a}; z \right\rangle$, which are also ground states of the final Hamiltonian. However, as long as sites 1 and 2 do not have topological interactions that transfer topological charge to other parts of the system, the collective fusion channel of the quasiparticles at sites 1 and 2 cannot be altered, i.e. the fusion channels constitute superselection sectors that are prevented from mixing by topological charge conservation. As such, the local Hamiltonian of the subsystem comprising sites 1 and 2 can be written as a direct product ${\bf H} = \bigoplus_{z} {\bf H}_{z}$ (with the understanding that occupation of a state with $z$ requires the complementary subsystem to have charge $\bar{z}$), which makes it clear that sectors with different values of $z$ cannot evolve into each other.

I now generalize the pair-creation discussion to a general splitting process. A representative configuration localizing a quasiparticle with charge $a$ at site 1 and a quasiparticle with charge $b$ at site 2, while similarly maintaining an energy gap of $\Delta$ at the sites, is obtained by taking $E_{a}^{(1)} =0$ and $E_{x}^{(1)} \geq \Delta$ for $x \neq a$ for site 1, and $E_{b}^{(2)} =0$ and $E_{x}^{(2)} \geq \Delta$  for $x \neq b$ for site 2.

If $a$ and $b$ are non-Abelian topological charges, the joint fusion channel of these two quasiparticles may potentially take any value $c$ with $N_{ab}^{c} \neq 0$. However, in order to have $c\neq 0$, the complement of the subsystem comprising sites 1 and 2 must have collective topological charge $\bar{c}$ to compensate and satisfy the overall constraint of the full system having collective topological charge $0$. The simplest way to do this is with one additional site localizing a quasiparticle with charge $\bar{c}$, however this fixes the value $c$ and does not accommodate for superpositions. The simplest way to allow for superpositions of fusion channels $c$ is to have two additional sites localizing quasiparticles carrying topological charges $\bar{a}$ and $\bar{b}$, respectively.
I denote the corresponding ground states for the site 1 and 2 subsystem as $\left| a, b; c \right\rangle$.
When there are no topological interactions between the quasiparticles, the energies of states $\left| a, b; c \right\rangle$ with different fusion channels $c$ are degenerate.

Topological interactions between sites $j$ and $k$ correspond to processes that transfer topological charges between the two sites with some amplitude. These can be divided into two classes of interest for the purposes considered here. The first class of topological interaction are those that transfer topological charge between the two sites while leaving the localized topological charge values at the sites unchanged. For quasiparticles of charge $a$ and $b$, this corresponds to the transfer of topological charge values $e \in \mathcal{C}_{ab}$ with amplitude $\Gamma_{e}$, where $\mathcal{C}_{ab} = \{ x \in \mathcal{C} \, | \, N_{ax}^{a} \neq 0, \, N_{bx}^{b} \neq 0 \}$. The effect of such interaction terms is generically to fully split the energy degeneracy of the fusion channels $c$, as they lead to fusion channel dependent shifts in the energy of~\cite{Bonderson09}
\begin{equation}
E_{a,b;c} = \sum_{e \in \mathcal{C}_{ab}} \left( \Gamma_{e} \left[ F^{a e b}_{c} \right]_{a b} + \Gamma_{e}^{\ast} \left[ F^{a e b}_{c} \right]^{\ast}_{a b} \right)
.
\end{equation}

The second class of topological interaction are those that change the localized topological charge values at the sites, which can be viewed as hopping terms corresponding to the transfer of charge values $e \in \mathcal{C}$ with amplitude $\Gamma_{e}$. The changes of topological charge in these terms are required to occur in a manner consistent with the fusion rules. In particular, such a transfer of topological charge $e$ can only give rise to transitions between states $\left| a, b; c \right\rangle$ and $\left| a', b'; c \right\rangle$ if $N_{ae}^{a'} \neq 0$ and $N_{be}^{b'} \neq 0$. As these terms generate mixing between states with different localized topological charge values, they play a crucial role for the splitting process and are the interactions that need to be controlled in a tunable manner. At a detailed level, it is useful to consider specific processes where one may focus only on the terms mixing the states that are taken below the gap.

I now focus on the process that splits a quasiparticle of charge $c$ into two quasiparticles of charge $a$ and $b$, respectively. As discussed, this situation will require additional quasiparticles elsewhere in the system to allow for the nontrivial fusion channel $c$, but those will not interact with the subsystem comprising sites 1 and 2. The initial configuration will have $E_{c}^{(1)} =0$ and $E_{x}^{(1)} \geq \Delta$ for $x \neq c$ for site 1, and $E_{0}^{(2)} =0$ and $E_{x}^{(2)} \geq \Delta$  for $x \neq 0$ for site 2. The final configuration will have $E_{a}^{(1)} =0$ and $E_{x}^{(1)} \geq \Delta$ for $x \neq a$ for site 1, and $E_{b}^{(2)} =0$ and $E_{x}^{(2)} \geq \Delta$  for $x \neq b$ for site 2. An adiabatic tuning between these configurations, while appropriately utilizing interactions between the sites allows one to realize the corresponding splitting operation. The low energy basis states for the subsystem during this process are $\left| c,0;c \right\rangle$ and $\left| a,b;c \right\rangle$. While $\left| a,b;z \right\rangle$ for $z \neq c$ (and $N_{ab}^{z} \neq 0$) should also be considered to be low energy states, as they are also ground states of the final configuration, the initial state cannot evolve into them, as long as the operations are localized within the subsystem of sites 1 and 2. Again, this is because some form of topological interaction between the subsystem and its complement is required to modify the subsystem's collective topological charge, and without such interactions the Hamiltonian for subsystem comprising sites 1 and 2 takes the form ${\bf H} = \bigoplus_{z} {\bf H}_{z}$.

Focusing on the fusion channel $c$, the model for the two site subsystem can now be simplified to a two-dimensional low-energy effective Hamiltonian that can (locally) be written as
\begin{equation}
{\bf H}_{c} = \left[
\begin{array}{cc}
E_{c}^{(1)}+ E_{0}^{(2)} & \Gamma_{b}^{\ast} \\
\Gamma_{b} & E_{a}^{(1)} + E_{b}^{(2)} + E_{a,b;c}
\end{array}
\right]
.
\end{equation}
Again, the adiabatic quasiparticle splitting process is seen to be locally equivalent to an adiabatic tuning process for a two-level system.
The fusion channel energy splitting energy $E_{a,b;c}$ is unimportant to this process, since the fusion channel is fixed by topological considerations.
As such, I set this splitting energy to zero (at least for the initial and final configurations, where interactions are not needed). In order to adiabatically tune between the initial and final configurations
\begin{eqnarray}
{\bf H}_{c}(t_{\rm i}) &=& \left[
\begin{array}{cc}
0  & 0 \\
0 & (2-\delta_{ac}) \Delta
\end{array}
\right]
,
\\
{\bf H}_{c}(t_{\rm f}) &=& \left[
\begin{array}{cc}
(2-\delta_{ac}) \Delta  & 0 \\
0 & 0
\end{array}
\right]
,
\end{eqnarray}
one must turn on the hopping amplitude $\Gamma_b$ during the process. More specifically, $\Gamma_b$ will give rise to an avoided crossing of size $2 |\Gamma_b |$, which must be large compared to the rate at which the parameters are varied in order to persist in the instantaneous ground state. If one has sufficient control over the topological interactions and the ability to induce amplitudes as large as $|\Gamma_b | = \Delta$, one could, in principle, fully maintain the energy gap $\Delta$ throughout the process. Thus, with appropriate adiabatic tuning of the system parameters, one has a quasiparticle splitting process, effecting evolution from the initial state $\left| c, 0; c \right\rangle$ to the final state $\left| a, b; c \right\rangle$.

I note that this same discussion for the splitting process applies for transporting a quasiparticle from one site to another. This is described by setting $a=0$ and $b=c$ for the splitting process, which yields evolution from the initial state $\left| c, 0; c \right\rangle$ with quasiparticle $c$ at site 1 to the final state $\left| 0, c; c \right\rangle$ with the quasiparticle at site 2.

One may also consider a quasiparticle fusion process going from a state with two quasiparticles of charge $a$ and $b$, respectively to a single quasiparticle of charge $c$. The initial configuration of the system will have $E_{a}^{(1)} =0$ and $E_{x}^{(1)} \geq \Delta$ for $x \neq a$ for site 1, and $E_{b}^{(2)} =0$ and $E_{x}^{(2)} \geq \Delta$  for $x \neq b$ for site 2.
If the initial state has the definite value $c$ for the collective fusion channel of the two quasiparticle, i.e. the initial state is $\left| a, b; c \right\rangle$, then the final configuration can be taken to have $E_{c}^{(1)} =0$ and $E_{x}^{(1)} \geq \Delta$ for $x \neq c$ for site 1, and $E_{0}^{(2)} =0$ and $E_{x}^{(2)} \geq \Delta$  for $x \neq 0$ for site 2.
Simply reversing the splitting process described above will result in a single quasiparticle of charge $c$.

However, for well-separated non-Abelian quasiparticles, the collective fusion channel is topologically protected, so one may envision starting with an initial state that has a superposition of the different fusion channels $z \in \mathcal{Z}_{ab}$, where I define $\mathcal{Z}_{ab} = \{ x \in \mathcal{C} | N_{ab}^{x} \neq 0  \}$. For this, I let the initial state be
\begin{equation}
\left| \Psi(t_{\text{i}}) \right\rangle = \sum_{z \in \mathcal{Z}_{ab}} \Psi_{z} \left| a, b; z \right\rangle
,
\end{equation}
where I have left terms from the complement of the two site subsystem implicit.
In this case, a measurement of the collective fusion channel should be performed at some point in the quasiparticle fusion process. Such a measurement can be included at any point where it is physically possible. If the measurement can be performed while in the two quasiparticle configuration, then a post-measurement state of $\left| a, b; c \right\rangle$ is obtained with probability $p(c) = |\Psi_{c}|^{2}$, and the discussion above for fusing quasiparticles that have a definite fusion channel value can be applied. It is useful to instead consider the case where the measurement is performed after the adiabatic tuning process, rather than before it. In this case, one must consider all the fusion channels that occur in the initial superposition, and the corresponding low-energy Hamiltonian ${\bf H} = \bigoplus_{z \in \mathcal{Z}_{ab}} {\bf H}_{z}$, where
\begin{equation}
{\bf H}_{z} = \left[
\begin{array}{cc}
E_{z}^{(1)}+ E_{0}^{(2)} & \Gamma_{b}^{\ast} \\
\Gamma_{b} & E_{a}^{(1)} + E_{b}^{(2)} + E_{a,b;z}
\end{array}
\right]
.
\end{equation}
The different $z$-sectors do not need to be separated from each other by an energy gap, as they are prevented from mixing with each other by topological charge conservation. However, the process within each $z$-sector should be treated adiabatically, and separated from other states (with the same collective fusion channel $z$) in the spectrum by an energy gap. As such, I assume the final system configuration will have $E_{z}^{(1)} = \mathcal{E}_{z}$  for $z \in \mathcal{Z}_{ab}$ and $E_{x}^{(1)} \geq \Delta$ for $x \notin \mathcal{Z}_{ab}$ for site 1; and $E_{0}^{(2)} =0$ and $E_{x}^{(2)} \geq \Delta$  for $x \neq 0$ for site 2. In order to (nearly) maintain the gap, I let $0 \leq \mathcal{E}_{z} \ll \Delta$ (though it can be done in other ways).

The initial and final configurations for the adiabatic tuning step can be taken to be
\begin{eqnarray}
{\bf H}_{z}(t_{\rm i}) &=& \left[
\begin{array}{cc}
(2 - \delta_{az}) \Delta  & 0 \\
0 & 0
\end{array}
\right]
, \\
{\bf H}_{z}(t_{\rm fa}) &=& \left[
\begin{array}{cc}
\mathcal{E}_{z}  & 0 \\
0 &  2\Delta + (\mathcal{E}_{a} - \Delta) \, {\bf 1}_{\mathcal{Z}_{ab}}(a)
\end{array}
\right]
,
\end{eqnarray}
where ${\bf 1}_{\mathcal{Z}_{ab}}$ is the indicator function for the set $\mathcal{Z}_{ab}$.
Adiabatically tuning between these configurations will evolve the initial state $\left| \Psi(t_{\text{i}}) \right\rangle$ to the final state
\begin{equation}
\label{eq:fusion_mid_state}
\left| \Psi(t_{\text{fa}}) \right\rangle = \sum_{z \in  \mathcal{Z}_{ab}} \Psi_{z} e^{i \phi_{z}} \left| z, 0; z \right\rangle
.
\end{equation}
Here, the relative phases $e^{i \phi_{z}}$ arise as a result of the different energies $E_{z}^{(1)}$ and $E_{a,b;z}$ associated with the different fusion channels, and how these parameters are varied during the process. Their precise values may be computed, in principle, e.g. using Berry phase methods, but they will not be important for the full quasiparticle fusion process that includes a fusion channel measurement (which is, in fact, the next step). It is worth reemphasizing that the coherent superposition in Eq.~(\ref{eq:fusion_mid_state}) will only be maintained in an idealized system. The inclusion of realistic phenomena, such as local noise, will cause rapid decoherence of superpositions of different localized topological charges, so the final state would be expected to actually become a mixed state with density matrix
\begin{equation}
\label{eq:fusion_mid_mixed_state}
\rho(t_{\text{fa}})= \sum_{z \in \mathcal{Z}_{ab}} |\Psi_{z}|^{2} \left| z, 0; z \right\rangle \left\langle z, 0; z \right|
.
\end{equation}
The distinction between the coherent and incoherent states is not important in light of the following fusion channel measurement.

With a single quasiparticle on site 1, the collective topological charge of the two site subsystem is carried by the single quasiparticle, and so can be measured by a local probe. Performing the measurement of the topological charge of this quasiparticle, represented by the either the state in Eq.~(\ref{eq:fusion_mid_state}) or (\ref{eq:fusion_mid_mixed_state}), the measurement outcome will find charge $c$ with probability $p(c) = |\Psi_{c}|^{2}$, and (assuming projective measurement) will yield the corresponding post-measurement state $\left| c, 0; c \right\rangle$.

Following this topological charge measurement, it is desirable to tune the system parameters to single out topological charge $c$ as the only energetically favored state (below the gap) on site 1. In other words, the final configuration of the entire fusion process is taken to have $E_{c}^{(1)} =0$ and $E_{x}^{(1)} \geq \Delta$ for $x \neq c$ for site 1, and $E_{0}^{(2)} =0$ and $E_{x}^{(2)} \geq \Delta$  for $x \neq 0$ for site 2. This corresponds to the low-energy Hamiltonian
\begin{eqnarray}
{\bf H}_{z}(t_{\rm ff}) &=& \left[
\begin{array}{cc}
(1- \delta_{zc}) \Delta  & 0 \\
0 &  (2- \delta_{ac}) \Delta
\end{array}
\right]
.
\end{eqnarray}
The tuning in this step only needs to be adiabatic in the $c$-sector, as changing the sectors with collective charge $z \neq c$ are no longer pertinent to the post-measurement state.
In summary, this fusion process evolves (non-deterministically) from an initial two quasiparticle ground state $\left| \Psi(t_{\text{i}}) \right\rangle = \sum_{z \in  \mathcal{Z}_{ab}} \Psi_{z} \left| a, b; z \right\rangle$ to a final single quasiparticle ground state $\left| \Psi(t_{\text{ff}}) \right\rangle = \left| c, 0; c \right\rangle$ with probability $p(c) = |\Psi_{c}|^{2}$.

\section{Measurement-Only Braiding}
\label{sec:MObraiding}

In this appendix, I explain why measurement-only braiding applied to quasiparticles carrying distinct topological charges does not provide access to the same braiding properties as does transport. For this, I consider two anyons to be braided $a_1$ and $a_4$, using two ancillary anyons $a_2$ and $a_3$. Applying the measurement-only protocols~\cite{Bonderson08a,Bonderson08b,Bonderson12a} for this scenario amounts to generating the following sequence of projectors
\begin{eqnarray}
X &=& C \, \, \Pi^{(23)}_{b_{23}} \Pi^{(24)}_{b_{24}} \Pi^{(12)}_{b_{12}} \Pi^{(23)}_{b_{23}} \notag \\
&=& C^{\prime}
\psscalebox{0.8}{
\pspicture[shift=-3.7](-1.4,-1.0)(2.0,6.0)
  \psset{linewidth=0.9pt,linecolor=black,arrowscale=1.5,arrowinset=0.15}
  \psline(-0.8,-0.5)(-0.8,1.0)
  \psline(0.4,0.0)(0.4,0.5)
  \psline(0.4,0.5)(0.0,1.0)
  \psline(0.4,0.5)(0.8,1.0)
  \psline(0.4,0.0)(0.8,-0.5)
  \psline(0.4,0.0)(0.0,-0.5)
  \psline(1.6,-0.5)(1.6,1.0)
    \psline{->}(0.4,0.0)(0.4,0.375)
    \psline{->}(0.4,0.5)(0.1,0.875)
    \psline{->}(0.4,0.5)(0.7,0.875)
    \psline{-<}(0.4,0.0)(0.1,-0.375)
    \psline{-<}(0.4,0.0)(0.7,-0.375)
    \psline{-<}(-0.8,0.0)(-0.8,-0.375)
    \psline{-<}(1.6,0.0)(1.6,-0.375)
  \psline(-0.8,1.0)(-0.4,1.5)
  \psline(0.0,1.0)(-0.4,1.5)
  \psline(-0.4,1.5)(-0.4,2.0)
  \psline(-0.8,2.5)(-0.4,2.0)
  \psline(0.0,2.5)(-0.4,2.0)
  \psline(0.8,1.0)(0.8,2.5)
  \psline(1.6,1.0)(1.6,2.5)
    \psline{->}(-0.4,1.5)(-0.4,1.875)
    \psline{->}(-0.4,2.0)(-0.1,2.375)
    \psline{->}(-0.4,2.0)(-0.7,2.375)
  \psline(-0.8,2.5)(-0.8,4.0)
  \psline(0.0,2.5)(1.2,3.0)
  \psline(1.6,2.5)(1.2,3.0)
  \psline(1.2,3.0)(1.2,3.5)
  \psline(0.0,4.0)(1.2,3.5)
  \psline(1.6,4.0)(1.2,3.5)
    \psline{->}(1.2,3.5)(0.3,3.875)
    \psline{->}(1.2,3.5)(1.5,3.875)
    \psline{->}(1.2,3.0)(1.2,3.375)
  \psline[border=2.0pt](0.8,2.5)(0.8,4.0)
  \psline(-0.8,4.0)(-0.8,5.5)
  \psline(1.6,4.0)(1.6,5.5)
  \psline(0.4,4.5)(0.4,5.0)
  \psline(0.8,4.0)(0.4,4.5)
  \psline(0.0,4.0)(0.4,4.5)
  \psline(0.8,5.5)(0.4,5.0)
  \psline(0.0,5.5)(0.4,5.0)
    \psline{->}(0.4,4.5)(0.4,4.875)
    \psline{->}(0.4,5.0)(0.1,5.375)
    \psline{->}(0.4,5.0)(0.7,5.375)
    \psline{->}(-0.8,5.0)(-0.8,5.375)
    \psline{->}(1.6,5.0)(1.6,5.375)
  \rput[bl]{0}(-0.95,-0.8){$a_{1}$}
  \rput[bl]{0}(-0.25,-0.8){$a_{2}$}
  \rput[bl]{0}(0.7,-0.8){$a_{3}$}
  \rput[bl]{0}(1.5,-0.8){$a_{4}$}
  \rput[bl]{0}(0.55,0.05){$b_{23}$}
  \rput[bl]{0}(-0.25,0.6){$a_{2}$}
  \rput[bl]{0}(0.7,0.6){$a_{3}$}
  \rput[bl]{0}(-0.25,1.55){$b_{12}$}
  \rput[bl]{0}(-0.95,2.0){$a_{1}$}
  \rput[bl]{0}(-0.15,2.0){$a_{2}$}
  \rput[bl]{0}(1.35,3.0){$b_{24}$}
  \rput[bl]{0}(1.5,3.55){$a_{4}$}
  \rput[bl]{0}(0.05,3.55){$a_{2}$}
  \rput[bl]{0}(0.55,4.55){$b_{23}$}
  \rput[bl]{0}(-0.95,5.6){$a_{1}$}
  \rput[bl]{0}(-0.25,5.6){$a_{2}$}
  \rput[bl]{0}(0.7,5.6){$a_{3}$}
  \rput[bl]{0}(1.5,5.6){$a_{4}$}
 \endpspicture
}
\end{eqnarray}
where $C$ and $C^{\prime}$ are constants that give the proper normalizations. Projectors with the desired fusion channels may be generated in these protocols through the use of ``forced-measurements'' or tunable interactions.

For this sequence of projectors to reduce to a unitary operator acting on the state subspace of $a_1$ and $a_4$ anyons, the fusion channels $b_{12}$, $b_{23}$, and $b_{24}$ of the projectors must all be Abelian topological charges.  When these fusion channels are all Abelian, it implies that the localized quasiparticles' charges are all related to each other through fusion with Abelian charges, i.e. they are either the same or closely related topological charge types. Specifically, $a_1 = b_{12} \times \bar{a}_2$, $a_3 = b_{23} \times \bar{a}_2$, and $a_4 = b_{24} \times \bar{a}_2$. In this case, the resulting operator can be written as~\cite{Bonderson12a}
\begin{equation}
X = \hat{X}^{(14)} \otimes \Pi^{(23)}_{b_{23}}
,
\end{equation}
where the operator on anyons $1$ and $4$ is
\begin{eqnarray}
\hat{X}^{(14)} &=& e^{i \phi}
\pspicture[shift=-1.3](-1.4,-0.7)(2.0,2.0)
  \psset{linewidth=0.9pt,linecolor=black,arrowscale=1.5,arrowinset=0.15}
  \psline(-0.8,0.0)(1.6,1.5)
  \psline(1.6,0.0)(-0.8,1.5)
  \psline(-0.2,0.375)(-0.2,1.125)
  \psline(1.0,0.375)(1.0,1.125)
     \psline{->}(0.0,0.5)(0.8,1.0)
     \psline{->}(0.8,0.5)(0.0,1.0)
     \psline{->}(0.8,1.0)(1.4,1.375)
     \psline{->}(0.0,1.0)(-0.6,1.375)
     \psline{-<}(0.8,0.5)(1.4,0.125)
     \psline{-<}(0.0,0.5)(-0.6,0.125)
     \psline{->}(-0.2,0.375)(-0.2,0.875)
     \psline{->}(1.0,0.375)(1.0,0.875)
  \psline[border=2.0pt](0.0,0.5)(0.6,0.875)
  \rput[bl]{0}(-0.95,-0.3){$a_{1}$}
  \rput[bl]{0}(1.5,-0.3){$a_{4}$}
  \rput[bl]{0}(-0.95,1.6){$a_{1}$}
  \rput[bl]{0}(1.5,1.6){$a_{4}$}
  \rput[bl]{0}(-0.05,1.1){$\bar{a}_{2}$}
  \rput[bl]{0}(0.45,1.1){$\bar{a}_{2}$}
  \rput[bl]{0}(-0.8,0.65){$b_{12}$}
  \rput[bl]{0}(1.15,0.65){$b_{24}$}
 \endpspicture
\label{eq:X_1}
\\
&=& e^{i \phi^{\prime}}
\pspicture[shift=-1.3](-1.4,-0.7)(2.0,2.0)
  \psset{linewidth=0.9pt,linecolor=black,arrowscale=1.5,arrowinset=0.15}
  \psline(-0.8,0.0)(1.6,1.5)
  \psline(1.6,0.0)(-0.8,1.5)
  \psline(-0.2,0.375)(-0.2,1.125)
     \psline{->}(0.8,1.0)(1.4,1.375)
     \psline{->}(0.0,1.0)(-0.6,1.375)
     \psline{-<}(0.8,0.5)(1.4,0.125)
     \psline{-<}(0.0,0.5)(-0.6,0.125)
     \psline{->}(-0.2,0.375)(-0.2,0.875)
  \psline[border=2.0pt](0.0,0.5)(0.6,0.875)
  \rput[bl]{0}(-0.95,-0.3){$a_{1}$}
  \rput[bl]{0}(1.5,-0.3){$a_{4}$}
  \rput[bl]{0}(-0.95,1.6){$a_{1}$}
  \rput[bl]{0}(1.5,1.6){$a_{4}$}
  \rput[bl]{0}(-0.5,0.65){$g$}
 \endpspicture
\label{eq:X_2}
\\
\label{eq:X_3}
&=& e^{i \phi^{\prime \prime}} \sum_{c}  \left[ F^{a_{4} g a_{4} }_{c} \right]_{a_{1} a_{1}} R^{a_{1} a_{4}}_{c} \,\, \Pi^{(14)}_{c} \\
\label{eq:X_4}
&=& e^{i \phi } \sum_{c}  R^{\bar{a}_{2} \bar{a}_{2}}_{\hat{c} } \,\, \Pi^{(14)}_{c}
\end{eqnarray}
where $g = b_{12} \times \bar{b}_{24}$, $\hat{c} = c \times \bar{b}_{12} \times \bar{b}_{24}$, and $e^{i \phi}$, $e^{i \phi^{\prime}}$, and $e^{i \phi^{\prime \prime}}$ are unimportant overall phase factors (which may depend on $b_{12}$ and $b_{24}$). Thus, instead of generating the $R$-symbol $R^{a_{1} a_{4}}_{c}$ (or $R^{a_{4} a_{1}}_{c}$) associated with braiding anyons $a_1$ and $a_4$, the measurement-only protocol generates the $R$-symbols $R^{\bar{a}_{2} \bar{a}_{2}}_{\hat{c} }$, which are equivalent to a modification of $R^{a_{1} a_{4}}_{c}$ by the phase factors $\left[ F^{a_{4} g a_{4} }_{c} \right]_{a_{1} a_{1}}$.

This analysis shows that: (1) measurement-only methods of generating braiding transformations are only applicable to distinct topological charges if they are related through fusion with Abelian charges, and (2) when the topological charges are distinct, the resulting measurement-only generated transformation is not actually that of braiding the distinct topological charges, but rather of braiding related identical charges. Generally (though not always), this implies a restriction on the braiding properties that can be extracted from the experiments using measurement-only methods, as compared with what may be obtained by actually transporting the quasiparticles.

\section{Experiments Probing Topological Protection}
\label{sec:Rabi_experiment}

The setup of the fusion rules experiment of Sec.~\ref{sec:FusionExp} can be used to perform an anyonic version of a Rabi oscillation experiment. Such experiments may be used to verify that the system is operating in the topological limit by determining the energy splittings between degenerate ground states and extracting the related correlation lengths.
This is performed by the following steps:
\begin{enumerate}
  \item Pair-create quasiparticles carrying topological charges $\bar{a}$ and $a$  from vacuum, and move them apart.
  \item Pair-create quasiparticles carrying topological charges $b$ and $\bar{b}$  from vacuum, and move them apart.
  \item Measure the collective topological charge of the $a$-$b$ pair of quasiparticles.
  \item Wait for a time interval of length $t$.
  \item Go to step 3.
\end{enumerate}

In the ideal topological limit, as long as nothing else is done other than repeating the measurement, it should always return the same measurement outcome.
However, interactions between quasiparticles $a$ and $\bar{a}$ and/or between quasiparticles $b$ and $\bar{b}$ will split the energy degeneracy of the different fusion channels of these quasiparticle pairs.
Specifically, for topological charge values $e$ with $N_{\bar{a}a}^{e} \neq 0$ and $N_{b\bar{b}}^{\bar{e}} \neq 0$, the basis states $\left| \bar{a}, a ; e \right\rangle \left| b, \bar{b} ; \bar{e} \right\rangle \left| e, \bar{e}; 0 \right\rangle$ will have energies $E_{e}$.
Generically, the degeneracy between these different fusion channel states will be fully split.
The resulting time evolution of this energy splitting generates rotations between the different fusion channels of the $a$-$b$ pair, and will give rise to nonzero probabilities of observing different measurement outcomes, which depend on the energy splitting and time intervals between measurements.
In particular, the probability of obtaining a measurement outcome $c'$ for the $a$-$b$ pair following a previous measurement outcome of $c$, with time interval $t$ between measurements, is
\begin{equation}
p_{ab}(c'|c ; t) = \left| \sum_{e} \left[F_{b}^{\bar{a} a b}\right]_{ec} e^{-i E_{e} t} \left[F_{b}^{\bar{a} a b}\right]_{ec'}^{\ast}  \right|^{2}
.
\end{equation}
Thus, by repeating this experiment for different values of $t$, one can infer the energy differences between the different fusion channels $e$.

Moreover, by varying the distances separating quasiparticles, these experiments may also be used to extract the topological correlation lengths of the system.
The fusion channel energies may be written as~\cite{Bonderson09}
\begin{eqnarray}
E_{e} &=& \sum_{g \in \mathcal{C}_{\bar{a}a}} \left( \Gamma_{g}^{(\bar{a}a)} \left[ F^{\bar{a} g a}_{e} \right]_{\bar{a}a} + \Gamma_{g}^{(\bar{a}a)\ast} \left[ F^{\bar{a}ga}_{e} \right]^{\ast}_{\bar{a}a} \right)
\notag \\
&& +\sum_{h \in \mathcal{C}_{b\bar{b}}} \left( \Gamma_{h}^{(b\bar{b})} \left[ F^{bh\bar{b}}_{\bar{e}} \right]_{b\bar{b}} + \Gamma_{h}^{(b \bar{b})\ast} \left[ F^{bh\bar{b}}_{\bar{e}} \right]^{\ast}_{b\bar{b}} \right)
, \quad
\end{eqnarray}
where $\Gamma_{g}^{(\bar{a}a)}$ is the tunneling amplitude of a charge $g$ from the $\bar{a}$ quasiparticle to the $a$ quasiparticle, and similarly for $\Gamma_{h}^{(b\bar{b})}$. The restriction of tunneling charges $g$ to the set $\mathcal{C}_{\bar{a}a} = \{ x \in \mathcal{C} \, | \, N_{ax}^{a} \neq 0 \}$ is needed for the topological charges of the localized quasiparticles to remain fixed, and similarly for $h \in \mathcal{C}_{b\bar{b}}$.

When a pair of quasiparticles are separated by a distance $r$ that is large with respect to the correlation length, the tunneling amplitude of a topological charge $g$ between these quasiparticles is are exponentially suppressed as $\Gamma_{g} = O(e^{- r/\xi_{g}})$. Thus, by repeating this experiment for different values of $r$, one may extract the correlation lengths $\xi_{g}$.

%

\end{document}